\documentclass[a4paper,12pt,openright]{book}
\usepackage[utf8]{inputenc}
\usepackage[margin=1.05in]{geometry}
\usepackage{amsthm,amssymb,amsmath}
\usepackage{enumitem}
\usepackage{graphicx}
\usepackage{algorithm}
\usepackage{listings}
\usepackage{placeins}
\usepackage{wrapfig}
\floatname{algorithm}{Listing}
\usepackage{tikz}
\usepackage{pgfplots}
\usetikzlibrary{shapes,snakes,automata,arrows}
\usetikzlibrary{external}
\usetikzlibrary{patterns}
\newcommand{\force}{\tikzset{external/force remake}}
\newcommand{\unforce}{\tikzset{external/force remake=false}}
\tikzset{defnode/.style={draw, circle, fill, black!80, inner sep=0, minimum size=3pt}}
\tikzset{mis/.style={draw, circle, red, inner sep=1.5pt}}
\tikzset{circ/.style={fill=none,draw, circle, red, semithick, inner sep=3pt}}
\definecolor{darkgreen}{rgb}{0,0.5,0}
\definecolor{darkdarkgray}{gray}{0.20}
\definecolor{red}{rgb}{0.6,0,0} 
\graphicspath{{./figs-finding/}}

\usepackage{varioref}
\usepackage{hyperref}
\hypersetup{
    colorlinks,
    citecolor=blue,
    filecolor=blue,
    linkcolor=black,
    urlcolor=yellow
}
\usepackage{minitoc}
\usepackage{xcolor}
\usepackage{xspace}
\usepackage{multirow}
\usepackage{subfiles}
\usepackage{subfigure}
\usepackage{mathrsfs}

\usepackage{varwidth}
\usepackage{setspace}

\newcommand{\tclocks}{{\sc \small T-Clocks}\xspace}

\newcommand{\C}{\ensuremath{\mathcal{C}}}
\newcommand{\F}{\ensuremath{\mathcal{F}}}

\newcommand{\G}{\ensuremath{\mathcal{G}}\xspace}
\newcommand{\J}{\ensuremath{\mathcal{J}}\xspace}

\newcommand{\T}{\ensuremath{\mathcal{T}}}
\newcommand{\lifetime}{\T\xspace}

\newcommand{\temp}[1]{\hat{#1}}

\newcommand{\leadstoo}{\overset{st}{\leadsto}}
\newcommand{\roundby}{
    \begin{tikzpicture}
      \draw[->,rounded corners=0pt] (0,0)--(.12,.06)--(.22,0)--(.22,-.08)--(.14,-.12)--(.1,-.08)--(0,-.08);
      \path (.27,0) coordinate (bidon);
    \end{tikzpicture}
}
\newcommand{\circlenode}{
    \begin{tikzpicture}
      \path (-.05,-.09) coordinate (bidon);
      \path (.05,.09) coordinate (bidon);
      \path (0,0) node[draw,circle,fill,inner sep=.8pt] {};
    \end{tikzpicture}
}

\newcommand{\PP}{\ensuremath{\mathscr{P}}\xspace}
\newcommand{\tdbroadcast}{{\textsc{TDB}}\xspace}
\newcommand{\foremost}{{\textsc{TDB}}\ensuremath{[foremost]}\xspace}
\newcommand{\fastest}{{\textsc{TDB}}\ensuremath{[fastest]}\xspace}
\newcommand{\shortest}{{\textsc{TDB}}\ensuremath{[shortest]}\xspace}

\newcommand{\TVG}{\ensuremath{{\G=(V,E,\T,\rho,\zeta)}}\xspace}

\newcommand{\TVGA}{\ensuremath{{\cal A}}}

\newcommand{\vs}{{\em vs.}\xspace}
\newcommand{\ie}{{\em i.e.,}\xspace}

\newcommand{\eg}{{\em e.g.,}\xspace}

\newcommand{\forallMIS}{\ensuremath{{\cal RMIS^\forall}}\xspace}
\newcommand{\existsMIS}{\ensuremath{{\cal RMIS^\exists}}\xspace}

\newcommand{\ER}{\ensuremath{{\cal E^R}}\xspace}
\newcommand{\EB}{\ensuremath{{\cal E^B}}\xspace}
\newcommand{\EP}{\ensuremath{{\cal E^P}}\xspace}
\newcommand{\ERrep}{\ensuremath{\circlenode$$\overset{{\cal R}}{\mbox{--}}$$\circlenode}\xspace}
\newcommand{\EBrep}{\ensuremath{\circlenode$$\overset{{\cal B}}{\mbox{--}}$$\circlenode}\xspace}
\newcommand{\EPrep}{\ensuremath{\circlenode$$\overset{{\cal P}}{\mbox{--}}$$\circlenode}\xspace}
\newcommand{\JOA}{\ensuremath{{\cal J}^{1\forall}}\xspace}
\newcommand{\JOArep}{\ensuremath{1$$\leadsto$$*}\xspace}
\newcommand{\JAO}{\ensuremath{{\cal J}^{\forall 1}}\xspace}
\newcommand{\JAOrep}{\ensuremath{*$$\leadsto$$1}\xspace}
\newcommand{\JOAST}{\ensuremath{{\cal J}^{1\forall{>}}}\xspace}
\newcommand{\JOASTrep}{\ensuremath{1$$\leadstoo$$*}\xspace}
\newcommand{\JAOST}{\ensuremath{{\cal J}^{\forall1{>}}}\xspace}
\newcommand{\JAOSTrep}{\ensuremath{*$$\leadstoo$$1}\xspace}
\newcommand{\RT}{\ensuremath{{\cal TC^{\circlearrowleft}}}\xspace}
\newcommand{\RTrep}{\ensuremath{*\,\roundby\,*}\xspace}
\newcommand{\RTST}{\ensuremath{{\cal TC^{\circlearrowleft >}}}\xspace}

\newcommand{\TC}{\ensuremath{{\mathcal{TC}}}\xspace}
\newcommand{\TCrep}{\ensuremath{*$$\leadsto$$*}\xspace}
\newcommand{\TCST}{\ensuremath{{\mathcal{TC}^>}}\xspace}
\newcommand{\TCSTrep}{\ensuremath{*$$\leadstoo$$*}\xspace}
\newcommand{\TCR}{{\ensuremath{{\cal TC^{\cal R}}}}\xspace}
\newcommand{\TCB}{{\ensuremath{{\cal TC^{\cal B}}}}\xspace}
\newcommand{\TCRrep}{\ensuremath{*$$\overset{{\cal R}}{\leadsto}$$*}\xspace}
\newcommand{\TCBrep}{\ensuremath{*$$\overset{{\cal B}}{\leadsto}$$*}\xspace}
\newcommand{\AC}{\ensuremath{{\cal C^*}}\xspace}
\newcommand{\ACrep}{\ensuremath{*$$\overset{*}{\mbox{--\hspace{1pt}--}}$$*}\xspace}
\newcommand{\TINT}{\ensuremath{{\cal C}^{\cap}}\xspace}
\newcommand{\TINTrep}{\ensuremath{${\footnotesize T-}$*$$\overset{*}{\mbox{--\hspace{1pt}--}}$$*}\xspace}
\newcommand{\PR}{\ensuremath{\cal P^R}\xspace}
\newcommand{\PRrep}{\ensuremath{\circlenode$$\overset{{\cal R}}{\mbox{--\hspace{1pt}--}}$$\circlenode}\xspace}
\newcommand{\CR}{\ensuremath{{\cal C^R}}\xspace}
\newcommand{\CRrep}{\ensuremath{*$$\overset{{\cal R}}{\mbox{--\hspace{1pt}--}}$$*}\xspace}
\newcommand{\KG}{\ensuremath{{\cal K}}\xspace}
\newcommand{\KGrep}{\ensuremath{*$--$*}\xspace}
\newcommand{\EOA}{\ensuremath{{\cal E}^{1\forall}}\xspace}
\newcommand{\EOArep}{\ensuremath{1$--$*}\xspace}
\newcommand{\KR}{\ensuremath{{\cal K^R}}\xspace}
\newcommand{\KRrep}{\ensuremath{*$$\overset{{\cal R}}{\mbox{--}}$$*}\xspace}

\newenvironment{myitem}{\vspace{-5pt}\itemize\addtolength{\itemsep}{-5pt}}{\enditemize\vspace{-5pt}}

\newcommand{\boite}[2]{
  \begin{discussion}[#1]
    \textup{#2}
  \end{discussion}
}
\newcommand{\IGNORE}[1]{}
\newcommand{\article}[1]{\vspace{-6pt}\begin{flushright}{\it Main article: #1}\end{flushright}\vspace{-6pt}}
\newcommand{\articles}[1]{\vspace{-6pt}\begin{flushright}{\it Main articles: #1}\end{flushright}\vspace{-6pt}}

\newcommand{\mypar}[1]{\noindent{\it #1.}}

\newtheorem{definition}{Definition}
\newtheorem{discussion}{Discussion}

\newtheorem{remark}{Remark}

\newtheorem{open}{Open question}
\newtheorem{avenue}[open]{Research avenue}

\newtheoremstyle{named}{}{}{\itshape}{}{\bfseries}{.}{.5em}{#1 \thmnote{#3}}
\theoremstyle{named}
\newtheorem*{class*}{Class}
\newtheorem*{subclass*}{Subclasses of}

\pgfdeclareimage[width=.8cm]{smartphone}{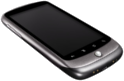}
\pgfdeclareimage[width=.4cm]{smartphone-small}{./figs-finding/smartphone.png}
\pgfdeclareimage[width=.9cm]{voiture}{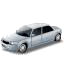}
\pgfdeclareimage[width=.45cm]{voiture-small}{./figs-finding/voiture.png}
\pgfdeclareimage[width=.8cm]{satellite}{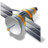}
\pgfdeclareimage[width=.4cm]{satellite-small}{./figs-finding/satellite.png}
\pgfdeclareimage[width=.8cm]{robot}{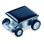}
\pgfdeclareimage[width=.4cm]{robot-small}{./figs-finding/robot.png}
\pgfdeclareimage[width=.9cm]{drone}{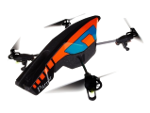}
\pgfdeclareimage[width=.45cm]{drone-small}{./figs-finding/drone.png}
\pgfdeclareimage[width=.58cm]{soldier}{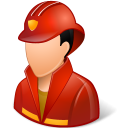}
\pgfdeclareimage[width=.3cm]{soldier-small}{./figs-finding/soldier.png}
\pgfdeclareimage[width=.9cm]{duck}{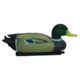}
\pgfdeclareimage[width=.45cm]{duck-small}{./figs-finding/duck.png}

\def\wgraph (#1,#2){%
  \tikzstyle{every node}=[draw,fill,circle,inner sep=#1]
  \path (0,0) node (c){};
  \path (c)+(-1,.4) node (a){};
  \path (c)+(1,.4) node (e){};
  \path (c)+(-.6,-.6) node (b){};
  \path (c)+(.6,-.6) node (d){};
  \tikzstyle{every node}=[font=\footnotesize]
  \path (a)+(-#2,0) node (la){$a$};
  \path (b)+(-#2,0) node (lb){$b$};
  \path[above] (c) node (lc){$c$};
  \path (d)+(#2,0) node (ld){$d$};
  \path (e)+(#2,0) node (le){$e$};
}
\def\wgraphanonymous (#1){%
  \tikzstyle{every node}=[draw,fill,circle,inner sep=#1]
  \path (0,0) node (c){};
  \path (c)+(-1,.4) node (a){};
  \path (c)+(1,.4) node (e){};
  \path (c)+(-.6,-.6) node (b){};
  \path (c)+(.6,-.6) node (d){};
}

\begin{document}

\title{Finding Structure in Dynamic Networks\thanks{This document is the first part of the author's habilitation thesis (HDR)~\cite{Cas18-hdr}, defended on June 4, 2018 at the University of Bordeaux. Given the nature of this document, the contributions that involve the author have been emphasized; however, these four chapters were specifically written for distribution to a larger audience. We hope they can serve as a broad introduction to the domain of highly dynamic networks, with a focus on temporal graph concepts and their interaction with distributed computing.}}
\author{{\Large Arnaud Casteigts}\smallskip\\LaBRI, University of Bordeaux\\{\tt arnaud.casteigts@labri.fr}}
\date{June 4, 2018\\\vspace{3cm}~\\}
\maketitle
\pagenumbering{gobble}

\newpage~\newpage

\pagenumbering{arabic}

\setstretch{1.1}
\setcounter{tocdepth}{1}
\tableofcontents
\setstretch{1}

\dominitoc

\chapter{Dynamic networks?}
\label{chap:dynamic-networks}

Behind the terms ``dynamic networks'' lies a rich diversity of contexts ranging from near-static networks with occasional changes, to networks where changes occur continuously and unpredictably. In the past two decades, these highly-dynamic networks gave rise to a profusion of research activities, resulting (among others) in new concepts and representations based on graph theory. This chapter reviews some of these emerging notions, with a focus on our own contributions. The content also serves as a definitional resource for the rest of the document, restricting ourselves mostly to the notions effectively used in the subsequent chapters. 

\section{A variety of contexts}
A network is traditionally defined as a set of entities together with their mutual relations. It is dynamic if these relations {\em change} over time.
There is a great variety of contexts in which this is the case. Here, we mention two broad categories, communication networks and complex systems, which despite an important overlap, usually capture different (and complementary) motivations.

\mypar{(Dynamic) communication networks} These networks are typically made of wireless-enabled entities ranging from smartphones to laptops, to drones, robots, sensors, vehicles, satellites, {\it etc.} As the entities move, the set of communication links evolve, at a rate that goes from occasional (\eg laptops in a managed Wi-Fi network) to nearly unrestricted (\eg drones and robots). Networks with communication faults also fall in this category, albeit with different concerns (on which we shall return later).

\mypar{(Dynamic) complex networks} This category comprises networks from a larger range of contexts such as social sciences, brain science, biology, transportation, and communication networks (as a particular case). 
An extensive amount of real-world data is becoming available in all these areas, making it possible to characterize various phenomena.

While the distinction between both categories may appear somewhat artificial, it shed some light to the typically different motivations underlying these areas. In particular, research in communication networks is mainly concerned with what can be done {\em from within} the network through distributed interactions among the entities, while research in complex networks is mainly concerned with defining mathematical models that capture, reproduce, or predict phenomena observed in reality, based mostly on the analysis of available data in which {\em centralized} algorithms play the key role. 

Interactions between both are strong and diverse. However, it seems that for a significant period of time, both communities (and sub-communities) remained essentially unaware of their respective effort to develop a conceptual framework to capture the network dynamics using graph theoretical concepts. In way of illustration, let us mention the diversity of terminologies used for even the most basic concepts in dynamic networks, \eg that of {\em journey}~\cite{BFJ03}, also called {\em schedule-conforming path}~\cite{Berman96}, {\em time-respecting path}~\cite{KKK00}, and {\em temporal path}~\cite{ChMMD08,TSM+09}; and that of {\em temporal distance}~\cite{BFJ03}, also called~{\em reachability time} \cite{Holme05}, {\em information latency} \cite{KKW08}, {\em propagation speed} \cite{JMR10} and  {\em temporal proximity}~\cite{Kostakos09}.

In the rest of this chapter, we review some of these efforts in a chronological order. Next, we present some of the main temporal concepts identified in the literature, with a focus on the ones used in this document. The early identification of these concepts, in a unifying attempt, was one of the components of our most influencial paper so far~\cite{CFQS12}. Finally, we present a general discussion as to some of the ways these new concepts impact the definition of combinatorial (or distributed) problems classically studied in static networks, some of which are covered in more depth in subsequent chapters.

\section{Graph representations}
\label{sec:models}

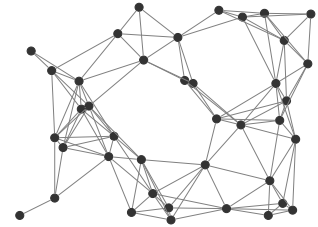
\begin{wrapfigure}{R}{5cm}
  \centering
  \vspace{2pt}
    \tikzsetnextfilename{tikz-udg}
    \begin{tikzpicture}[xscale=.5,yscale=.5]
      \tikzstyle{every node}=[defnode]
      \path (1.84,6.38) node (v6) {};
      \path (1.2,4.88) node (v3) {};
      \path (4.62,6.4) node (v20) {};
      \path (2.76,4.92) node (v10) {};
      \path (1.42,4.62) node (v5) {};
      \path (4.84,6.32) node (v21) {};
      \path (7.86,6.84) node (v38) {};
      \path (7.54,4.84) node (v37) {};
      \path (4.2,3.02) node (v17) {};
      \path (5.46,5.38) node (v23) {};
      \path (7.94,8.16) node (v39) {};
      \path (6.72,8.18) node (v28) {};
      \path (7.3,5.86) node (v35) {};
      \path (2.1,5.72) node (v8) {};
      \path (6.82,2.82) node (v29) {};
      \path (3.42,8.34) node (v13) {};
      \path (6.14,8.08) node (v27) {};
      \path (2.86,7.64) node (v11) {};
      \path (1.9,5.64) node (v7) {};
      \path (6.86,3.74) node (v30) {};
      \path (5.52,8.26) node (v24) {};
      \path (0.28,2.82) node (v0) {};
      \path (7.24,7.46) node (v34) {};
      \path (7.2,3.14) node (v33) {};
      \path (7.12,5.34) node (v32) {};
      \path (3.48,4.3) node (v14) {};
      \path (1.12,6.66) node (v2) {};
      \path (3.78,3.4) node (v16) {};
      \path (7.46,2.96) node (v36) {};
      \path (4.26,2.7) node (v18) {};
      \path (0.58,7.18) node (v1) {};
      \path (1.2,3.28) node (v4) {};
      \path (4.44,7.54) node (v19) {};
      \path (3.54,6.94) node (v15) {};
      \path (3.22,2.9) node (v12) {};
      \path (2.62,4.38) node (v9) {};
      \path (5.16,4.16) node (v22) {};
      \path (6.1,5.22) node (v26) {};
      \path (5.72,3.0) node (v25) {};
      \path (7.02,6.32) node (v31) {};

      \tikzstyle{every path}=[very thin, gray];
      \draw (v6)--(v3);
      \draw (v6)--(v10);
      \draw (v3)--(v10);
      \draw (v6)--(v5);
      \draw (v3)--(v5);
      \draw (v10)--(v5);
      \draw (v20)--(v21);
      \draw (v25)--(v17);
      \draw (v20)--(v23);
      \draw (v21)--(v23);
      \draw (v38)--(v39);
      \draw (v38)--(v28);
      \draw (v39)--(v28);
      \draw (v38)--(v35);
      \draw (v37)--(v35);
      \draw (v23)--(v35);
      \draw (v6)--(v8);
      \draw (v3)--(v8);
      \draw (v10)--(v8);
      \draw (v5)--(v8);
      \draw (v25)--(v29);
      \draw (v39)--(v27);
      \draw (v28)--(v27);
      \draw (v6)--(v11);
      \draw (v13)--(v11);
      \draw (v20)--(v26);
      \draw (v21)--(v26);
      \draw (v37)--(v26);
      \draw (v23)--(v26);
      \draw (v35)--(v26);
      \draw (v6)--(v7);
      \draw (v3)--(v7);
      \draw (v10)--(v7);
      \draw (v5)--(v7);
      \draw (v8)--(v7);
      \draw (v25)--(v30);
      \draw (v37)--(v30);
      \draw (v29)--(v30);
      \draw (v26)--(v30);
      \draw (v28)--(v24);
      \draw (v27)--(v24);
      \draw (v38)--(v34);
      \draw (v39)--(v34);
      \draw (v28)--(v34);
      \draw (v35)--(v34);
      \draw (v27)--(v34);
      \draw (v24)--(v34);
      \draw (v25)--(v33);
      \draw (v37)--(v33);
      \draw (v29)--(v33);
      \draw (v30)--(v33);
      \draw (v38)--(v32);
      \draw (v37)--(v32);
      \draw (v23)--(v32);
      \draw (v35)--(v32);
      \draw (v26)--(v32);
      \draw (v30)--(v32);
      \draw (v10)--(v14);
      \draw (v17)--(v14);
      \draw (v8)--(v14);
      \draw (v6)--(v2);
      \draw (v3)--(v2);
      \draw (v8)--(v2);
      \draw (v11)--(v2);
      \draw (v7)--(v2);
      \draw (v25)--(v16);
      \draw (v10)--(v16);
      \draw (v17)--(v16);
      \draw (v14)--(v16);
      \draw (v27)--(v31);
      \draw (v38)--(v31);
      \draw (v37)--(v31);
      \draw (v23)--(v31);
      \draw (v28)--(v31);
      \draw (v35)--(v31);
      \draw (v26)--(v31);
      \draw (v34)--(v31);
      \draw (v32)--(v31);
      \draw (v25)--(v36);
      \draw (v37)--(v36);
      \draw (v29)--(v36);
      \draw (v30)--(v36);
      \draw (v33)--(v36);
      \draw (v25)--(v18);
      \draw (v17)--(v18);
      \draw (v14)--(v18);
      \draw (v16)--(v18);
      \draw (v6)--(v1);
      \draw (v2)--(v1);
      \draw (v3)--(v4);
      \draw (v5)--(v4);
      \draw (v0)--(v4);
      \draw (v20)--(v19);
      \draw (v21)--(v19);
      \draw (v13)--(v19);
      \draw (v27)--(v19);
      \draw (v11)--(v19);
      \draw (v24)--(v19);
      \draw (v6)--(v15);
      \draw (v20)--(v15);
      \draw (v21)--(v15);
      \draw (v8)--(v15);
      \draw (v13)--(v15);
      \draw (v11)--(v15);
      \draw (v19)--(v15);
      \draw (v17)--(v12);
      \draw (v14)--(v12);
      \draw (v16)--(v12);
      \draw (v18)--(v12);
      \draw (v3)--(v9);
      \draw (v10)--(v9);
      \draw (v5)--(v9);
      \draw (v8)--(v9);
      \draw (v7)--(v9);
      \draw (v14)--(v9);
      \draw (v16)--(v9);
      \draw (v4)--(v9);
      \draw (v12)--(v9);
      \draw (v25)--(v22);
      \draw (v17)--(v22);
      \draw (v23)--(v22);
      \draw (v26)--(v22);
      \draw (v30)--(v22);
      \draw (v14)--(v22);
      \draw (v16)--(v22);
      \draw (v18)--(v22);
    \end{tikzpicture}
  \caption{\label{fig:udg}A graph depicting communication links in a wireless network.}
\end{wrapfigure} 
\paragraph{Standard graphs.} The structure of a network is classically given by a {\em graph}, which is a set of vertices (or nodes) $V$ and a set of edges (links) $E \subseteq V\times V$, \ie pairs of nodes. 
The edges, and by extension the graph itself, may be {\em directed} or {\em undirected}, depending on whether the relations are unidirectional or bidirectional (symmetrical). Figure~\ref{fig:udg} shows an example of undirected graph. Graph theory is a central topic in discrete mathematics. We refer the reader to standard books for a general introduction (\eg \cite{Trudeau13}), and we assume some acquaintance of the reader with basic concepts like paths, distance, connectivity, trees, cycles, cliques, {\it etc.}

\subsection{Graph models for dynamic networks}
\label{sec:graph-models}
\begin{figure}[h]
  \centering
\tikzset{leg/.style={very thick,red,->,shorten >=1pt,shorten <=1pt}}
\tikzsetnextfilename{tikz-G}
\begin{tikzpicture}
  \path (0,0) node (seq){
    \begin{minipage}[c]{9.2cm}
      \begin{tabular}{@{}c@{~~~}|@{~~~}c@{~~~}|@{~~~}c@{~~~}|@{~~~}c@{~}}
        \tikzsetnextfilename{tikz-G0}
        \begin{tikzpicture}[scale=.8]
          \wgraphanonymous (.6pt)
          \tikzstyle{every node}=[font=\footnotesize]
          \draw (a)--(c);
          \draw (b)--(c);
          \draw (b)--(d);
          \draw (c)--(d);
          \path (c)+(0,-1.1) node {$G_0$};
        \end{tikzpicture}
        &
        \tikzsetnextfilename{tikz-G1}
        \begin{tikzpicture}[scale=.8]
          \wgraphanonymous (.6pt)
          \tikzstyle{every node}=[font=\scriptsize]
          \draw (a)--(b);
          \draw (b)--(c);
          \draw (c)--(d);
          \path (c)+(0,-1.1) node {$G_1$};
        \end{tikzpicture}
        &
        \tikzsetnextfilename{tikz-G2}
        \begin{tikzpicture}[scale=.8]
          \wgraphanonymous (.6pt)
          \tikzstyle{every node}=[font=\scriptsize]
          \draw (a)--(b);
          \draw (c)--(d);
          \draw (c)--(e);
          \draw (d)--(e);
          \path (c)+(0,-1.1) node {$G_2$};
        \end{tikzpicture}
        &
        \tikzsetnextfilename{tikz-G3}
        \begin{tikzpicture}[scale=.8]
          \wgraphanonymous (.6pt)
          \tikzstyle{every node}=[font=\scriptsize]
          \draw (c)--(e);
          \draw (d)--(e);
          \path (c)+(0,-1.1) node {$G_3$};
        \end{tikzpicture}
      \end{tabular}
      \smallskip
    \end{minipage}
  };
  \draw[->,shorten <=1cm,shorten >=1cm] (seq.south west)--(seq.south east) node[right=-20pt] {time};
\end{tikzpicture}
\caption{\label{fig:sequence}A dynamic network seen as a sequence of graphs}
\end{figure}
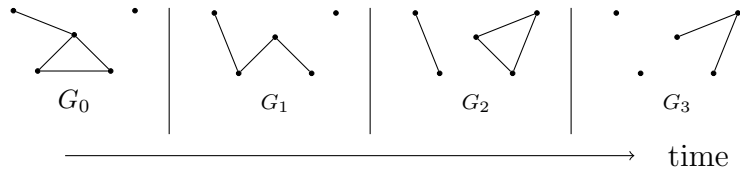

Dynamic networks can be modeled in a variety of ways using graphs and related notions. Certainly the first approach that comes to mind is to represent a dynamic network as a {\em sequence} of standard (static) graphs as depicted in Figure~\ref{fig:sequence}. Each graph of the sequence ({\em snapshot}, in this document) represents the relations among vertices at a given discrete time. This idea is natural, making it difficult to trace back its first occurrence in the literature. Sequences of graphs were studied (at least) back in the 1980s (see, \eg~\cite{SE81,frederickson85,reif87,eppstein90}) and simply called dynamic graphs. However, in a subtle way, the graph properties considered in these studies still refer to individual snapshots, and typical problems consist of updating standard information about the current snapshot, like connectivity~\cite{reif87}.

A {\em conceptual shift} occurred when researchers started to consider the whole sequence as being {\em the} graph. Then, properties of interest became related to temporal features of the network rather than to its punctual structure. In the distributed computing community, one of the first work in this direction was that of Awerbuch and Even in 1984~\cite{AE84}, who considered the broadcasting problem in a network that satisfies a temporal version of connectivity. On the modeling side, one of the first work considering explicitly a graph sequence as being {\em the} graph seem to be due to Harary and Gupta~\cite{HG97}. Then, a number of graph models were introduced to describe dynamic networks, reviewed here in chronological order.

In 2000, Kempe, Kleinberg, and Kumar~\cite{KKK00} defined a model called {\em temporal network}, where the network is represented by a graph $G$ (the {\em footprint}, in this document) together with a labeling function $\lambda: E\to \mathbb{R}$ that associates to every edge a number indicating when the two corresponding endpoints communicate. The model is quite basic: the communication is punctual and a single time exist for every edge. Both limitations can be circumvented by considering multiple edges and artificial intermediate vertices slowing down communication.

Independently, Ferreira and Viennot~\cite{FV02} represented a dynamic network by a graph $G$ together with a matrix that describes the {\em presence schedule} of edges (the same holds for vertices). The matrix is indexed in one dimension by the edges and in the other by integers that represent time. The authors offer an alternative point of view as a sequence of graphs where each graph $G_i(V_i,E_i)$ correspond to the $1$-entries of the matrix, this sequence being called an {\em evolving graph}, and eventually renamed as {\em untimed} evolving graph in subsequent work (see below).

\boite{Varying the set of vertices}{
\label{disc:varying-nodes}
Some graph models consider a varying set of vertices in addition to a varying set of edges, some do not. From a formal point of view, all models can easily be adapted to fit in one or the other category. In the present document, we are mostly interested in {\em edge} dynamics, thus we will ignore this distinction most of the time (unless it matters in the context). Also note that an absent node could sometimes be simulated by an isolated node using only edge dynamics.
}

A similar statement as that of Discussion~\ref{disc:varying-nodes} could be made for {\em directed} edges versus {\em undirected} edges. We call on the reader's flexibility to ignore non essential details in the graph models when these details are not important. We state them explicitly when they are.

Ferreira and his co-authors~\cite{Fer02,BFJ03,Fer04} further generalize evolving graphs by considering a sequence of graph $G_i=\{G_1, G_2, ...\}$ where each $G_i$ may span a {\em non unitary} period of time. The duration of each $G_i$ is encoded in an auxiliary table of times $t_1, t_2, ...$ such that every $G_i$ spans the period $[t_i,t_{i+1})$.\footnote{For French readers: $[a,b)$ is an equivalent notation to $[a,b[$.}
A consequence is that the times can now be taken from the real domain, with mild restrictions pertaining to theoretical limitations (\eg countability or accumulation points). In order to disambiguate both versions of evolving graphs, the earlier version from~\cite{FV02} is qualified as {\em untimed} evolving graphs.
Another contribution of~\cite{BFJ03} is to incorporate the latency (\ie time it takes to cross a given edge at a given time) through a function $\zeta$ called {\em traversal time} in~\cite{BFJ03}.

In 2009, Kostakos~\cite{Kostakos09} describes a model called {\em temporal graphs}, in which the whole dynamic graph is encoding into a single static (directed) graph, built as follows. Every entity (vertex) is duplicated as many times as there are time steps, then directed edges are added between consecutive copies of the same node. One advantage of this representation is that some temporal features become immediately available on the graph, \eg directed paths correspond to non-strict journeys (defined further down). Recently, the term ``temporal graph'' has also been used to refer to the temporal networks of~\cite{KKK00}, with some adaptations (see \eg~\cite{AGMS15}). This latter usage of the term seems to become more common than the one from~\cite{Kostakos09}.

In the complex network community, where researchers are concerned with the effective manipulation of data, different models of dynamic networks have emerged whose purpose is essentially to be used as a {\em data structure} (as opposed to being used only as a descriptive language, see Discussion~\ref{disc:models} below). In this case, the network history is typically recorded as a sequence of timed links (\eg time-stamped e-mail headers in~\cite{KKW08}) also called link streams. The reader is referred to~\cite{Graphstream,KKW08,WDCG12,H12} for examples of use of these models, and to~\cite{LVM17} for a recent survey.

As part of a unifying effort with Flocchini, Quattrociocchi, and Santoro from 2009 to 2012, we reviewed a vast body of research in dynamic networks~\cite{CFQS12}, harmonizing concepts and attempting to bridge the gap between the two (until then) mostly disjoint communities of complex systems and networking (distributed computing). Taking inspiration from several of the above models, we defined a formalism called TVG (for {\it time-varying graphs}) whose purpose was to favor expressivity and elegance over all practical aspects such as implementability, data structures, and efficiency. The resulting formalism is free from intrinsic limitation.

Let $V$ be a set of entities (vertices, nodes) and $E$ a set of relations (edges, links) between them. These relations take place over an interval $\T$ called the {\em lifetime} of the network, being a subset of $\mathbb{N}$ (discrete) or $\mathbb{R^+}$ (continuous), and more generally some time domain $\mathbb{T}$. In its basic version, a {\em time-varying graph} is a tuple \TVG such that

\begin{myitem}
\item $\rho: E \times \T \rightarrow \{0,1\}$, called {\em presence} function, indicates if a given edge is available at a given time.
\item $\zeta: E \times \T \rightarrow \mathbb{T}$, called {\em latency} function, indicates the time it takes to cross a given edge at a given start time (the latency of an edge could itself vary in time).
\end{myitem}

The latency function is optional and other functions could equally be added, such as a {\em node presence} function $\psi: V \times \T \rightarrow \{0,1\}$, a {\em node latency} function $\varphi: V\times \T \rightarrow \mathbb{T}$ (accounting \eg for local processing times), {\it etc.}

\boite{Model, formalism, language}{
\label{disc:models}
TVGs are often referred to as a {\em formalism}, to stress the fact that using it does not imply restrictions on the environment, as opposed to the usual meaning of the word {\em model}. However, the notion of formalism has a dedicated meaning in mathematics related to formal logic systems. With hindsight, we see TVGs essentially as a {\em descriptive language}. In this document, the three terms are used interchangeably.
}

The purely descriptive nature of TVGs makes them quite general and allows temporal properties to be easily expressible. In particular, the presence and latency functions are not {\it a priori} constrained and authorize theoretical constructs like accumulation points or uncountable 0/1 transitions. While often not needed in complex systems and offline analysis, this generality is relevant in distributed computing, \eg to characterize the power of an adversary controlling the environment (see~\cite{CFGSY15} and Section~\ref{sec:expressivity} for details).

\begin{figure}[h]
  \centering
  \tikzsetnextfilename{tikz-example}
    \begin{tikzpicture}[scale=3.8]
    \tikzstyle{every node}=[font=\scriptsize]
    \wgraph(.1pt,2pt)
    \draw (a)--(b);
    \draw (a)--(c);
    \draw (b)--(c);
    \draw (b)--(d);
    \draw (c)--(d);
    \draw (c)--(e);
    \draw (d)--(e);
    \tikzstyle{every node}=[font=\scriptsize,sloped, inner sep=1pt]
    \draw (a)--node[below]{$\{t \in \mathbb{R}: \lfloor t\rfloor$ prime$\}$}(b);
    \draw (b)--node[below=2pt]{$[0,1]\cup [2,5]$}(d);
    \draw (d)--node[below]{$[1,\pi]$}(e);
    \draw (a)--node[above,inner sep=1.5pt]{$[5,7]$}(c);
    \draw (b)--node[above]{$[99999,\infty)$}(c);
    \draw (c)--node[above=1pt]{$[0,\infty)$}(d);
    \draw (c)--node[above]{$\{[i, i+2] : i \mod 3 = 0\}$}(e);
  \end{tikzpicture}

  \caption{\label{fig:example-graph} A dynamic network represented by a labeled graph. The labels indicates when the corresponding edges are present.}
\end{figure}
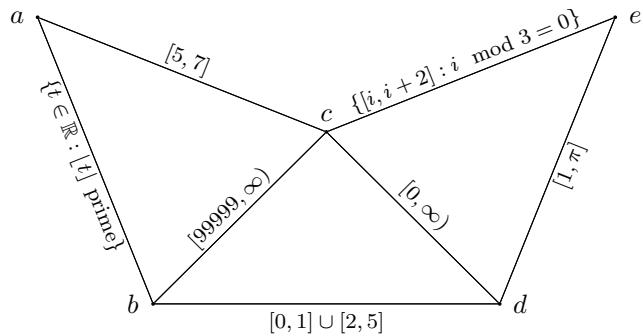

\paragraph{Visual representation.} Dynamic networks can be {\em depicted} in different ways. One of them is the sequence-based representation shown above in Figure~\ref{fig:sequence}. Another is a {\em labeled graph} like the one in Figure~\ref{fig:example-graph}, where labels indicate when an edge is present (either as intervals or as discrete times). Other representations include chronological diagrams of contacts~(see \eg~\cite{H12,GVFWL16,LVM17}).

\subsection{Moving the focus away from models (a plea for unity)}
\label{sec:away}

Up to specific considerations, the vast majority of temporal concepts transcend their formulation in a given graph model (or stream model), and the same holds for many algorithmic ideas. Of course, some models are more relevant than others depending on the uses. In particular, is the model being used as a data structure or as a descriptive language? Is time discrete or continuous? Is the point of view local or global? Is time synchronous or asynchronous? Do links have duration? Having several models at our disposal is a good thing. On the other hand, the diversity of terminology makes it harder for several sub-communities to track the progress made by one another. We hope that the diversity of models does not prevent us from acknowledging each others {\em conceptual} works accross communities.

In this document, most of the temporal concepts and algorithmic ideas being reviewed are independent from the model. Specific models are used to {\em formulate} these concepts, but once formulated, they can often be considered at a more general level. To stress independence from the models, we tend to refer to a graph representing a dynamic network as just a {\em graph} (or a network), using calligraphic letters like $\cal G$ or $\cal H$ to indicate their dynamic nature. In contrast, when referring to static graphs in a way not clear from the context, we add the adjective {\em static} or {\em standard} explicitly and use regular letters like $G$ and $H$.

\section{Some temporal concepts}
\label{sec:concepts}

\articles{ADHOCNOW'11~\cite{CFQS11} (long version IJPEDS'12~\cite{CFQS12}).}

This section presents a number of basic concepts related to dynamic networks. Many of them were independently identified in various communities using different names. We limit ourselves to the most central ones, alternating between time-varying graphs and untimed evolving graphs (\ie basic sequences of graphs) for their formulation (depending on which one is the most intuitive). Most of the terminology is in line with the one of our 2012 article~\cite{CFQS12}, in which a particular effort was made to identify the first use of each concept in the literature and to give proper credit accordingly. Most of these concepts are now becoming folklore, and we believe this is a good thing.

\subsubsection{Subgraphs}
\label{sec:subgraph}
There are several ways to restrict a dynamic network $\TVG$. Classically, one may consider a subgraph resulting from taking only a subset of $V$ and $E$, while maintaining the behavior of the presence and latency functions, specialized to their new domains. Perhaps more specifically, one may restrict the lifetime to a given sub-interval $[t_a,t_b] \subseteq \T$, specializing again the functions to this new domain without otherwise changing their behavior. In this case, we write $\G_{[t_a,t_b]}$ for the resulting graph and call it a {\em temporal} subgraph of \G.

\subsubsection{Footprints and Snapshots}
\label{sec:footprint}

Given a graph \TVG, the {\em footprint} of $\G$ is
 the standard graph consisting of the vertices and edges which are present at least once over ${\cal T}$. This notion is sometimes identified with the {\em underlying graph} $G=(V,E)$, that is, the {\em domain} of possible vertices and edges, although in general an element of the domain may not appear in a considered interval, making the distinction between both notions useful. We denote the footprint of a graph \G by $\mathit{footprint}(\G)$ or simply $\cup(\G)$. The {\em snapshot} of $\G$ at time $t$ is the standard graph $G_t=(V, \{e: \rho(e,t)=1\})$. 
The footprint can also be defined as the union of all snapshots. Conversely, we denote by $\cap(\G)$ the intersection of all snapshots of $\G$, which may be called {\em intersection graph} or {\em denominator} of $\G$. Observe that, so defined, both concepts make sense as well in discrete time as in continuous time. 
In a context of infinite lifetime, Dubois et al.~\cite{BDKP16} defined the {\em eventual footprint} of $\G$ as the graph $(V,E')$ whose edges reappear infinitely often; in other words, the {\em limsup} of the snapshots.

In the literature, snapshots have been variously called layers, graphlets, or instantaneous graphs; footprints have also been called underlying graphs, union graphs, or induced graphs.

\subsubsection{Journeys and temporal connectivity}
\label{sec:journeys}

\begin{figure}[h]
\centering
\tikzset{leg/.style={very thick,red,->,shorten >=1pt,shorten <=1pt}}
\begin{tabular}{@{}c@{~~~}|@{~~~}c@{~~~}|@{~~~}c@{~~~}|@{~~~}c@{~}}
  \tikzsetnextfilename{tikz-journey-G0}
  \begin{tikzpicture}[scale=.9]
    \wgraph (.6pt,8pt)
    \tikzstyle{every node}=[font=\footnotesize]
    \draw (a)--(c);
    \draw (b)--(c);
    \draw (b)--(d);
    \draw (c)--(d);
    \draw[leg] (a) edge[bend left=5](c);
    \path (c)+(0,-1.1) node {$G_0$};
  \end{tikzpicture}
  &
  \tikzsetnextfilename{tikz-journey-G1}
    \begin{tikzpicture}[scale=.9]
      \wgraph(.6pt,8pt)
      \tikzstyle{every node}=[font=\scriptsize]
      \draw (a)--(b);
      \draw (b)--(c);
      \draw (c)--(d);
      \draw[leg] (c) edge[bend left=5](d);
      \path (c)+(0,-1.1) node {$G_1$};
    \end{tikzpicture}
  &
  \tikzsetnextfilename{tikz-journey-G2}
    \begin{tikzpicture}[scale=.9]
      \wgraph(.6pt,8pt)
      \tikzstyle{every node}=[font=\scriptsize]
      \draw (a)--(b);
      \draw (c)--(d);
      \draw (c)--(e);
      \draw (d)--(e);
      \path (c)+(0,-1.1) node {$G_2$};
    \end{tikzpicture}
  &
  \tikzsetnextfilename{tikz-journey-G3}
    \begin{tikzpicture}[scale=.9]
      \wgraph(.6pt,8pt)
      \tikzstyle{every node}=[font=\scriptsize]
      \draw (c)--(e);
      \draw (d)--(e);
      \draw[leg] (d) edge[bend right=5] (e);
      \path (c)+(0,-1.1) node {$G_3$};
    \end{tikzpicture}\vspace{-5pt}
\end{tabular}
\caption{\label{fig:journey}Intuitive representation of a journey (from $a$ to $e$) in a dynamic network.}
\end{figure}
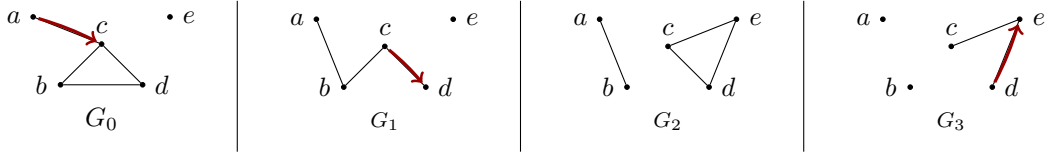

The concept of {\em journey} is central in highly dynamic networks. Journeys are the analogue of paths in standard graphs. 
In its simplest form, a journey in $\TVG$ is a sequence of ordered pairs $\J=\{(e_1,t_1),$ $(e_2,t_2) \dots,$ $(e_k,t_k)\}$, such that $e_1, e_2,...,e_k$ is a walk in $(V,E)$, $\rho(e_i,t_i)=1$, and $t_{i+1} \ge t_i$. 
An intuitive representation is shown on Figure~\ref{fig:journey}. The set of all journeys from $u$ to $v$ when the context of $\G$ is clear is denoted by $\J^*_{(u,v)}$.

Several versions can be formulated, for example taking into account the latency by requiring that $t_{i+1}\ge t_i + \zeta(e_i,t_i)$. 
In a communication network, it is often also required that $\rho(e_i,t)=1$ for all $t\in [t_i,t_i+\zeta(e_i,t_i))$, \ie the edge remains present during the communication period.
 When time is discrete, a more abstract way to incorporate latency is to distinguish between {\em strict} and {\em non-strict} journeys~\cite{KKK00}, strictness referring here to requiring that $t_{i+1} > t_i$ in the journey times. In other words, non-strict journeys correspond to neglecting latency.

 Somewhat orthogonally, a journey is
{\em direct} if every next hop occurs without delay at the intermediate nodes (\ie $t_{i+1}=t_i + \zeta(e_i,t_i)$); it is {\em indirect} if it makes a pause at least at one intermediate node. We showed that this distinction plays a key role for computing temporal distances among the nodes in {\em continuous} time (see~\cite{CFMS11,CFMS14}, reviewed in Section~\ref{sec:temporal-lags}). We also showed, using this concept, that the ability of the nodes to buffer a message before retransmission decreases dramatically the expressive power of an adversary controlling the topology (see~\cite{CFGSY12,CFGSY13,CFGSY15}, reviewed in Section~\ref{sec:expressivity}).

\begin{figure}[h]
  \centering
  \begin{tabular}{c|c|c}
    \tikzsetnextfilename{tikz-TC-G0}
  \begin{tikzpicture}[scale=.62]
    \tikzstyle{every node}=[draw, fill, circle, inner sep=1pt]
    \path (0,0) node (a) {};
    \path (a)+(-.6,-.2) coordinate (a');
    \path (3,0) node (b) {};
    \path (b)+(.6,.2) coordinate (b');
    \path (2.5,1) node (c) {};
    \path (c)+(-1,0) coordinate (c');
    \path (0.5,-1) node (d) {};
    \path (d)+(1,0) coordinate (d');
    \tikzstyle{every node}=[]
    \path[left] (a) node {$a$};
    \path[right] (b) node {$b$};
    \path[above] (c) node {$c$};
    \path[below] (d) node {$d$};
    \tikzstyle{every path}=[draw, semithick]
    \path (a)--(d);
    \path (b)--(c);
    \tikzstyle{every path}=[draw, dashed,->]
    \path (c)--(c');
    \path (d)--(d');
  \end{tikzpicture}
  &
    \tikzsetnextfilename{tikz-TC-G1}
  \begin{tikzpicture}[scale=.62]
    \tikzstyle{every node}=[draw, fill, circle, inner sep=1pt]
    \path (0,0) node (a) {};
    \path (a)+(-.6,-.2) coordinate (a');
    \path (3,0) node (b) {};
    \path (b)+(.6,.2) coordinate (b');
    \path (1.5,1) node (c) {};
    \path (c)+(-1,0) coordinate (c');
    \path (1.5,-1) node (d) {};
    \path (d)+(1,0) coordinate (d');
    \tikzstyle{every node}=[]
    \path[left] (a) node {$a$};
    \path[right] (b) node {$b$};
    \path[above] (c) node {$c$};
    \path[below] (d) node {$d$};
    \tikzstyle{every path}=[draw, dashed,->]
    \path (c)--(c');
    \path (d)--(d');
  \end{tikzpicture}
  &
    \tikzsetnextfilename{tikz-TC-G2}
  \begin{tikzpicture}[scale=.62]
    \tikzstyle{every node}=[draw, fill, circle, inner sep=1pt]
    \path (0,0) node (a) {};
    \path (a)+(-.6,-.2) coordinate (a');
    \path (3,0) node (b) {};
    \path (b)+(.6,.2) coordinate (b');
    \path (.5,1) node (c) {};
    \path (2.5,-1) node (d) {};
    \tikzstyle{every node}=[]
    \path[left] (a) node {$a$};
    \path[right] (b) node {$b$};
    \path[above] (c) node {$c$};
    \path[below] (d) node {$d$};
    \tikzstyle{every path}=[draw, semithick]
    \path (a)--(c);
    \path (b)--(d);
    \tikzstyle{every path}=[draw]
  \end{tikzpicture}
  \end{tabular}\\
  $\overset{\xrightarrow{\hspace*{4cm}}}{time}$
  \caption{\label{fig:TC}Connectivity over time and space. {\it (Dashed arrows denote movements.)}}
\end{figure}
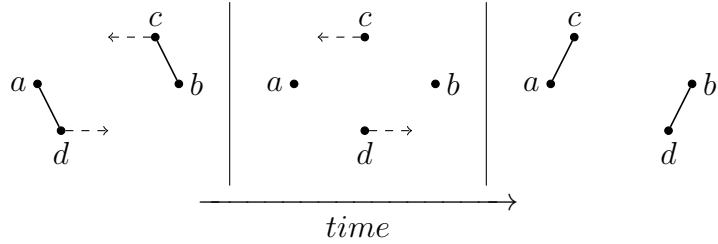

Finally, $departure(\J)$ and $arrival(\J)$ denote respectively the starting time $t_1$ and the last time $t_k$ (or $t_k+\zeta(e_k,t_k)$ if latency is considered) of journey~$\J$.
When the context of $\G$ is clear, we denote the {\em possibility} of a journey from $u$ to $v$ by $u \leadsto v$, which does not imply $v \leadsto u$ even if the links are undirected, because time induces its own level of direction (\eg $a \leadsto e$ but $e \not\leadsto a$ in Figure~\ref{fig:journey}). If forall $u$ and $v$, it holds that $u \leadsto v$, then $\G$ is {\em temporally connected} (Class~\TC in Chapter~\ref{sec:classes}). Interestingly, a graph may be temporally connected even if none of its snapshots are connected, as illustrated in Figure~\ref{fig:TC} (the footprint must be connected, though).

The example on Figure~\ref{fig:TC} suggests a natural extension of the concept of connected components for dynamic networks, which we discuss in a dedicated paragraph in Section~\ref{sec:curiosities}.

\subsubsection{Temporal distance and related metrics}
\label{sec:distance}

As observed in 2003 by Bui-Xuan {\em et al.}~\cite{BFJ03}, journeys have both a {\em topological} length (number of hops) and a {\em temporal} length (duration),
which gives rise to several concepts of distance and at least three types of optimal journeys ({\em shortest}, {\em fastest}, and {\em foremost}, covered in Section~\ref{sec:shfafo} and~\ref{sec:temporal-lags}).
Unfolding the concept of temporal distance leads to that of temporal diameter and temporal eccentricity~\cite{BFJ03}. 
Precisely, the temporal eccentricity of a node $u$ at time $t$ is the earliest time that $u$ can reach all other nodes through a journey starting after $t$. The temporal diameter at time $t$ is the maximum among all nodes eccentricities (at time $t$). Another characterization of the temporal diameter (at time $t$) is the smallest $d$ such that $\G_{[t,t+d]}$ is temporally connected.
These concepts are central in some of our contributions (\eg~\cite{CFMS11,CFMS14}, further discussed in Section~\ref{sec:temporal-lags}). Interestingly, all these parameters refer to time quantities, and these quantities themselves vary over time, making their study (and computation) more challenging.

\subsection{Further concepts}

\nocite{ACFQS11}

The number of definitions built on top of temporal concepts could grow large. Let us mention just a few additional concepts which we had compiled in~\cite{CFQS12,SQFCA11} and~\cite{CB10}. Most of these emerged in the area of complex systems, but are of general applicability.

{\em Small-world.} A temporal analogue of the small-world effect is defined in~\cite{TSM+09} based on the {\em duration} of journeys (as opposed to hop distance in the original definition in static graphs~\cite{WD98}). Perhaps without surprise, this property is observed in a number of more theoretical works considering stochastic dynamic networks (see \eg~\cite{ChMMD08,CP11}). An analogue of {\em expansion} for dynamic networks is defined in~\cite{CP11}.

{\em Network backbones.}
A temporal analogue of the concept of {\em backbone} was defined in~\cite{KKW08} as the ``subgraph consisting of edges on which information has the potential to flow the quickest.'' In fact, we observe in~\cite{CFMS14} that an edge belongs to the backbone relative to time $t$ 
iff it is used by a foremost journey starting at time $t$. As a result, the backbone consists exactly of the union of all foremost broadcast trees relative to initiation time $t$ (the computation of such structure is reviewed in Section~\ref{sec:temporal-lags}).

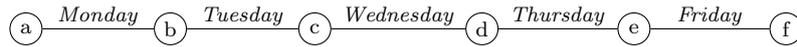
\begin{figure}[h]
  \begin{center}
    \tikzsetnextfilename{tikz-fairness}
    \begin{tikzpicture}[xscale=2]
      \scriptsize
      \tikzstyle{every node}=[draw,circle, minimum size=12pt, inner sep=0pt]
      \path (0,0) node (a){a};
      \path (a)+(.95,0) node (b){b};
      \path (a)+(1.9,0) node (c){c};
      \path (a)+(3,0) node (d){d};
      \path (a)+(4,0) node (e){e};
      \path (a)+(5,0) node (f){f};
      \tikzstyle{every node}=[below,font=\scriptsize,inner sep=1pt]
      \draw (a)--node[above]{\textit{Monday}}(b);
      \draw (b)--node[above]{\textit{Tuesday}}(c);
      \draw (c)--node[above]{\textit{Wednesday}}(d);
      \draw (d)--node[above]{\textit{Thursday}}(e);
      \draw (e)--node[above]{\textit{Friday}}(f);
       \end{tikzpicture}
    \caption{\label{fig:ex1} Weekly interactions between six people (from~\cite{CFQS12}).}
  \end{center}
\end{figure}

{\em Centrality.}
The structure (\ie footprint) of a dynamic network may not reflect how well interactions are balanced within. In~\cite{CFQS12}, we defined a metric of {\em fairness} as the standard deviation among temporal eccentricities. In the caricatural example of Figure~\ref{fig:ex1} (depicting weekly interaction among entities),
node $c$ or $d$ are structurally more central, but node $a$ is actually the most central in terms of temporal eccentricity: it can reach all other nodes within $5$ to $11$ days, compared to nearly three weeks for $d$ and more than a month for $f$.

Together with a sociologist, Louise Bouchard, in 2010~\cite{CB10}, we proposed to apply these measures (together with a stochastic version of network backbones) to the study of health networks in Canada. (The project was not retained and we started collaborating on another topic.)
This kind of concepts, including also {\em temporal betweenness} or {\em closeness} (which we expressed in the TVG formalism in~\cite{SQFCA11}), received a lot of attention lately (see \eg~\cite{tredan,RC17}).

Other temporal or dynamic extensions of traditional concepts, not covered here, include {\em treewidth}~\cite{treewidth}, {\em temporal flows}~\cite{flows},
and characteristic temporal distance~\cite{TSM+09}. Several surveys reviewed the conceptual shift induced by time in dynamic networks, including (for the distributed computing community) Kuhn and Oshman~\cite{KO11}, Michail and Spirakis~\cite{MS18}, and our own 2012 survey with Flocchini, Quattrociocchi, and Santoro~\cite{CFQS12}.

\section{Redefinition of problems}
\label{sec:redefinition}
\articles{arXiv'11~\cite{CMM11}, DRDC reports'13~\cite{CF13a,CF13b}, IJFCS~\cite{CFMS15}.}

The fact that a network is dynamic has a deep impact on the kind of tasks one can perform within. This impact ranges from making a problem harder, to making it impossible,
to redefining the metrics of interest, or even change the whole definition of the problem.
In fact, many standard problems {\em must} be redefined in highly dynamic networks. For example, what is a {\em spanning tree} in a partitioned (yet temporally connected) network? What is a maximal independent set, and a dominating set? Deciding which definition to adopt depends on the target application. We review here a few of these aspects through a handful of problems like {\em broadcast}, {\em election}, {\em spanning trees}, and classic {\em symmetry-breaking} tasks (independent sets, dominating sets, vertex cover), on which we have been involved.

\subsection{New optimality metrics in broadcasting}
\label{sec:defshfafo}

As explained in Section~\ref{sec:distance}, the length of a journey can be measured both in terms of the number of hops or in terms of duration, giving rise to two distinct notions of distance among nodes: the {\em topological distance} (minimum number of hop) and the {\em temporal distance} (earliest reachability), both being relative to a source $u$, a destination $v$, {\em and} an initiation time $t$. 
 
\begin{figure}[h]
  \centering
  \begin{tabular}{c|c}
      \begin{minipage}[c]{.28\linewidth}
        \begin{tikzpicture}[xscale=1.2, yscale=1.6]
          \tikzstyle{every node}=[draw,fill,circle, inner sep=1.2pt]
          \path (0,0) node (a){};
          \path (a)+(40:1) node (b){};
          \path (b)+(-40:1) node (e){};
          \path (e)+(40:1) node (c){};
          \path (e)+(-40:1) node (g){};
          \path (c)+(-40:1) node (d){};
          \path (a)+(-40:1) node (f){};
          \tikzstyle{every node}=[font=\footnotesize]
          \path (a) node[left] (la){$a$};
          \path (b) node[above] (lb){$b$};
          \path (c) node[right] (lc){$c$};
          \path (d) node[below] (ld){$d$};
          \path (e) node[above] (le){$e$};
          \path (f) node[below] (lf){$f$};
          \path (g) node[below] (lg){$g$};
          \tikzstyle{every node}=[below,font=\scriptsize,inner sep=1pt]
          \draw (a)--node[above left]{$[1,2]$}(b);
          \draw (b)--node[above]{$[3,4]$}(c);
          \draw (c)--node[above right]{$[5,6]$}(d);
          \draw (a)--node[above]{$[4,5]$}(e);
          \draw (e)--node[above]{$[9,10]$}(d);
          \draw (a)--node[below left]{$[6,8]$}(f);
          \draw (f)--node[below]{$[6,8]$}(g);
          \draw (g)--node[below right]{$[6,8]$}(d);
          \tikzstyle{every node}=[font=\large]
          \path (0,.6) coordinate (toplimit);
          \path (1.5,0) coordinate (rightlimit);
        \end{tikzpicture}
      \end{minipage}
      &
      \begin{minipage}[c]{.6\linewidth}
        Optimal journeys from $a$ to $d$ (starting at time $0$):\medskip\\
        \indent\quad- the shortest: a-e-d (only two hops)\medskip\\
        \indent\quad- the foremost: a-b-c-d (arriving at $5+\zeta$)\medskip\\
        \indent\quad- the fastest: a-f-g-d (no intermediate waiting)
      \end{minipage}
    \end{tabular}
    \caption{\label{fig:distance}Different meanings for length and distance {\it (assuming latency $\zeta \ll 1$)}.}
\end{figure}
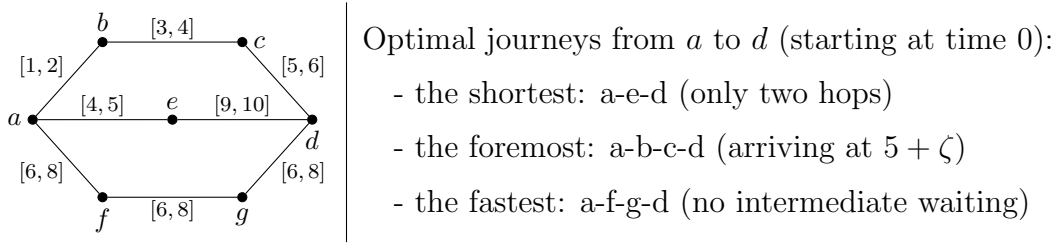

Bui-Xuan, Ferreira, and Jarry~\cite{BFJ03} define three optimality metrics based on these notions: 
{\em shortest} journeys minimize the number of hops, {\em foremost} journeys minimize reachability time, and {\em fastest} journeys minimize the duration of the journey (possibly delaying its departure). See Figure~\ref{fig:distance} for an illustration. 

The {\em centralized} problem of computing the three types of journeys given full knowledge of the graph $\G$ is introduced (and algorithms are proposed) in~\cite{BFJ03}. We investigated a distributed analogue of this problem; namely, the ability for the nodes to broadcast according to these metrics without knowing the underlying networks, but with various assumptions about its dynamics~\cite{CFMS10,CFMS15} (reviewed in Section~\ref{sec:shfafo}).

 \subsection{Election and spanning trees}
\label{sec:election}

While the definition of problems like broadcast or routing remains intuitive in highly dynamic networks, other problems are definitionally ambiguous. Consider {\em leader election} and {\em spanning trees}. In a static network, leader election consists of distinguishing a single node, the leader, for playing subsequently a distinct role. The spanning tree problem consists of selecting a cycle-free set of edges that interconnects all the nodes. Both problems are central and widely studied. How should these problems be defined in a highly dynamic network which (among other features) is possibly partitioned most of the time? 

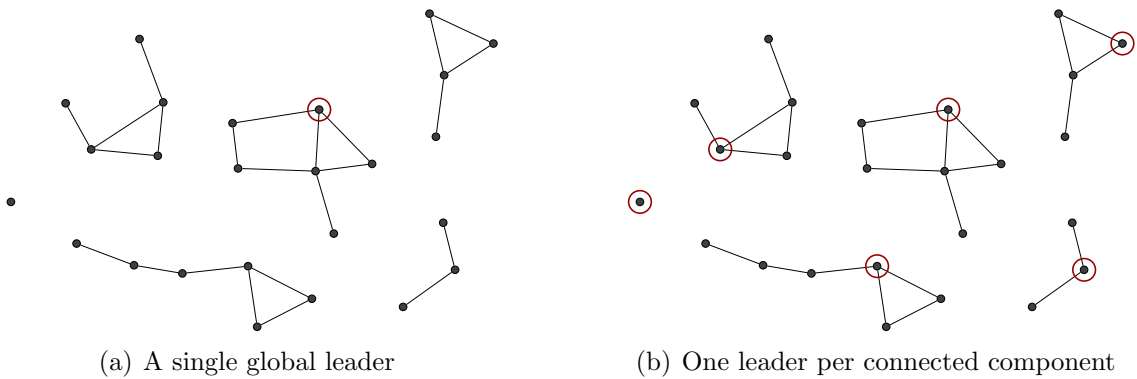
\begin{figure}[h]
  \subfigure[A single global leader]{
    \label{fig:temporal-leader}
    \tikzsetnextfilename{tikz-leader-1}
    \begin{tikzpicture}[scale=.6]
      \tikzstyle{every node}=[draw,fill=darkgray,circle,inner sep=1]
      \path (5.26,7.14) node (v6) {};
      \path (6.78,6.68) node (v9) {};
      \path (6.9,5.68) node (v10) {};
      \path (5.14,5.96) node (v5) {};
      \path (3.68,6.1) node (v3) {};
      \path (3.12,7.12) node (v1) {};
      \path (4.74,8.54) node (v8) {};
      \path (8.68,6.98) node (v15) {};
      \path (8.6,5.62) node (v14) {};
      \path (9.0,4.24) node (v16) {};
      \path (1.92,4.94) node (v0) {};
      \path (11.24,6.38) node (v20) {};
      \path (11.42,7.74) node (v22) {};
      \path (9.84,5.78) node (v17) {};
      \path (12.5,8.44) node (v24) {};
      \path (11.1,9.1) node (v19) {};

      \path (11.4,4.48) node (v21) {};
      \path (10.52,2.62) node (v18) {};
      \path (11.66,3.44) node (v23) {};
      \path (7.12,3.52) node (v11) {};
      \path (3.36,4.02) node (v2) {};
      \path (4.62,3.54) node (v4) {};
      \path (5.68,3.36) node (v7) {};
      \path (8.52,2.8) node (v13) {};
      \path (7.32,2.18) node (v12) {};
      \tikzstyle{every path}=[very thin];
      \draw (v15)--(v14);
      \draw (v13)--(v12);
      \draw (v15)--(v17);
      \draw (v6)--(v3);
      \draw (v24)--(v19);

      \draw (v9)--(v10);
      \draw (v6)--(v5);
      \draw (v3)--(v1);
      \draw (v6)--(v8);
      \draw (v9)--(v15);
      \draw (v10)--(v14);
      \draw (v14)--(v16);
      \draw (v2)--(v4);
      \draw (v11)--(v7);
      \draw (v4)--(v7);
      \draw (v11)--(v13);
      \draw (v11)--(v12);
      \draw (v23)--(v21);
      \draw (v23)--(v18);
      \draw (v3)--(v5);
      \draw (v20)--(v22);
      \draw (v14)--(v17);
      \draw (v22)--(v24);
      \draw (v22)--(v19);

      \tikzstyle{every node}=[draw, red, circle, semithick, inner sep=3pt]
      \path (v15) node {};
    \end{tikzpicture}
  }
  \hspace{20pt}
  \subfigure[One leader per connected component]{
    ~~
    \label{fig:instant-leaders}
    \tikzsetnextfilename{tikz-leader-2}
    \begin{tikzpicture}[scale=.6]
      \tikzstyle{every node}=[draw,fill=darkgray,circle,inner sep=1]
      \path (5.26,7.14) node (v6) {};
      \path (6.78,6.68) node (v9) {};
      \path (6.9,5.68) node (v10) {};
      \path (5.14,5.96) node (v5) {};
      \path (3.68,6.1) node (v3) {};
      \path (3.12,7.12) node (v1) {};
      \path (4.74,8.54) node (v8) {};
      \path (8.68,6.98) node (v15) {};
      \path (8.6,5.62) node (v14) {};
      \path (9.0,4.24) node (v16) {};
      \path (1.92,4.94) node (v0) {};
      \path (11.24,6.38) node (v20) {};
      \path (11.42,7.74) node (v22) {};
      \path (9.84,5.78) node (v17) {};
      \path (12.5,8.44) node (v24) {};
      \path (11.1,9.1) node (v19) {};

      \path (11.4,4.48) node (v21) {};
      \path (10.52,2.62) node (v18) {};
      \path (11.66,3.44) node (v23) {};
      \path (7.12,3.52) node (v11) {};
      \path (3.36,4.02) node (v2) {};
      \path (4.62,3.54) node (v4) {};
      \path (5.68,3.36) node (v7) {};
      \path (8.52,2.8) node (v13) {};
      \path (7.32,2.18) node (v12) {};
      \tikzstyle{every path}=[very thin];
      \draw (v15)--(v14);
      \draw (v13)--(v12);
      \draw (v15)--(v17);
      \draw (v6)--(v3);
      \draw (v24)--(v19);

      \draw (v9)--(v10);
      \draw (v6)--(v5);
      \draw (v3)--(v1);
      \draw (v6)--(v8);
      \draw (v9)--(v15);
      \draw (v10)--(v14);
      \draw (v14)--(v16);
      \draw (v2)--(v4);
      \draw (v11)--(v7);
      \draw (v4)--(v7);
      \draw (v11)--(v13);
      \draw (v11)--(v12);
      \draw (v23)--(v21);
      \draw (v23)--(v18);
      \draw (v3)--(v5);
      \draw (v20)--(v22);
      \draw (v14)--(v17);
      \draw (v22)--(v24);
      \draw (v22)--(v19);

      \tikzstyle{every node}=[draw, red, circle, semithick, inner sep=3pt]
      \path (v0) node {};
      \path (v3) node {};
      \path (v11) node {};
      \path (v15) node {};
      \path (v23) node {};
      \path (v24) node {};
    \end{tikzpicture}
    ~~
  }
  \caption{\label{fig:leader}Two possible definitions of the leader election problem.}
\end{figure}

At least two (somewhat generic) versions emerge. Starting with {\em election}, is the objective still to distinguish a {\em unique} leader? This option makes sense if the leader can influence other nodes reasonably often. However, if the temporal connectivity within the network takes a long time to be achieved, then it may be more relevant to elect a leader in {\em each} component, and maintain a leader per component when components merge and split. Both options are depicted on Figure~\ref{fig:leader}.

The same declination holds for spanning trees, the options being (1) to build a unique tree whose edges are {\em intermittent}, or (2) to build a different tree in each component, to be updated when the components split and merge. Together with Flocchini, Mans, and Santoro, we considered the first option in~\cite{CFMS10,CFMS15} (reviewed in Section~\ref{sec:shfafo}), building (distributedly) a fixed but intermittent broadcast tree in a network whose edges are all recurrent (Class~\ER). In a different line of work (with a longer list of co-authors)~\cite{Cas06,PCGC10,CCGP13,BCCJN14b}, we proposed and studied a ``best effort'' principle for maintaining a set of spanning trees of the second type, while guaranteeing that some properties hold {\em whatever} the dynamics (reviewed in Section~\ref{sec:forest}). A by-product of this algorithm is to maintain a single {\em leader} per tree (the root).

\subsection{Covering problems}
\label{sec:covering}

\tikzset{every node/.style={defnode}}
\def\carre (#1,#2){
  \path (#1,#2+1) node (a){};
  \path (#1+1,#2+1) node (b){};
  \path (#1,#2) node (c){};
  \path (#1+1,#2) node (d){};
}

With Mans and Mathieson in 2011~\cite{CMM11}, we explored three canonical ways of redefining combinatorial problems in highly-dynamic networks, with a focus on covering problems like {\em dominating set}. A dominating set in a (standard) graph $G=(V,E)$ is a subset of nodes $S\subseteq V$ such that each node in the network either is in $S$ or has a neighbor in $S$. The goal is usually to minimize the size (or cost) of $S$. Given a dynamic network $\G=\{G_1,G_2,\dots\}$, the problem admits three natural declinations:
\vspace{-5pt}

\begin{itemize}
\item {\em Temporal} version: domination is achieved {\em over time} -- every node outside the set must share an edge {\em at least once} with a node in the set, as illustrated below.\vspace{-5pt}
  \begin{center}
    \begin{tabular}{c@{\hspace{20pt}}c@{\hspace{20pt}}|@{\hspace{20pt}}c@{\hspace{20pt}}|@{\hspace{20pt}}c}
      &$G_1$&$G_2$&$G_3$\\
      &&&\\
      \begin{minipage}[c]{3cm}
        Temporal\\dominating set\\\\
      \end{minipage}
      &
      \tikzsetnextfilename{tikz-dominating-temporal-1}
      \begin{tikzpicture}
        \carre (0,0)
        \draw (b)--(d);
        \draw (c)--(d);
        \path (d) node[circ]{};
      \end{tikzpicture}
      &
        \tikzsetnextfilename{tikz-dominating-temporal-2}
      \begin{tikzpicture}
        \carre (0,0)
        \draw (a)--(d);
        \draw (c)--(d);
        \path (d) node[circ]{};
      \end{tikzpicture}
      &
    \tikzsetnextfilename{tikz-dominating-temporal-3}
      \begin{tikzpicture}
        \carre (0,0)
        \draw (a)--(c);
        \draw (c)--(d);
        \path (d) node[circ]{};
      \end{tikzpicture}
    \end{tabular}
    \vspace{-3pt}
  \end{center}

\item {\em Evolving} version: domination is achieved in every snapshot, but the set can vary between them. This version is commonly referred to as ``dynamic graph algorithms'' in the algorithmic literature (also called ``reoptimization'').
\vspace{-5pt}
  \begin{center}
    \centering
    \begin{tabular}{c@{\hspace{20pt}}c@{\hspace{20pt}}|@{\hspace{20pt}}c@{\hspace{20pt}}|@{\hspace{20pt}}c}
      &$G_1$&$G_2$&$G_3$\\
      &&&\\
      \begin{minipage}[c]{3cm}
        Evolving\\dominating set\\\\
      \end{minipage}
      &
        \tikzsetnextfilename{tikz-dominating-evolving-1}
      \begin{tikzpicture}
        \carre (0,0)
        \draw (b)--(d);
        \draw (c)--(d);
        \path (a) node[circ]{};
        \path (d) node[circ]{};
      \end{tikzpicture}
      &
        \tikzsetnextfilename{tikz-dominating-evolving-2}
      \begin{tikzpicture}
        \carre (0,0)
        \draw (a)--(d);
        \draw (c)--(d);
        \path (b) node[circ]{};
        \path (d) node[circ]{};
      \end{tikzpicture}
      &
        \tikzsetnextfilename{tikz-dominating-evolving-3}
      \begin{tikzpicture}
        \carre (0,0)
        \draw (a)--(c);
        \draw (c)--(d);
        \path (b) node[circ]{};
        \path (c) node[circ]{};
      \end{tikzpicture}
    \end{tabular}
    \vspace{-3pt}
  \end{center}

\item {\em Permanent} version: domination is achieved in every snapshot, with a {\em fixed} set.
\vspace{-5pt}

  \begin{center}
    \centering
    \begin{tabular}{c@{\hspace{20pt}}c@{\hspace{20pt}}|@{\hspace{20pt}}c@{\hspace{20pt}}|@{\hspace{20pt}}c}
      &$G_1$&$G_2$&$G_3$\\
      &&&\\
      \begin{minipage}[c]{3cm}
        Permanent\\dominating set\\\\
      \end{minipage}
      &
        \tikzsetnextfilename{tikz-dominating-permanent-1}
      \begin{tikzpicture}
        \carre (0,0)
        \draw (b)--(d);
        \draw (c)--(d);
        \path (a) node[circ]{};
        \path (b) node[circ]{};
        \path (d) node[circ]{};
      \end{tikzpicture}
      &
        \tikzsetnextfilename{tikz-dominating-permanent-2}
      \begin{tikzpicture}
        \carre (0,0)
        \draw (a)--(d);
        \draw (c)--(d);
        \path (a) node[circ]{};
        \path (b) node[circ]{};
        \path (d) node[circ]{};
      \end{tikzpicture}
      &
        \tikzsetnextfilename{tikz-dominating-permanent-3}
      \begin{tikzpicture}
        \carre (0,0)
        \draw (a)--(c);
        \draw (c)--(d);
        \path (a) node[circ]{};
        \path (b) node[circ]{};
        \path (d) node[circ]{};
      \end{tikzpicture}
    \end{tabular}
    \vspace{-3pt}
  \end{center}
\end{itemize}

The three versions are related. Observe, in particular, that the temporal version consists of computing a dominating set in $\cup \G$ (the footprint), and the permanent version one in $\cap \G$. Solutions to the permanent version are valid (but possibly sub-optimal) for the evolving version, and those for the evolving version are valid (but possibly sub-optimal) for the temporal version~\cite{CMM11}. 
In fact, the solutions to the \emph{permanent} and the \emph{temporal} versions actually form upper and lower bounds for the \emph{evolving} version. 
Note that the permanence criterion may {\em force} the addition of some elements to the solution, which explains why some problems like spanning tree and election do not admit a permanent version.

\begin{open}
  Make a more systematic study of the connexions between the temporal, the evolving, and the permanent versions. Characterize the role these versions play with respect to each other both in terms of lower bound and upper bound, and design algorithms exploiting this triality (``threefold duality'').
\end{open}

In a distributed (or online) setting, the permanent and the temporal versions are not directly applicable because the future of the network is not known a priori. 
Nonetheless, if the network satisfies other forms of regularity, like \emph{periodicity} \cite{FMS09,FKMS12a,Kel12} (Class~\EP) or \emph{recurrence of edges}~\cite{CFMS10} (Class~\ER), then such solutions can be built despite lack of information about the future. 
In this perspective, Dubois {\em et al.}~\cite{DKP15} define a variant of the temporal version in infinite lifetime networks, requiring that the covering relation holds {\em infinitely often} (\eg for dominating sets, every node not in the set must be dominated infinitely often by a node in the set). We review in Section~\ref{sec:robustness} a joint work with Dubois, Petit, and Robson~\cite{CDPR17}, where we define a new hereditary property in graphs called {\em robustness} that captures the ability for a solution to have such features in a large class of highly-dynamic networks (Class~\TCR, all classes are reviewed in Chapter~\ref{sec:classes}). The robustness of maximal independent sets (MIS) is investigated in particular and the locality of finding a solution in various cases is characterized.

This type of problems have recently gained interest in the algorithmic and distributed computing communities. For example, Mandal et al.~\cite{permanent} study approximation algorithms for the permanent version of {\em dominating sets}. Akrida {\it et al.}~\cite{AMSZ18} (2018) define a variant of the temporal version in the case of vertex cover, in which a solution is not just a set of nodes (as it was in~\cite{CMM11}) but a set of pairs $(nodes, times)$, allowing different nodes to cover the edges at different times (and within a sliding time window). Bamberger {\it et al.}~\cite{kuhn} (2018) also define a temporal variant of two covering problems (vertex coloring and MIS) relative to a sliding time window.

\subsection*{Concluding remark}
\label{sec:mobile-agents}

Given the reporting nature of this document, we reviewed here the definition of concepts and problems in which we have been effectively involved (through contributions). As such, the content does not aim at comprehensiveness and many concepts and problems were not covered. In particular, the {\em network exploration} problem received a lot of attention recently in the context of dynamic networks, for which we refer the reader to a number of dedicated works~\cite{FMS13,FKMS12d,IW11,BDP17,LDFS16}.

\chapter{Feasibility of distributed problems}
\label{sec:feasibility}

A common approach to analyzing distributed algorithms is the characterization of necessary and sufficient conditions to their success.
These conditions commonly refer to the communication model, synchronicity, or 
structural properties of the network (\eg is the topology a tree, a grid, a ring, {\it etc.}) 
In a dynamic network, the topology changes {\em during} the computation, and this has
 a dramatic effect on computations.
In order to understand this effect, the engineering community has developed a number of
{\em mobility models}, which govern how the nodes move and make it possible to compare results on a relatively fair basis and enable their reproducibility (the most famous, yet unrealistic model is {\em random waypoint}).

In the same way as mobility models enable the {\em experimental} investigations
of algorithms and protocols in highly dynamic networks, logical properties on the network
dynamics, \ie {\em classes of dynamic graphs}, have the potential
to guide a more formal exploration of their qualities and limitations.
This chapter reviews our contributions in this area, which has acted as a general driving force through most of our works in dynamic networks. 
The resulting classes of dynamic graphs are then revisited, compiled, and independently discussed in Chapter~\ref{sec:classes}.

\section{Basic conditions}
\label{sec:conditions}
\article{SIROCCO'09~\cite{CCF09}}

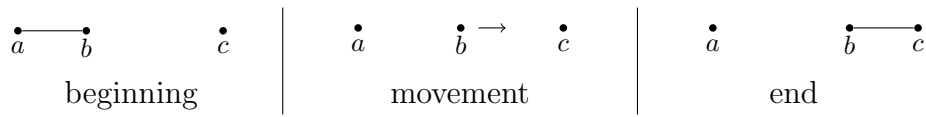
\begin{figure}[h]
  \begin{center}
    \begin{tabular}{c|c|c}
      \scriptsize
      \tikzsetnextfilename{tikz-mobile-scenario-1}
      \begin{tikzpicture}[scale=.9]
        \tikzstyle{every node}=[fill=black, circle, inner sep=1pt]
        \path (0,0) node (a) {};
        \path (1,0) node (b) {};
        \path (3,0) node (c) {};
        \path (0,.3) coordinate (tanchor) {};
        \tikzstyle{every node}=[font=\footnotesize]
        \path (a)+(0,-.25) node (la) {$a$};
        \path (b)+(0,-.25) node (lb) {$b$};
        \path (c)+(0,-.25) node (lc) {$c$};
        \draw (a)--(b);
      \end{tikzpicture}
      \quad\quad&\quad
      \tikzsetnextfilename{tikz-mobile-scenario-2}
      \begin{tikzpicture}[scale=.9]
        \tikzstyle{every node}=[fill=black, circle, inner sep=1pt]
        \path (0,0) node (a) {};
        \path (1.5,0) node (b) {};
        \path (3,0) node (c) {};
        \path (0,.3) coordinate (tanchor) {};
        \tikzstyle{every node}=[font=\footnotesize]
        \path (a)+(0,-.25) node (la) {$a$};
        \path (b)+(0,-.25) node (lb) {$b$};
        \path (c)+(0,-.25) node (lc) {$c$};
        
        \draw[->, shorten >=20pt, shorten <=5pt] (b)--(c);
      \end{tikzpicture}
      \quad\quad&\quad
      \tikzsetnextfilename{tikz-mobile-scenario-3}
      \begin{tikzpicture}[scale=.9]
        \tikzstyle{every node}=[fill=black, circle, inner sep=1pt]
        \path (0,0) node (a) {};
        \path (2,0) node (b) {};
        \path (3,0) node (c) {};
        \path (0,.3) coordinate (tanchor) {};
        \tikzstyle{every node}=[font=\footnotesize]
        \path (a)+(0,-.25) node (la) {$a$};
        \path (b)+(0,-.25) node (lb) {$b$};
        \path (c)+(0,-.25) node (lc) {$c$};
        \draw (b)--(c);
      \end{tikzpicture}
      \\
      beginning&movement&end
    \end{tabular}
  \end{center}
  \caption{\label{fig:abc}A basic mobility scenario.}
\end{figure}

In way of introduction, consider the broadcasting of a piece of information in the dynamic network depicted on Figure~\ref{fig:abc}. The ability to complete this task depends on which node is the initial emitter: $a$ and $b$ may succeed, while $c$ is guaranteed to fail. The obvious reason is that no journey (thus no chain of causality) exists from $c$ to $a$. 
  
Together with Chaumette and Ferreira in 2009~\cite{CCF09}, we identified a number of such requirements relative to a few basic tasks, namely broadcasting, counting, and election. For the sake of abstraction, the algorithms were described in a high-level model called {\em graph relabelings systems}~\cite{LMS99}, in which interactions consist of {\em atomic} changes in the state of neighboring nodes. Different scopes of action were defined in the literature for such models, ranging from a single edge to a closed star where the state of all vertices changes atomically~\cite{chalopin,paulusma}. The {\em expression} of algorithms in the edge-based version is close to that of population protocols~\cite{AAD+06}, but the context of execution is different (especially in our case), and perhaps more importantly, the type of {\em questions} which are investigated are different.

In its simplest form, the broadcasting principle is captured by a single rule, represented as
\tikzsetnextfilename{tikz-broadcast-rule}
\begin{tikzpicture}[xscale=.6]
  \tikzstyle{every node}=[defnode,minimum size=2pt]
  \path (1,1) node (a) {};
  \path (a)+(1,0) node (b) {};
  \path (a)+(3,0) node (c) {};
  \path (a)+(4,0) node (d) {};
  \tikzstyle{every node}=[font=\scriptsize, inner sep=3pt]
  \path (a) node[above] (la) {$I$};
  \path (b) node[above] (lb) {$N$};
  \path (c) node[above] (lc) {$I$};
  \path (d) node[above] (ld) {$I$};
  \draw (a)--(b);
  \draw (c)--(d);
  \draw[->, shorten >= 10pt, shorten <= 10pt] (b)--(c);
\end{tikzpicture}
and meaning that if an informed node ``interacts'' (in a sense that will become clear) with a non-informed node, then the latter becomes informed. This node can in turn propagate information through the same rule. Graph relabeling systems\footnote{This model is sometimes referred to as ``local computations'', but with a different meaning to that of Naor and Stockmeyer~\cite{NS93}, therefore we do not use this term.} typically take place over a fixed graph, which unlike the interaction graph of population protocols, {\em is} thought of as a static network. In fact, the main purpose of graph relabeling systems is to abstract {\em communications}, while that of population protocols is to abstract {\em dynamism} (through scheduling). In our case, dynamism is not abstracted by the scheduling; relabeling operations take place over a support graph that itself changes, which is fundamentally different and makes it possible to inject properties of the network dynamics into the analysis. 

\paragraph{The model.} Given a dynamic network $\G=\{G_1,G_2,...\}$, the model we proposed in~\cite{CCF09}\footnote{In~\cite{CCF09}, we relied on the general (\ie timed) version of evolving graphs. With hindsight, untimed evolving graphs, \ie basic sequences of graphs, are sufficient and make the description simpler.} considers relabeling operation taking place {\em on top of the sequence}. Precisely, every $G_i$ may support a number of interactions (relabelings), then at some point, the graph transitions from $G_i$ to $G_{i+1}$. The adversary (scheduler, or daemon) controls both the selection of edges on which interactions occur, and the moment when the system transitions from $G_i$ to $G_{i+1}$, subject to the constraint that every edge of every $G_i$ is selected {\em at least once} before transitioning to $G_{i+1}$. 
This form of fairness is reasonable, as otherwise some edges may be present or absent without incidence on the computation.
The power of the adversary mainly resides in the {\em order} in which the edges are selected and the number of times they are selected in each $G_i$. 
The adversary does {\em not} control the sequence of graph itself (contrary to common models of message adversaries).

\subsection{Necessary or sufficient conditions in terms of dynamics}

The above model allows us to define formally the concept of necessary and sufficient conditions for a given algorithm, in {\em terms of network dynamics}. For a given algorithm ${\cal A}$ and network ${\cal G}$, ${\cal X}$ is the set of all possible executions and $X \in {\cal X}$ one of them corresponding to the adversary choices.

\begin{definition}[Necessary condition]
A property $P$ (on a graph sequence) is {\em necessary} (in the context of ${\cal A}$) if its non-satisfaction on $\G$ implies that no sequence of relabelings can transform the initial states to desired final states. ($\neg P(\G) \implies \forall X \in {\cal X}, {\cal A}$ fails.)
\end{definition}

\begin{definition}[Sufficient condition]
A property $P$ (on graph sequences) is {\em sufficient} if its satisfaction on $\G$ implies that all possible sequences of relabelings take the initial states into desired final states. ($P(\G) \implies \forall X \in {\cal X}, {\cal A}$ succeeds.)
\end{definition}

In other words, if a necessary condition is not satisfied by $\G$, then the execution will fail {\em whatever} the choices of the adversary; if a sufficient condition is satisfied, the execution will succeed whatever the choices of the adversary. In between lies the actual power of the adversary. In the case of the afore-mentioned broadcast algorithm, this space is not empty: it is {\em necessary} that a journey exist from the emitter to all other nodes, and it is {\em sufficient} that a {\em strict} journey exist from the emitter to all other nodes.

\begin{discussion}[Sensitivity to the model]
By nature, sufficient conditions are dependent on additional constraints imposed to the adversary, namely here, of selecting every edge at least once. No property on the sequence of graph could, {\em alone}, guarantee that the nodes will effectively interact, thus sufficient conditions are intrinsically {\em model-sensitive}. On the other hand, necessary conditions on the graph evolution do not depend on the particular model, which is one of the advantages of considering high-level computational model without abstracting dynamism.
\end{discussion}

We considered three other algorithms in~\cite{CCF09} which are similar to some protocols in~\cite{AAD+06}, albeit considered here in a different model and with different questions in mind. The first is a counting algorithm in which a designated counter node increments its count when it interacts with a node for the first time (\tikzsetnextfilename{tikz-counting-v1}\begin{tikzpicture}[xscale=.7]
  \tikzstyle{every node}=[defnode]
  \path (1,1) node (a) {};
  \path (a)+(1,0) node (b) {};
  \path (a)+(3,0) node (c) {};
  \path (a)+(4,0) node (d) {};
  \tikzstyle{every node}=[font=\scriptsize, inner sep=3pt]
  \path (a) node[above] (la) {$k$};
  \path (b) node[above] (lb) {$N$};
  \path (c) node[above] (lc) {$k$$+$$1$};
  \path (d) node[above] (ld) {$F$};
  \draw (a)--(b);
  \draw (c)--(d);
  \draw[->, shorten >= 10pt, shorten <= 10pt] (b)--(c);
\end{tikzpicture}).
An obvious necessary condition to count all nodes is the existence of a direct edge between the counter and every other node (possibly at the same time or at different times). In fact, this condition is also sufficient in the considered model, leaving no power at all to the adversary: either the condition holds and success is guaranteed, or it does not and failure is certain.

The second counting algorithm has {\em uniform} initialization: every node has a variable initially set to $1$. When two nodes interact, one of them cumulates the count of the other, which is eliminated (\tikzsetnextfilename{tikz-counting-v2}\begin{tikzpicture}[xscale=.7]
  \tikzstyle{every node}=[defnode,minimum size=2pt]
  \path (1,1) node (a) {};
  \path (a)+(1,0) node (b) {};
  \path (a)+(3,0) node (c) {};
  \path (a)+(4,0) node (d) {};
  \tikzstyle{every node}=[font=\scriptsize, inner sep=3pt]
  \path (a) node[above] (la) {$i$};
  \path (b) node[above] (lb) {$j$};
  \path (c) node[above] (lc) {$i$$+$$j$};
  \path (d) node[above] (ld) {$0$};
  \draw (a)--(b);
  \draw (c)--(d);
  \draw[->, shorten >= 10pt, shorten <= 10pt] (b)--(c);
\end{tikzpicture}).
Here, a necessary condition to complete the process is that at least one node can be reached by all others through a journey. A sufficient condition due to Marchand de Kerchove and Guinand~\cite{MG12} is that all pairs of nodes interact {\em at least once} over the execution
(\ie the footprint of $\G$ is complete).
The third counting algorithm adds a circulation rule to help surviving counters to meet, with same 
necessary condition as before. 

\begin{open}
  Find a sufficient condition for this version of the algorithm in terms of network dynamics.
\end{open}

\subsection{Tightness of the conditions}
\label{sec:tightness}

\begin{figure}[h]
  \centering
  \subfigure[Necessary condition ${\cal C}_{N}$]{
  \begin{tikzpicture}
    \tikzstyle{every node}=[font=\small]
    \draw[rounded corners=12pt] (0,0) node[above right,yshift=3pt] {~$\neg {\cal C_N}(\G) \rightarrow$ failure guaranteed} rectangle (6.4,3.5);
    \draw[rounded corners=8pt] (.5,1) rectangle (5.8,3);
    \path (.5,2) node[right] (in){${\cal C_N}(\G)$};
    \path (2.4,2.5) node[right, text width=4cm] (ntight){no information};
    \path (2.4,1.5) node[right, text width=4cm] (tight){successful possible};
    \draw[->] (in)-- node[above,sloped,font=\scriptsize]{$\neg$\,tight}(ntight.west);
    \draw[->] (in)-- node[below,sloped,font=\scriptsize]{tight}(tight.west);
  \end{tikzpicture}
  }
  \hspace{20pt}
  \subfigure[Sufficient condition ${\cal C}_{S}$]{
  \begin{tikzpicture}
    \tikzstyle{every node}=[font=\small]
    \draw[rounded corners=12pt] (0,0) rectangle (6,3.5);
    \draw[rounded corners=8pt] (.5,1.9) rectangle (5.5,3);
    \path (.5,2.5) node[right] {${\cal C_S}(\G) \rightarrow$ success guaranteed};
    \path (.2,1) node[right] (in){$\neg{\cal C_S}(\G)$};
    \path (2.3,1.5) node[right, text width=3.3cm] (ntight){no information};
    \path (2.3,.5) node[right, text width=3.3cm] (tight){$\exists$ failure possible};
    \draw[->] (in)-- node[above,sloped,font=\scriptsize]{$\neg$\,tight}(ntight.west);
    \draw[->] (in)-- node[below,sloped,font=\scriptsize]{tight}(tight.west);
  \end{tikzpicture}
  }
  \caption{\label{fig:tightness}Tightness of conditions.}
\end{figure}
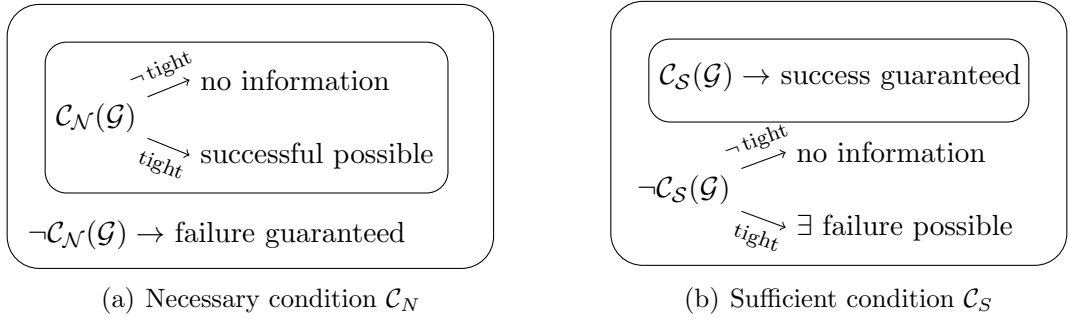

Given a condition (necessary or sufficient), an important question is whether it is {\em tight} for the considered algorithm. Marchand de Kerchove and Guinand~\cite{MG12} 
defined a {\em tightness} criterion as follows. Recall that a necessary condition is one whose non-satisfaction implies failure; it is {\em tight} if, in addition, its satisfaction does make success {\em possible} ({\ie} a nice adversary {\em could} make it succeed). Symmetrically, a sufficient condition is tight if its non-satisfaction does make failure possible (the adversary can make it fail). This is illustrated on Figure~\ref{fig:tightness}.

It was observed~\cite{MG12} that all of the conditions in~\cite{CCF09} are tight. Interestingly, this implies that, while the adversary has no power at all in the case of the first counting algorithm, it has a lot in the case of the second.

\subsection{Relating conditions to graph classes}

Each of the above conditions naturally induces a {\em class} of dynamic networks in which the corresponding property is satisfied. In fact, if a node plays a distinguished role in the algorithm (non-uniform initialization), then at least two classes of graphs are naturally induced by each property, based on quantification (one existential, one universal). These classes correspond to graphs in which \dots
\begin{myitem}
\item \dots at least one node can reach all the others through a journey ($1\leadsto *$),
\item \dots all nodes can reach each other through journeys ($*\leadsto *$),
\item \dots at least one node shares at some point an edge with every other ($1-*$),
\item \dots all pairs of nodes share at some point an edge ($*-*$),
\item \dots at least one node can be reached by all others through a journey ($*\leadsto 1$),
\item \dots at least one node can reach all the others through a strict journey ($1 \leadstoo *$),
\item \dots all nodes can reach each other through strict journeys ($* \leadstoo *$).
\end{myitem}

These classes were assigned an ``$\cal F_i$'' number in~\cite{Cas07,CCF09} and for some a distinct ``$\cal C_i$'' number in~\cite{CFQS12}. We revisit all the classes and their names in Chapter~\ref{sec:classes}, and we review algorithms for testing membership of a given graph to each of them.

\subsection{Towards formal proofs}

As explained above, one of the advantages of working at a high level of abstraction with atomic communication (here, graph relabelings) is that impossibility results obtained in these models are general, \ie, they apply automatically to lower-level models like message passing. Another important feature of such models is that they are well suited for formalization, and thereby for formal proofs.

From 2009 to 2012~\cite{coq09,coq11}, Cast{\'e}ran {\it et al.} developed an early set of tools and methodologies for formalizing graph relabeling systems within the framework of the {\em Coq} proof assistant, materializing as the {\small \sc Loco} library. More recently, Corbineau {\it et al.} developed the {\small \sc Padec} library~\cite{padec}, which allows one to build certified proofs (again with {\em Coq}) in a computational model called {\em locally shared memory model with composite atomicity}, much related to graph relabeling systems. This library is being developed intensively and may in the mid term incorporate other models and features. Besides the {\em Coq} realm, recent efforts were made by Fakhfakh {\em et al.} to prove the correctness of algorithms in dynamic networks based on the Event-B framework~\cite{FatenGeneral,FatenForest}. These algorithms are formalized using graph relabeling systems (in particular, one of them is our {\em spanning forest} algorithm presented in Section~\ref{sec:forest}). Another work pertaining to certifying distributed algorithms for mobile robots in the {\em Coq} framework, perhaps less directly relevant here, is that of Balabonski {\it et al.}~\cite{robots-coq}.

\begin{avenue}
  Building on top of these plural (and related) efforts, formalize in the framework of Coq or Event-B the main objects involved in this section, namely sequences of graphs, relabeling algorithms, and the fairness condition for edge selection. Use them to prove {\em formally} that a given assumption on the network dynamics is necessary or sufficient to a given algorithm.
\end{avenue}

\section{Shortest, fastest, and foremost broadcast}
\label{sec:shfafo}
\articles{IFIP-TCS'10~\cite{CFMS10} and IJFCS'15~\cite{CFMS15}}

We reviewed in Section~\ref{sec:redefinition} different ways in which the time dimension impacts the formulation of distributed and combinatorial problems. One of them is the declination of optimal journeys into three metrics: shortest, fastest, and foremost defined in~\cite{BFJ03} in a context of centralized offline problems.
In a series of work with Flocchini, Mans, and Santoro~\cite{CFMS10,CFMS15}, we studied a distributed version of these problems, namely shortest, fastest, and foremost {\em broadcast} with termination detection at the emitter, in which the evolution of the network is not known to the nodes, but must obey various forms of regularities (\ie be in various classes of dynamic networks). Some of the findings were surprising, for example the fact that the three variants of the problems require a gradual set of assumptions, each of which is strictly included in the previous one.

\subsection{Broadcast with termination detection (TDB)}
 The problem consists of broadcasting a piece of information from a given source (emitter) to all other nodes, with termination detection at the source (TDB, for short). Only the broadcasting phase is required to be optimal, not the termination phase. The metrics were adapted as follows:

\begin{itemize}
\item \foremost: every node is informed at the earliest possible time,
\item \shortest: the number of hops relative to every node is minimized,
\item \fastest: the time between first emission and last reception is globally minimized.
\end{itemize}

These requirements hold relative to a given initiation time $t$, which is triggered externally at the initial emitter.
Since the schedule of the network is not known in advance, we examine what minimal {\em regularity} in the network make TDB feasible, and so for each metric. Three cases are considered:
\begin{itemize}
  \item Class \ER ({\em recurrent} edges): graphs whose {\em footprint is connected} (not necessarily complete) and every edge in it re-appears infinitely often. In other words, if an edge is available once, then it will be available recurrently.
\item Class $\EB \subset \ER$ ({\em bounded-recurrent} edges): graphs in which every edge of the footprint cannot be absent for more than $\Delta$ time, for a fixed $\Delta$.

\item Class $\EP \subset \EB$ ({\em periodic} edges) where every edge of the footprint obeys a periodic schedule, for some period $p$. (If every edge has its own period, then $p$ is their least common multiple.)
\end{itemize}

As far as {\em inclusion} is concerned, it holds that 
 $\EP \subset \EB \subset \ER$ and the containment is strict. However, we show that being in either class only helps if additional {\em knowledge} is available.
The argument appears in different forms in~\cite{CFMS10,CFMS15} and proves quite ubiquitous -- let us call it a ``late edge'' argument.

{\bf Late edge argument.}
If an algorithm is able to decide termination of the broadcast in a network $\G$ by time $t$, then one can design a second network $\G'$ indistinguishable from $\G$ up to time $t$, with an edge appearing for the first time after time $t$. Depending on the needs of the proof, this edge may (1) reach new nodes, (2) create shortcuts, or (3) make some journeys faster.

In particular, this argument makes it clear that the nodes cannot decide when the broadcast is complete, unless additional knowledge is available. We consider various combinations of knowledge among the following: number of nodes $n$, a bound $\Delta$ on the reappearance time (in $\EB$), and the period $p$ (in $\EP$), the resulting settings are referred to as $\ER_n$, $\EB_\Delta$, etc. 

\paragraph{The model.} A message passing model in continuous time is considered, where the latency $\zeta$ is fixed and known to the nodes. When a link appears, it lasts sufficiently long for transmitting at least one message. If a message is sent less than $\zeta$ time before the edge disappears, it is lost. 
Nodes are notified immediately when an incident link appears ({\tt onEdgeAppearance()}) or disappears ({\tt onEdgeDisappearance()}), which we call a {\em presence oracle} in the present document. 
Together with $\zeta$ and the fact that links are bidirectional, the immediacy of the oracle implies that a node transmitting a message can detect if it was indeed successfully received (if the corresponding edge is still present $\zeta$ time after the emission). Based on this observation, we introduced in~\cite{CFMS10} a special primitive {\tt sendRetry()} that re-sends a message upon the next appearance of an edge if the transmission fails (or if the edge is absent when called), which simplifies the expression of algorithms w.l.o.g. Finally, a node can identify an edge over multiple appearances, and we do not worry about interferences. 

In a subsequent work with G\'omez-Calzado, Lafuente, and Larrea~\cite{GCLL15} (reviewed in Section~\ref{sec:carlos}), we explored various ways of relaxing these assumptions, in particular the presence oracle. Raynal {\it et al.}~\cite{RSCW14} also explored different variants of this model.

\subsection{Main results}
\label{sec:basic-results}
\label{sec:foremost}

We review here only the most significant results from~\cite{CFMS10,CFMS15}, referring the reader to these articles for missing details. The first problem, \foremost, can be solved already in $\ER_n$ by a basic flooding technique: every time a new edge appears locally to an informed node, information is sent onto it. Knowledge of $n$ is not required for the broadcast itself, but for termination detection due to a late edge argument.
Using the parent-child relations resulting from the broadcasting phase, termination detection proceeds by sending acknowledgments up the tree back to the emitter every time a new node is informed, which is feasible thanks to the recurrence of edges. The emitter detects termination after $n-1$ acknowledgments have been received.

\shortest and \fastest are not feasible in $\ER_n$ because of a late edge argument (in its version 2 and 3, respectively).
Moving to the more restricted class $\EB$, observe first that being in this class without knowing $\Delta$ is indistinguishable from being in $\ER$. Knowing $\Delta$ makes it possible for a node to learn its incident edges {\em in the footprint}, because these edges {\em must} appear at least once within any window of duration $\Delta$. 
The main consequence is that the nodes can perform a {\em breadth-first search (BFS)} relative to the footprint, which guarantees the shortest nature of journeys. It also makes it possible for a parent to know its definitive list of children, which enables recursive termination using a linear number of messages (against $O(n^2)$ in the termination described above).
Knowing both $n$ and $\Delta$ further improves the termination process, which now becomes implicit after $\Delta n$ time units.

\begin{remark} 
\label{rem:delta-n}
$\EB_\Delta$ is strictly stronger than $\EB_n$ in the considered model, because $n$ can be inferred from $\Delta$, the reverse being untrue~\cite{CFMS15}. In fact, the whole footprint can be learned in $\EB_\Delta$ with potential consequences on many problems.
\end{remark}

\fastest remains unsolvable in $\EB$.
In fact, one may design a network in $\EB$ where fastest journeys do not exist because the journeys keep improving infinitely many times, despite $\Delta$, exploiting here the continuous nature of time. (We give such a construct in~\cite{CFMS15}.)
Being in $\EP_p$ prevents such constructs and makes the problem solvable. In fact, the whole schedule becomes learnable in $\EP_p$ with great consequences. In particular, the source can learn the exact time (modulo $p$) when it has minimum {\em temporal eccentricity} (\ie when it takes the smallest time to reach all nodes), and initiate a foremost broadcast at that particular time (modulo $p$), which {\em will} be fastest. Temporal eccentricities can be computed using \tclocks~\cite{CFMS12b,CFMS14}, reviewed in Section~\ref{sec:temporal-lags}.

\begin{remark}
  A potential risk in continuous time (identified by E. Godard in a private communication) is that the existence of accumulation points in the presence function might prevent fastest journeys to exist at all even in $\EP$. Here, we are on the safe side thanks to the fact that every edge appears at least for $\zeta$ time units, which is constant is the considered model.
\end{remark}

Missing results are summarized through Tables~\ref{tab:results-feasibility} and \ref{tab:results-complexity}. Besides {\em feasibility}, we characterized the time and message complexity of all algorithms, distinguishing between information messages and other (typically smaller) control messages.
We also considered the {\em reusability} of structures from one broadcast to the next, \eg the underlying paths in the footprint. Interestingly, while some versions of the problem are harder to solve, they offer a better reusability. Some of these facts are discussed further in Section~\ref{sec:relations}.

\begin{table}[t]
  \begin{center}
    \renewcommand{\arraystretch}{1}
    \begin{tabular}{|c|c|c||c|c|l|}
      \hline
      Metric & Class &Knowledge& Feasibility &Reusability&~~~~~~Result from\\
      \hline
      \multirow{4}{*}{Foremost}&\ER&     $\emptyset$& no & --&$\verb+\+$\\
      ~&       \ER&             $n$& yes & no&~\cite{CFMS10} (long \cite{CFMS15})\\
      ~& \EB  &        $\Delta$ & yes & no&$/$\\
      ~&\EP&             $p$& yes & yes&\cite{CFMS11}  (long \cite{CFMS14})\\
      \hline 
      \multirow{3}{*}{Shortest}&\ER&     $\emptyset$&no &--&$\verb+\+$\\
      ~&       \ER&             $n$   &  no &--&~\cite{CFMS10}  (long \cite{CFMS15})\\
      ~& \EB &        $\Delta$ & yes & yes&$/$\\
      \hline
      \multirow{4}{*}{Fastest}&\ER&     $\emptyset$&no &--&$\verb+\+$\\
      ~&       \ER&             $n$   &  no &--&~\cite{CFMS10}  (long \cite{CFMS15})\\
      ~& \EB  &        $\Delta$ & yes & no&$/$\\
      ~&\EP&             $p$& yes & yes&\cite{CFMS12b}  (long \cite{CFMS14})\\
      \hline 
    \end{tabular}
    \renewcommand{\arraystretch}{1.2}
  \end{center}
  \caption{\label{tab:results-feasibility}Feasibility and reusability of \tdbroadcast in different classes of dynamic networks (with associated knowledge).}
\end{table}

 \begin{table}[t]
   \begin{center}
     \footnotesize
 \begin{tabular}{|@{~}c@{~}|@{~}c@{~}|@{~}c@{~}|@{~}c@{~}|@{~}c@{~}|@{~}c@{~}||@{~}c@{~}|@{~}c@{~}|}
\hline
Metric&Class&Knowl.&Time&Info. msgs&Control msgs&Info. msgs&Control msgs\\
~&~&~&~&($1^{st}$ run)&($1^{st}$ run)&(next runs)&(next runs)\\\hline
  Foremost&$\ER$&     $n$&unbounded&$O(m)$&$O(n^2)$&$O(m)$ &$O(n)$\\
         ~&$\EB$&     $n$&$O(n\Delta)$&$O(m)$&$O(n^2)$&$O(m)$&$O(n)$\\
         ~&       ~&        $\Delta$&$O(n\Delta)$&$O(m)$&$O(n)$&$O(m)$&0\\
         ~&       ~&   $n \& \Delta$&$O(n\Delta)$&$O(m)$&0&$O(m)$&0\\
         \hline
  Shortest&$\EB$&        $\Delta$&$O(n\Delta)$&$O(m)$&$2n-2$&$O(n)$&0\\
         \multirow{2}{*}{{\it either of} {\Large \{}}&       ~&   $n \& \Delta$&$O(n\Delta)$&$O(m)$&$n-1$&$O(n)$&0\\
         &       ~&   $n \& \Delta$&$O(n\Delta)$&$O(m)$&0&$O(m)$&0\\
  \hline 
\end{tabular}
\end{center}
\caption{\label{tab:results-complexity}Complexity of \tdbroadcast in $\ER$ and $\EB$ with related knowledge (Table from~\cite{CFMS15}). Control messages are typically much smaller than information messages, and thus counted separately.
}
\end{table}

\begin{open}
  While $\ER$ is more general (weaker) than $\EB$ and $\EP$, it still represents a strong form of regularity. The recent characterization of Class~$\TCR$ in terms of eventual footprint~\cite{BDKP16} (see Section~\ref{sec:robustness} in the present chapter) makes the inner structure of this class more apparent. It seems plausible to us (but without certainty) that sufficient structure may be found in \TCR to solve \foremost with knowledge $n$, which represents a significant improvement over $\ER$.
\end{open}

\section{Bounding the temporal diameter}
\label{sec:carlos}
\article{EUROPAR'15~\cite{GCLL15}}

Being able to bound communication delays in a network is instrumental in solving a number of distributed tasks. It makes it possible, for instance, to distinguish between a crashed node and a slow node, or to create a form of synchronicity in the network. In highly-dynamic networks, the communication delay between two (remote) nodes may be arbitrary long, and so, even if the communication delay between every two neighbors are bounded. This is due to the disconnected nature of the network, which de-correlates the global delay ({\em temporal diameter}) from local delays ({\em edges latencies}).

In a joint work with G\'omez, Lafuente, and Larrea~\cite{GCLL15}, we explored different ways of bounding the temporal diameter of the network and of exploiting such a bound. This work was first motivated by a problem familiar to my co-authors, namely the agreement problem, for which it is known that a subset of sufficient size (typically a majority of the nodes) must be able to communicate timely. For this reason, the work in~\cite{GCLL15} considers properties that apply among {\em subsets} of nodes (components). Here, we give a simplified account of this work, focusing on the case that these properties apply to the {\em whole network}. One reason is to make it easier to relate these contributions to the other works presented in this document, while avoiding many details. The reader interested more specifically in the agreement problem, or to finer (component-based) versions of the properties discussed here is referred to~\cite{GCLL15}. The agreement problem in highly-dynamic networks has also been studied in a number of recent works, including for example~\cite{GAS11,matthias}.

\paragraph{The model.}
The model is close to the one of Section~\ref{sec:shfafo}, with some relaxations. Namely, time is continuous and the nodes communicate using message passing. Here, the latency $\zeta$ is not a constant, which induces a partial asynchrony, but it remains bounded by some $\zeta_{MAX}$.
Different options are considered regarding the awareness that nodes have of their incident links, starting with the use of a {\em presence oracle} as before (nodes are immediately notified when an incident link appears or disappears). Then, we explore possible replacements for such an oracle, which are described gradually. As before, a node can identify a same edge over multiple appearances, and we do not worry about interferences.

\subsection{Temporal diameter}
\label{sec:temporal-diameter}

Let us recall that the temporal diameter of the network, at time $t$, is the smallest duration $d$ such that $\G_{[t,t+d]}$ is temporally connected (\ie $\G_{[t,t+d]} \in \TC$). The objective is to guarantee the existence of a bound $\Delta$ such that $\G_{[t,t+\Delta]}\in \TC$ for all $t$, which we refer to as having a {\em bounded temporal diameter}.
The actual definitions in~\cite{GCLL15} rely on the concept of $\Delta$-journeys, which are journeys whose duration is bounded by $\Delta$. Based on these journeys, a concept of $\Delta$-component is defined as a set of nodes able to reach each other within every time window of duration $\Delta$.

Networks satisfying a bounded temporal diameter are said to be in class $\TC(\Delta)$ in~\cite{GCLL15}. For consistency with other class names in this document, we rename this class as $\TCB$, mentioning the $\Delta$ parameter only if need be. (A more general distinction between parametrized and non-parametrized classes, inspired from~\cite{GCLL15}, is discussed in Chapter~\ref{sec:classes}.)
At an abstract level, being in \TCB with a known $\Delta$ makes it possible for the nodes to make decisions that depend on (potentially) the whole network following a wait of $\Delta$ time units.
However, if no additional assumptions are made, then one must ensure that no single opportunity of journey is missed by the nodes. Indeed, membership to \TCB may rely on specific journeys whose availability are transient. In discrete time, a feasible (but costly) way to circumvent this problem is to send a message in each time step, but this makes no sense in {\em continuous} time. 

This impossibility motivates us to write a first version of our algorithms in~\cite{GCLL15} using the presence oracles from~\cite{CFMS10,CFMS15}; \ie primitives of the type {\tt onEdgeAppearance()} and {\tt onEdgeDisappearance()}. However, we observe that these oracles have no realistic implementations and thus we explore various ways of avoiding them, possibly at the cost of stronger assumptions on the network dynamics (more restricted classes of graphs).

\subsection{Link stability}
\label{sec:stability}
Instead of {\em detecting} edges, we study how $\TCB$ could be specialized for enabling a similar trick to the one in discrete time, namely that if the nodes send messages at {\em regular} interval, at least {\em some} of the possible $\Delta$-journeys will be effectively used. The precise condition quite specific and designed to this sole objective. Precisely, we require the existence of particular kinds of $\Delta$-journeys (called $\beta$-journeys in~\cite{GCLL15}) in which every next hop can be performed with some flexibility as to the exact transmission time, exploiting a stability parameter $\beta$ on the duration of edge presences.
The name $\beta$ was inspired from a similar stability parameter used by Fern\'andez-Anta et al.~\cite{AMMZ12} for a different purpose.

The graphs satisfying this requirement (for some $\beta$ and $\Delta$) form a strict subset of $\TCB$ for the same $\Delta$. The resulting class was denoted by $\TC'(\beta)$ in~\cite{GCLL15}, and called {\em oracle-free}. 
We now believe this class is quite specific, and may preferably be formulated in terms of communication model within the more general class $\TCB$. This matter raises an interesting question as to whether and when a set of assumptions should be stated as a class of {\em graphs} and when it should not. No such ambiguity arises in static networks, where computational aspects and synchronism are not captured by the graph model itself, whereas it becomes partially so with graph theoretical models like TVGs, through the latency function. These aspects are discussed again in Section~\ref{disc:multi-dimensional}.

\subsection{Steady progress}

While making the presence oracle unnecessary, the stability assumption still requires the nodes to send the message regularly over potentially long periods of time.
Fern\'andez-Anta {\em et al.} consider another parameter called $\alpha$ in~\cite{AMMZ12}, in a discrete time setting. The parameter is formulated in terms of partitions within the network, as the largest number of consecutive steps, for every partition $(S,\bar{S})$ of $V$, without an edge between $S$ and $\bar{S}$. This idea is quite general and extends naturally to continuous time. Reformulated in terms of journeys, parameter $\alpha$ is a bound on the time it takes for {\em every next hop} of {\em some} journeys to appear, ``some'' being here at least one between every two nodes (and in our case, within every $\Delta$-window). 

One of the consequences of this parameter is that a node can {\em stop} retransmitting a message $\alpha$ time units after it received it, adding to the global communication bound a second {\em local} one of practical interest. In~\cite{GCLL15}, we considered $\alpha$ only in conjunction with $\beta$, resulting in a new class based on $(\alpha,\beta)$-journeys and called $\TC''(\alpha,\beta) \subset \TC'(\beta) \subset \TCB$ (of which the networks in~\cite{AMMZ12} can essentially be seen as discrete versions).
With hindsight, the $\alpha$ parameter deserves to be considered independently. In particular, a concept of $\alpha$-journey where the time of every next hop is bounded by some duration is of great independent interest, and it is perhaps of a more {\em structural} nature than $\beta$. As a result, we do consider an $\alpha$-$\TCB$ class in Chapter~\ref{sec:classes} while omitting classes based on $\beta$.

{\bf Impact on message complexity.} Intuitively, the $\alpha$ parameter makes it possible to reduce drastically the number of messages. However, its precise effects in an adversarial context remain to be understood. 
In particular, a node has no mean to decide which of several journeys {\em prefixes} will eventually lead to an $\alpha$-journey. As a result, even though a node can stop re-transmitting a message after $\alpha$ time units, it will have to retransmit the same message again if it receives it again in the future (\eg possibly through a different route). 

\begin{open}
Understand the real effect of $\alpha$-journeys in case of an adversarial (\ie worst-case) edge scheduling $\rho$. In particular, does it significantly reduces the number of messages?
\end{open}

In conclusion, the properties presented above helped us propose in~\cite{GCLL15} different versions of a same algorithm (here, a primitive called {\em terminating reliable broadcast} in relation to the agreement problem). Each version lied at different level of abstraction and with a gradual set of assumptions, offering a tradeoff between realism and assumptions on the network dynamics.

\section{Minimal structure and robustness}
\article{arXiv'17~\cite{CDPR17} (submitted)}
\label{sec:robustness}

As we have seen along this chapter, highly dynamic networks may possess various types of internal structure that an algorithm can exploit. Because there is a natural interplay between generality and structure, an important question is whether general types of dynamic networks possess sufficient structure to solve interesting problems.
Arguably, one of the weakest assumptions in dynamic networks is that every pair of nodes is able to communicate recurrently (infinitely often) through journeys. This property was identified more than three decades ago by Awerbuch and Even~\cite{AE84}. The corresponding class of dynamic networks (Class~5 in~\cite{CFQS12} -- hereby referred to as \TCR) is indeed the most general among all classes of infinite-lifetime networks discussed in this document. This means that, if a structure is present in \TCR, then it can be found in virtually any network, including always-connected networks ($\AC$), networks whose edges are recurrent or periodic ($\ER$, $\EB$, $\EP$), and networks in which all pairs of nodes are infinitely often neighbors ($\KR$). (All classes are reviewed in Chapter~\ref{sec:classes}.) Therefore, we believe that the question is important. We review here a joint work with Dubois, Petit, and Robson~\cite{CDPR17}, in which we exploit the structure of \TCR to built stable intermittent structures.

\subsection{Preamble}

In Section~\ref{sec:covering}, we presented three ways of interpreting standard combinatorial problems in highly-dynamic networks~\cite{CMM11}, namely a {\em temporal}, an {\em evolving}, and a {\em permanent} version. The {\em temporal} interpretation requires that the considered property (\eg in case of dominating sets, being either in the set, or adjacent to a node in it) is realized {\em at least once} over the execution.

Motivated by a distributed setting, Dubois {\it et al.}~\cite{DKP15} define an extension of the temporal version, in which the property must hold not only once, but {\em infinitely often} -- in the case of dominating sets, this means being either in the set or {\em recurrently} neighbor to a node in the set. 
Aiming for generality, they focus on \TCR and exploit the fact that this class also corresponds to networks whose {\em eventual footprint} is connected~\cite{BDKP16}. In other words, if a network is in \TCR, then its footprint contains a connected spanning subset of edges which {\em are} recurrent, and vice versa. 

This observation is perhaps simple, but has profound implications.
While some of the edges incident to a node may disappear forever, some others {\em must} reappear infinitely often. Since a distributed algorithm has no mean to distinguish between both types of edges, Dubois {\it et al.}~\cite{DKP15} call a solution {\em strong} if it remains valid relative to the actual set of recurrent edges, whatever they be.

\subsection{Robustness and the case of the MIS}

In a joint work with Dubois, Petit, and Robson~\cite{CDPR17}, we revisited these notions, employing the terminology of ``robustness'' (suggested by Y. Métivier), and defining a new form of heredity in {\em standard} graphs, motivated by these considerations about {\em dynamic} networks.

\begin{definition}[Robustness]
  A solution (or property) is said to be {\em robust} in a graph $G$ iff it is valid in {\em all} connected spanning subgraphs of $G$ (including $G$ itself).
\end{definition}

This notion indeed captures the uncertainty of not knowing which of the edges are recurrent and which are not in the footprint of a network in $\TCR$. Then we realized that it also has a very natural motivation in terms of static networks, namely that some edges of a given network may crash definitely and the network is used so long as it is connected.
This duality makes the notion quite general and its study compelling. It also makes it simpler to think about the property.

In~\cite{CDPR17}, we focus on {\em maximal independent sets} (MIS), which is a maximal set of non-neighbor nodes. Interestingly, robust MISs may or may not exist depending on the considered graph. 
\tikzsetnextfilename{tikz-robust}
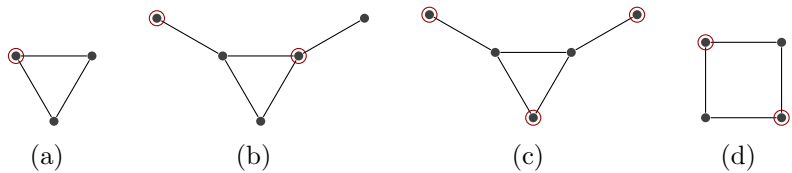
\begin{figure}[h]
  \centering
  \subfigure[]{
    \label{fig:mis-a}
    \begin{tikzpicture}
      \tikzstyle{every node}=[circle, inner sep=1.2pt, fill=darkgray]
      \path (0,0) node (a) {};
      \path (a)+(0:1) node (b) {};
      \path (a)+(-60:1) node (c) {};
      
      \draw (a)--(b)--(c)--(a);
      
      \tikzstyle{every node}=[circle, draw, red, inner sep=2pt]
      \path (a) node {};
    \end{tikzpicture}
  }
  ~
  \subfigure[]{
    \label{fig:mis-b}
    \begin{tikzpicture}
      \tikzstyle{every node}=[circle, inner sep=1.2pt, fill=darkgray]
      \path (0,0) node (a) {};
      \path (a)+(-30:1) node (b) {};
      \path (b)+(0:1) node (c) {};
      \path (b)+(-60:1) node (d) {};
      \path (c)+(30:1) node (e) {};
      
      \draw (a)--(b)--(c)--(e);
      \draw (b)--(d)--(c);
      
      \tikzstyle{every node}=[circle, draw, red, inner sep=2pt]
      \path (a) node {};
      \path (c) node {};
    \end{tikzpicture}
  }
  ~
  \subfigure[]{
    \label{fig:mis-c}
    \begin{tikzpicture}
      \tikzstyle{every node}=[circle, inner sep=1.2pt, fill=darkgray]
      \path (0,0) node (a) {};
      \path (a)+(-30:1) node (b) {};
      \path (b)+(0:1) node (c) {};
      \path (b)+(-60:1) node (d) {};
      \path (c)+(30:1) node (e) {};
      
      \draw (a)--(b)--(c)--(e);
      \draw (b)--(d)--(c);
      
      \tikzstyle{every node}=[circle, draw, red, inner sep=2pt]
      \path (a) node {};
      \path (d) node {};
      \path (e) node {};
    \end{tikzpicture}
  }
  ~
  \subfigure[]{
    \label{fig:mis-d}
    \begin{tikzpicture}
      \tikzstyle{every node}=[circle, inner sep=1.2pt, fill=darkgray]
      \path (0,0) node (a) {};
      \path (a)+(0:1) node (b) {};
      \path (b)+(-90:1) node (c) {};
      \path (c)+(-180:1) node (d) {};
      
      \draw (a)--(b)--(c)--(d)--(a);
      
      \tikzstyle{every node}=[circle, draw, red, inner sep=2pt]
      \path (a) node {};
      \path (c) node {};
    \end{tikzpicture}
  }
\vspace{-0.25cm}
  \caption{\label{fig:examples} Four examples of MISs in various graphs (or footprints).}
\end{figure}
For example, if the graph is a triangle (Figure~\ref{fig:mis-a}), then only one MIS exists up to isomorphism, consisting of a single node. However, this set is no longer maximal in one of the possible connected spanning subgraphs: the triangle graph
admits no robust MIS. Some graphs do admit a robust MIS, but not all of the MISs are robust. 
Figures~\ref{fig:mis-b} and~\ref{fig:mis-c} show two MISs
in the bull graph, only one of which is robust. Finally,
 some graphs are such that {\em all} MISs are robust (\eg the square on Figure~\ref{fig:mis-d}). 

We characterize exactly the set of graphs in which all MISs are robust~\cite{CDPR17}, denoted \forallMIS, and prove that it consists {\em exactly} of the union of complete bipartite graphs and a new class of graphs called {\em sputniks}, which contains among others things all the trees (for which any property is trivially robust). 
Graphs not in \forallMIS may still {\em admit} a robust MIS, \ie be in \existsMIS, such as the bull graph on Figure~\ref{fig:examples}. However, the characterization of \existsMIS proved quite complex, and instead of a closed characterization, we presented in~\cite{CDPR17} an algorithm that finds a robust MIS if one exists, and rejects otherwise.
Interestingly, our algorithm has low polynomial complexity despite the fact that exponentially many MISs {\em and} exponentially many connected spanning subgraphs may exist.

We also turn to the distributed version of the problem, and prove that finding a robust MIS in \forallMIS is a {\em local} problem, namely a node can decide whether or not it belongs to the MIS by considering only information available within $\frac{\log n}{\log \log n} = o(\log n)$ hops in the graph (resp. in the footprint in case of dynamic networks). On the other hand, we show that finding a robust MIS in \existsMIS (or deciding if one exists in general graphs) is {\em not} local, as it may require information up to $\Omega(n)$ hops away,
which implies a separation between the MIS problem and the robust MIS problem in general graphs, since the former is solvable within $2^{\mathcal{O}(\sqrt{\log n})}$ rounds in the LOCAL model~\cite{PS96}.

\subsubsection{Some remarks}

Whether a closer characterization of \existsMIS exists in terms of natural graph properties remains open. (It might be that it does not.) 
It would be interesting, at least, to understand how large this {\em existential} class is compared to its {\em universal} counterpart \forallMIS. Another general question is whether large {\em universal} classes exist for other combinatorial problems than MIS, or if the notion is somewhat too restrictive for the universal versions of these classes. Of particular interests are other symmetry breaking problems like MDS or k-coloring, which play an important role in communication networks.

\newcommand{\nocontentsline}[3]{}
\newcommand{\tocless}[2]{\bgroup\let\addcontentsline=\nocontentsline#1{#2}\egroup}
\tocless\section{Conclusion and perspectives}

Identifying necessary or sufficient conditions for distributed problems has been a recurrent theme in our research. It has acted as a driving force and sparked off many of our investigations. Interestingly, the structure revealed through them often turn out to be quite general and of a broader applicability. The next chapter reviews all the classes of graphs found through these investigations, together with classes inferred from other assumptions found in the literature.

We take the opportunity of this conclusion to discuss a matter which we believe is important and perhaps insufficiently considered by the distributed computing community: that of structures available a finite number of times. Some distributed problems may require a certain number $k$ of occurrences of a structural property, like an edge appearance of a journey. If an algorithm requires $k_1$ such occurrences and another algorithm requires $k_2$ such occurrences, an instinctive reaction is to discard the significance of their difference, especially if $k_1$ and $k_2$ are of the same order, on the basis that constant factors among various complexity measures is not of utmost importance. The missed point here is that, in a dynamic network, such difference may not only relate to complexity, but also to feasibility! A dynamic network $\G$ may typically enable only a finite number of occurrences of some desired structure, \eg three round-trip journeys between all the nodes. If an algorithm requires only three such journeys and another requires four, then the difference between both is highly significant.

For this reason, we call for the definition of structural metrics in dynamic networks which may be used to characterize {\em fine-grained} requirements of an algorithm. Early efforts in this direction, motivated mainly by complexity aspects (but with similar effects), include Bramas and Tixeuil~\cite{BT16}, Bramas {\it et al.}~\cite{BMT16}, and Dubois {\it et al.}~\cite{DKP15}.

\begin{avenue}
  \label{open:complexity-measure} 
  Systematize the definition of complexity measures based on temporal features which may be available on a non-recurrent basis. Start comparing algorithms based on the number of occurrences they require of these structures.
\end{avenue}

\chapter{Around classes of dynamic networks}
\label{sec:classes}

In the same way as standard graph theory identifies a large number of special classes of graphs (trees, planar graphs, grids, complete graphs, {\it etc.}), we review here a collection of {\em classes of dynamic graphs} introduced in various works. Many of these classes were compiled in the context of a joint work with Flocchini, Quattrociocchi, and Santoro in 2012~\cite{CFQS12}, some others defined in a joint work with Chaumette and Ferreira in 2009~\cite{CCF09}, with G\'omez-Calzado, Lafuente, and Larrea 2015~\cite{GCLL15}, and with Flocchini, Mans, and Santoro~\cite{CFMS10,CFMS15}; finally, some are original generalizations of existing classes. We resist the temptation of defining a myriad of classes by limiting ourselves to properties {\em used effectively in the literature} (often in the form of necessary or sufficient conditions for distributed algorithms, see Chapter~\ref{sec:feasibility}).

Here, we get some distance from distributed computing and consider the intrinsic features of the classes and their inter-relations from a set-theoretic point of views. Some discussions are adapted from the above papers, some are new; the existing ones are revisited with (perhaps) more hindsight. In a second part, we review our efforts related to testing automatically properties related to these classes, given a {\em trace} of a dynamic network. Finally, we discuss the connection between classes of graphs and real-world mobility, with an opening on the emerging topic of movement synthesis.

\section{List of classes}

\articles{
SIROCCO'09~\cite{CCF09},
IJPEDS'12~\cite{CFQS12},
IJFCS'15~\cite{CFMS15},
EUROPAR'15~\cite{GCLL15}.}

The classes listed below are described mostly in the language of time-varying graphs (see definitions in Section~\ref{sec:graph-models}). However, the corresponding properties are conceptually general and may be expressed in other models as well. For generality, we formulate them in a continuous time setting (nonetheless giving credit to the works introducing their discrete analogues). Common restrictions apply in the whole section. In particular, we consider networks with a {\em fixed} number of nodes $n$. We consider only edge appearances which lasts sufficiently long for the edge to be used (typically in relation to the latency); for example, when we say that an edge ``appears at least once'', we mean implicitly for a sufficient duration to be used. The exact meaning of {\em being used} is vague to accomodate various notions of interaction (\eg atomic operations or message exchanges).
Finally, we limit ourselves to undirected edges, which impacts inclusion relations among classes. 

\paragraph{Classes names.} Opportunity is taken to give new names to the classes. Some classes were assigned a ${\cal C}_{x}$ name in some of our previous works and some others a ${\cal F}_x$ name (sometimes for the same class). We hope the new names convey mnemonic information which will prove more convenient. In particular, the {\em base} letter indicates the subject of the property applies: journeys ($\cal J$), edges ($\cal E$), paths ($\cal P$), connectivity ($\cal C$), temporal connectivity ($\cal TC$). The superscript provides information about the property itself: recurrent ($\cal R$), bounded recurrent ($\cal B$), periodic ($\cal P$), round-trip ($\circlearrowleft$), one to all ($1\forall$), all to one ($\forall1$), and so on. Note that the letter ${\cal P}$ is used twice in this convention, but its position (base or superscript) makes it non-ambiguous.

\subsection{Classes based on {\em finite} properties}
\label{sec:finite-properties}

We say that a graph property $Prop$ is {\em finite}
 if $Prop(\G_{[0,t]}) \implies Prop(\G)$ for some $t$. In other words, if by some time the property has been satisfied, then the subsequent evolution of the graph with respect to this property is irrelevant. Among other consequences, this makes the properties satisfiable by graphs whose lifetime is {\em finite} (with advantages for offline analysis).
Below are a number of classes based on finite properties. The contexts of their introduction in briefly reminded in Section~\ref{sec:literature}.

\begin{class*}[\JOA (Temporal source)]
  $\exists u\in V, \forall v\in V, u \leadsto v$. At least one node can reach all the others through a journey.
\end{class*}

\begin{class*}[\JAO (Temporal sink)]
  $\exists v\in V, \forall u\in V, u \leadsto v$. At least one node can be reached by all others through a journey. 
\end{class*}

\begin{class*}[\TC (Temporal connectivity)]
  $\forall u,v\in V, u \leadsto v$. Every node can reach all the others through a journey.
\end{class*}

\begin{class*}[\RT\ (Round-trip temporal connectivity)]
  $\forall u,v\in V, \exists \J_1\in \J^*_{(u,v)}, \exists \J_2\in \J^*_{(v,u)}, arrival(\J_1) \le departure(\J_2)$. Every node can reach every other node and be reached from that node afterwards.
\end{class*}

\begin{class*}[\EOA (Temporal star)]
  $\exists u\in V, \forall v\in V, \exists t\in \T, (u,v)\in G_t$. At least one node will share an edge at least once with every other node (possibly at different times).
\end{class*}

\begin{class*}[\KG (Temporal clique)]
  $\forall u,v\in V, \exists t\in \T, (u,v)\in G_t$. Every pair of nodes will share an edge at least once (possibly at different times).
\end{class*}

\begin{subclass*}[\KG (Counsils)]
Three subclasses of~\KG are defined by Laplace~\cite{Laplace12}, 
called $\F_{OC}$ (open counsil), $\F_{PCC}$ (punctually-closed counsil), and $\F_{CC}$ (closed counsil) corresponding to gradual structural constraints in~\KG. For instance, $\F_{OC}$ corresponds to network in which the nodes gather incrementally as a clique that eventually forms a complete graph, each node joining the clique at a {\em distinct} time.

\end{subclass*}

\boite{\label{disc:strict}On the relevance of strict journeys}{
Some of the classes presented here were sometimes separated into a {\em strict} and a {\em non-strict} version, depending on which kind of journey is guaranteed to exist. 
However, the very notion of a {\em strict} journey is more relevant in discrete time, where a clear notion of time step exists. In constrast, an explicit latency is commonly considered in continous time and unifies both versions elegantly (non-strict journeys simply correspond to assuming a latency of $0$). 
}

In some of our works (\eg~\cite{CCF09}), we considered strict versions of some classes. We maintain the distinction between both versions in some of the content below, due to (1) the reporting nature of this document, and (2) the fact that many works target discrete time. In particular, we suggest to denote strict versions of the classes using a superscript ``$>$'' applied to the class name, such as in \JOAST, \TCST, and \RTST. Later in the section, we will ignore the disctinction between both, in particular in the updated hierarchy given in Figure~\ref{fig:hierarchy-complete}.

\subsection{Classes based on {\em recurrent} properties}
\label{sec:recurrent-properties}

When the lifetime is {\em infinite}, temporal connectivity is often taken for granted and more elaborate properties are considered. We review below a number of such classes, most of which were compiled in~\cite{CFQS12}. 
For readability, most domains of the variables are not repeated in each definition. By convention, variable $t$ always denotes a time in the lifetime of the network (\ie $t \in \T$), variables $\Delta$ and $p$ denote durations (in $\mathbb{T}$), variable $e$ denotes an edge (in $E$) and variables $u$ and $v$ denote vertices (in $V$), all relative to a network $\TVG$.

\begin{class*}[\TCR\ (Recurrent temporal connectivity)]
  \label{cl:recurrent-connectivity}
  For all $t$, $\G_{[t,+\infty)} \in \TC$. At any point in time, the network will again be temporally connected in the future.
\end{class*}

\begin{class*}[\TCB\ (Bounded temporal diameter)]
  \label{cl:recurrent-connectivity}
  $\exists \Delta, \forall t$, $\G_{[t,t+\Delta)} \in \TC$. At any point in time, the network is temporally connected within the next $\Delta$ units of time.
\end{class*}

\begin{class*}[\ER\ (Recurrent edges)]
  \label{cl:recurrent-edges}
  The footprint $(V,E)$ is connected and $\forall e\in E, \forall t, \exists t'>t, \rho(e,t')=1$. If an edge appears once, it appears infinitely often (or remains present).
\end{class*}

\begin{class*}[\EB\ (Bounded edge recurrence)]
  \label{cl:bounded-recurrent-edges}
  The footprint $(V,E)$ is connected and there is a $\Delta$ such that $\forall e \in E, \forall t, \exists t' \in [t, t+\Delta), \rho(e,t')=1$. If an edge appears once, then it always re-appears within bounded time (or remains present).
\end{class*}

\begin{class*}[\EP\ (Periodic edges)]
  \label{cl:periodic-edges}
  The footprint $(V,E)$ is connected and $\forall e\in E, \forall t, \rho(e,t)=\rho(e,t+kp)$, for some $p$ and all integer $k$. The schedule of every edge repeats modulo some period. (If every edge has its own period, then $p$ is the least common multiple.)
\end{class*}

\begin{class*}[\PR\ (Recurrent paths)]
  \label{cl:eventually-routable}
  $\forall u,v, \forall t, \exists t'>t$, a path from $u$ to $v$ exists in $G_{t'}$. A classical path will exist infinitely many times between every two vertices.
\end{class*}

\begin{class*}[\CR\ (Recurrently-connected snapshots)]
  \label{cl:eventually-connected}
  $\forall t, \exists t'>t, G_{t'}$ is connected. At any point in time, there will be a connected snapshot in the future.
\end{class*}
 
\begin{class*}[\AC\ (Always-connected snapshots)]
  \label{cl:always-connected}
  $\forall t, G_t$ is connected. At any point in time, the snapshot is connected.
\end{class*}

\begin{class*}[$\TINT$\ (T-interval connectivity)]
  \label{cl:t-interval-connected}
  For a given $T \in \mathbb{T}$, $\forall t$, $\cap\G_{[t,t+T)}$ is connected. For all period of length $T$, there is a spanning connected subgraph which is stable.
\end{class*}

\begin{class*}[$\KR$\ (Complete graph of interaction)]
  \label{cl:complete}
  The footprint $G=(V,E)$ is complete, and $\forall e, \forall t, \exists t'>t : \rho(e,t')$$=$$1$. Every pair of vertices share an edge infinitely often. (In other words, the eventual footprint is complete.)
\end{class*}

\begin{class*}[$\alpha$-$\TCB$ (Steady progress)]
  There is a duration $\alpha$ such that for all starting time $t$ and vertices $u$ and $v$, at least one (elementary) journey from $u$ to $v$ is such that every next hop occurs within $\alpha$ time units.
\end{class*}

The next two classes are given for completeness. We argued in Section~\ref{disc:relevance-classes} that they may preferably not be stated as independent classes, due to their high-degree of specialization, and rather stated as extra stability constraints within \TCB or $\alpha$-\TCB.

\begin{class*}[$\beta$-$\TCB$ (Bounded temporal diameter with stable links)]
  There is a maximal latency $\zeta_{MAX}$, durations $\beta\ge 2\zeta_{MAX}$ and $\Delta$ such that for all starting time $t$ and vertices $u$ and $v$, at least one journey from $u$ to $v$ within $[t, t+\Delta)$ is such that its edges are crossed in disjoint periods of length $\beta$, each edge remaining present in the corresponding period.
\end{class*}

\begin{class*}[$(\alpha,\beta)$-$\TCB$] $\alpha$-$\TCB$ $\cap$ $\beta$-$\TCB$.
\end{class*}

\subsection{Dimensionality of the assumptions}
  \label{disc:relevance-classes}
  \label{disc:multi-dimensional}

The complexity of the above description of $\beta$-$\TCB$ raises a question as to the {\em prerogatives} of a class. 
In static graphs, the separation between structural properties (\ie the network topology) and computational and communicational aspects (\eg synchronicity) are well delineated. The situation is more complex in dynamic networks, and especially in continuous time, due to the injection of time itself into the graph model.
For example, in time-varying graphs, the latency function encodes assumptions about synchronicity directly within the graph, making the distinction between structure and communication more ambiguous.

The down side of this expressivity is a temptation to turn any combination of assumptions into a dedicated class. We do not think this is a reasonable approach, neither do we have a definite opinion on this matter. Some types of assumptions remain non-ambiguously orthogonal to graph models (\eg size of messages, knowledge available to the nodes, unique identifiers), but it is no longer true that dynamic graph models capture only structural information. 
A discussion on related topics can also be found in Santoro~\cite{Santoro16}.

\begin{avenue}
  Clarify the prerogatives of (dynamic) graph models with respect to the multi-dimensionality of assumptions in distributed computing.
\end{avenue}

\subsection{Parametrized classes} 
\label{sec:parametrized}
With G\'omez-Calzado, Lafuente, and Larrea, we distinguish in~\cite{GCLL15} between two versions of some classes which admit parameters. Let us extend the discussion to parametrized classes in general, including for example \TCB, \EB, and \TINT, which are all parametrized by a {\em duration}. Given such a class, one should distinguish between the {\em instantiated} version (\ie for a given value of the parameter) and the {\em universal} version (union of instantiated versions over all finite parameter values). Taking class $\EB$ as an example, three notations can be defined: \EB is the instantiated version with an {\em implicit} parameter (as used above); $\EB(x)$ is the instantiated version with an {\em explicit} parameter; and $\cup\EB$ is the universal version. Some inclusions relations among classes, discussed next, are sensitive to this aspect, for example the fact that $\TINT \subseteq \AC \subseteq \cup\TINT$ brought some confusion in previous versions of the hierarchy (where this distinction was absent).

\subsection{Background in distributed computing}
\label{sec:literature}

We review here some of the contexts in which the above classes were introduced. We mainly focus on recurrent properties, the case of finite properties being already covered in Section~\ref{sec:conditions}.

\TCR is perhaps the most general class among the ones having infinite lifetime. This property corresponds to class ${\cal C}_5$ in our 2012 article~\cite{CFQS12}. It seems to have been first considered by Awerbuch and Even in the 80s~\cite{AE84}. Being in this class is often implicitly assumed in mobile ad hoc networks, as it captures the ability for the nodes to influence each other recurrently. We describe in Section~\ref{sec:robustness} a joint work with Dubois, Petit, and Robson~\cite{CDPR17}, in which the structure of \TCR is explored.

\TCB corresponds to the subset of \TCR in which the communication time is bounded by some value $\Delta$. This class corresponds in essence to classical assumptions made in static distributed computing, when the communication is asynchronous, but interpreted here as a {\em bounded temporal diameter} of a highly-dynamic network. Together with Gomez-Calzado, Lafuente, and Larrea, we explored in~\cite{GCLL15} (reviewed in Section~\ref{sec:carlos}) gradual restrictions of this class that make it possible to exploit the communication bound in realistic and efficient ways. One of the restriction corresponds to Class~$\alpha$-$\TCB$, where $\alpha$ stands for a parameter introduced by Fernandez-Anta {\it et al.} in~\cite{AMMZ12}, which we re-interpret in terms of the existence of journeys whose wait at {\em every} intermediate nodes is bounded (steady progress). Interestingly, this property also manifests with high probability in a wide range of edge-markovian dynamic graphs~\cite{BCF09}, making it possible to stop re-transmission after some time ({\it parsimonious} broadcast).

Classes \ER, \EB, and \EP were introduced in a joint work with Flocchini, Mans, and Santoro~\cite{CFMS10} (journal version~\cite{CFMS15}). These classes were shown to have a tight relation with the problems of foremost, shortest, and fastest broadcasts (with termination detection). The reader is referred to
Section~\ref{sec:shfafo} for details.
These three classes were later considered by Aaron {\it et al.} in~\cite{AKM14}, who show that the {\em dynamic map visitation} problem admit different complexities depending on the class, namely severe inapproximability in \ER, limited approximability in \EB, and tractability in \EP. A number of works on network exploration~\cite{FMS09,IW11,FKMS12a} and \cite{KerO09,LW09b} also considered periodical networks.

Classes \PR and \CR were introduced by Ramanathan {\it et. al} in~\cite{RBK07}; these classes capture, among other things, the ability to wait, for any two given nodes, that a snapshot of the network occurs where these nodes are connected by a standard {\em path} (Class~\PR); or to wait for a snapshot where all the nodes are connected (Class~\CR). The original paper does not mention that these features must hold infinitely often (only ``one or more'' times).

\KR are graphs from \ER whose {\em footprint} is a complete graph. In other words, every two nodes in the network are infinitely often neighbors. 
The network resulting from the random scheduler from~\cite{AAD+06} {\em almost surely} belongs to this class and inspired its definition. However, we take the opportunity to correct here a statement from~\cite{CFQS12}, in which we misattributed the results from~\cite{AAD+06} to~\KR. The results from~\cite{AAD+06} apply when a fairness criterion is satisfied, namely that every global state (configuration) that is infinitely often reachable is infinitely often reached. The random scheduler in~\cite{AAD+06} is only a {\em possible} way of satisfying this constraint, as it is one of the possible ways of generating graphs in~\KR. As for the relation between \KR and fair executions of population protocols, both are in fact {\em incomparable}.

The characteristic property of \AC (every snapshot is connected) was considered in~\cite{OW05} in the context of information dissemination.
For exemple, this property implies that at any point in time, at least one non-informed node has an informed neighbor (unless dissemination is complete), which bounds the propagation time by $n-1$ rounds in synchronous systems. 

Class \TINT corresponds to a variant of Class~\AC with an additional stability parameter. It was introduced by Kuhn {\it et al.} in~\cite{KLO10} to study problems such as counting, token dissemination, and computation of functions whose input is spread over all the nodes (with adversarial edge scheduling). The running time of some algorithms for these problems in class $\TINT(T)$ is sped up by a factor of $T$. In other words, the algorithms profit from stability. Godard and Mazauric~\cite{GM14} consider the particular case that a connected spanning subgraph is {\em always} present (\ie $\TINT(\infty)$), which they refer to as a {\em statically-connected} dynamic network (see also Section~\ref{sec:diameter}.)

\paragraph{Message adversaries.}

A significant line of work in distributed computing has considered dynamic transmission faults occurring on top of an (otherwise) static network. These systems are typically synchronous, and an execution is modeled by sequence of graphs, each element of which represents successful transmissions in the corresponding round. The main difficuly is to deal with uncertainty as to what faults will occur in what round, and a typical way of restricting uncertainty is to restrict the {\em set} of possible graphs, while having little or no control over the order in which these graphs will occur. The reader is referred to Santoro and Widmayer~\cite{widmayer} for one of the seminal works in this area and to~\cite{message} for a more recent survey. In the past decade, there has been interesting convergences between these models and the kind of highly-dynamic networks discussed here (\eg~\cite{heard-of,matthias}). In particular, while the common approach was mainly to restrict the {\em set} of possible graphs, some works (like the {\em heard-of}~\cite{heard-of} model) have considered constraints in the form of predicates that apply to the sequence {\em as a whole}, which is what some of the above classes of dynamic networks essentially are.

\begin{avenue}
  Establish a methodical comparison between predicates {\it ``à la heard-of''} and classes of dynamic networks discussed above. The usual restrictions on the set of allowed graphs (\eg standard connectivity) have been relaxed in recent works (\eg~\cite{squad}), taking the assumptions down to pure temporal reachability achieved on top of possibly disconnected graphs (close in spirit to the above classes). Explicit efforts to unify both areas have also been made by Coulouma {\it et al.}~\cite{coulouma} and Godard~\cite{godard}.
\end{avenue}

\subsection{Synthetic view of the classes}

The classes are summarized in Table~\ref{tab:classes}. Besides being dichotomized into {\em finite} or {\em recurrent}, some are based on the concept of journeys, others on standard paths (within snapshots), and others on the individual behavior of edges. Some are {\em uniform}, meaning that every node plays the same role in the definition; some are not. The distinction between strict and non-strict journeys is preserved in this table (it will be dropped subsequently based on Discussion~\vref{disc:strict}).

\begin{table}[h]
  \begin{center}
    \small
    \newcommand{\cm}{${\color{darkgreen}\checkmark}$}
    \newcommand{\dash}{{\color{red}--}}
    \begin{tabular}{|@{\,}c@{\,}|@{\,}c@{\,}|@{\,}c@{\,}|@{\,}c@{\,}|@{\,}c@{~}@{\,}c@{~}@{\,}c@{\,}@{\,}c@{\,}@{\,}c@{\,}|@{\,}c@{\,}|}
      \hline
      Proposed&In~\cite{CFQS12}&In~\cite{CCF09}&Other&Journey-&Path-&Edge-&Finite&Uniform&Visual\\
      name&&&names&based&based&based&&&\\\hline
      \JOA&$\C_1$&$\F_1$&-&\cm&\dash&\dash&\cm&\dash&\JOArep\\
      \JOAST&-&$\F_{3}$&-&\cm&\dash&\dash&\cm&\dash&\JOASTrep\\
      \JAO&$\C_2$&$\F_7$&-&\cm&\dash&\dash&\cm&\dash&\JAOrep\\
      \JAOST&-&-&-&\cm&\dash&\dash&\cm&\dash&\JAOSTrep\\
      \TC&$\C_3$&$\F_2$&\footnotesize \TC~\cite{AGMS15}&\cm&\dash&\dash&\cm&\cm&\TCrep\\
      \TCST&-&$\F_{4}$&\footnotesize \TC~\cite{AGMS15}&\cm&\dash&\dash&\cm&\cm&\TCSTrep\\
      \RT&$\C_4$&-&-&\cm&\dash&\dash&\cm&\cm&\RTrep\\
      \EOA&-&$\F_{5}$&-&\dash&\dash&\cm&\cm&\dash&\EOArep\\
      \KG&-&$\F_{6}$&-&\dash&\dash&\cm&\cm&\cm&\KGrep\\
      {\small $Counsils$}&-&-&\footnotesize ${\F}_{\scriptscriptstyle OC/[P]CC}$\,\cite{Laplace12}&\dash&\dash&\cm&\cm&\dash&\dash\\\hline
      \TCR&$\C_5$&-&\footnotesize ETDN\cite{RBK07}, ${\cal COT}$\cite{BDKP16}&\cm&\dash&\dash&\dash&\cm&\TCRrep\\
      \TCB&-&-&\footnotesize ${\cal TC}${\small$(\Delta)$}\,\cite{GCLL15}&\cm&\dash&\dash&\dash&\cm&\TCBrep\\
      $\alpha$-\TCB&-&-&-&\cm&\dash&\dash&\dash&\cm&\dash\\
      \ER&$\C_6$&-&\footnotesize $\F_9$\,\cite{Cas07}, ${\cal R}$\cite{CFMS15}&\dash&\dash&\cm&\dash&\cm&\ERrep\\
      \EB&$\C_7$&-&\footnotesize ${\cal B}$\cite{CFMS15}&\dash&\dash&\cm&\dash&\cm&\EBrep\\
      \EP&$\C_8$&-&\footnotesize ${\cal P}$\cite{CFMS15}&\dash&\dash&\cm&\dash&\cm&\EPrep\\
      \KR&$\C_{13}$&-&\footnotesize $\F_8$~\cite{Cas07}&\dash&\dash&\cm&\dash&\cm&\KRrep\\
      \PR&$\C_{12}$&-&\footnotesize ERDN~\cite{RBK07}&\dash&\cm&\dash&\dash&\cm&\PRrep\\
      \CR&$\C_{11}$&-&\footnotesize ECDN~\cite{RBK07}&\dash&\cm&\dash&\dash&\cm&\CRrep\\
      \AC&$\C_9$&-&-&\dash&\cm&\dash&\dash&\cm&\ACrep\\
      \TINT&$\C_{10}$&-&-&\dash&\cm&\dash&\dash&\cm&\TINTrep\\
      \hline
    \end{tabular}
  \end{center}
  \caption{\label{tab:classes}Summary of key properties of the classes.}
\end{table}

\begin{avenue}
  Revisit the expression of the classes using {\em temporal logic}. Many of the definitions make extensive use of temporal adjectifs and quantifiers (\eg ``eventually'', ``over time'', ``at least once'' {\it etc.}) making temporal logic a natural choice.
\end{avenue}

\paragraph{Subclasses of dynamic networks induced by footprints}
Dubois et al.~\cite{DKP15} define an notion of {\em induced subclass} in which one restricts a class of dynamic network to the ones whose footprints fall into a given class of (static) graphs $\cal F$. For example, $\TCR_{|{\cal F}}$ is the subset of $\TCR$ whose footprint is in ${\cal F}$. Fluschnik et al.~\cite{nieder2} also suggest a classification of dynamic networks based mostly on properties of the footprint (underlying graph) that impact the computational complexity of computing separators~\cite{KKK00,nieder}.

\paragraph{Typesetting the classes in \LaTeX}~\smallskip\\
To all intents and purposes, here are some typeset commands:\smallskip

{\small
\begin{minipage}{\dimexpr\textwidth-3cm}
\begin{verbatim}
\newcommand{\ER}{\ensuremath{{\cal E^R}}\xspace}
\newcommand{\EB}{\ensuremath{{\cal E^B}}\xspace}
\newcommand{\EP}{\ensuremath{{\cal E^P}}\xspace}
\newcommand{\JOA}{\ensuremath{{\cal J}^{1\forall}}\xspace}
\newcommand{\JAO}{\ensuremath{{\cal J}^{\forall 1}}\xspace}
\newcommand{\RT}{\ensuremath{{\cal TC^{\circlearrowleft}}}\xspace}
\newcommand{\TC}{\ensuremath{{\cal TC}}\xspace}
\newcommand{\TCR}{{\ensuremath{{\cal TC^R}}}\xspace}
\newcommand{\TCB}{{\ensuremath{{\cal TC^B}}}\xspace}
\newcommand{\AC}{\ensuremath{{\cal C^*}}\xspace}
\newcommand{\TINT}{\ensuremath{{\cal C}^{\cap}}\xspace}
\newcommand{\PR}{\ensuremath{\cal P^R}\xspace}
\newcommand{\CR}{\ensuremath{{\cal C^R}}\xspace}
\newcommand{\KG}{\ensuremath{{\cal K}}\xspace}
\newcommand{\EOA}{\ensuremath{{\cal E}^{1\forall}}\xspace}
\newcommand{\KR}{\ensuremath{{\cal K^R}}\xspace}
\end{verbatim}
\end{minipage}
}

\section{Relations between classes}
\label{sec:relations}

\articles{
SIROCCO'09~\cite{CCF09},
IJPEDS'12~\cite{CFQS12},
IJFCS'15~\cite{CFMS15},
EUROPAR'15~\cite{GCLL15}.}

Formally, each class is a {\em set} of graphs, and these sets are related to each other through inclusion relations. We review here a number of such relations among the known classes. Most of the relations concerning finite properties are from~\cite{Cas07,CCF09} and most of the ones concerning recurrent properties are from~\cite{CFQS12}. Here, we give a unified account including also new classes and relations. Then, we discuss relations of a more computational nature which we characterized with Flocchini, Mans, and Santoro in~\cite{CFMS15}. Finally, some possible uses of such hierarchy are reviewed.

\subsection{Inclusion relations among classes}

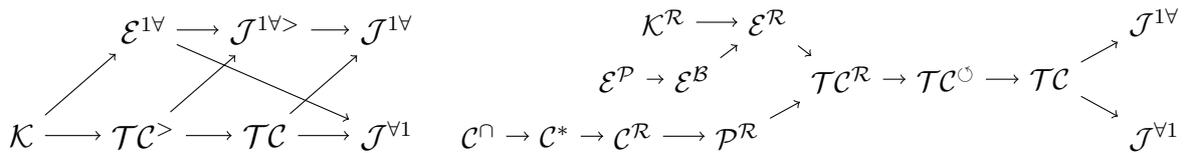
\begin{figure}[h]
    \tikzsetnextfilename{tikz-hierarchy-finite}
      \begin{tikzpicture}[xscale=1.6, yscale=1.4]
        \tikzstyle{every node}=[]
        \path (0,0) node (f6) {\KG};
        \path (1,0) node (f4) {\TCST};
        \path (1,1) node (f5) {\EOA};
        \path (2,0) node (f2) {\TC};
        \path (2,1) node (f3) {\JOAST};
        \path (3,0) node (f7) {\JAO};
        \path (3,1) node (f1) {\JOA};
        \tikzstyle{every path}=[->]
        \draw (f6)--(f4);
        \draw (f6)--(f5);
        \draw (f4)--(f2);
        \draw (f4)--(f3);
        \draw (f5)--(f3);
        \draw (f5)--(f7);
        \draw (f2)--(f7);
        \draw (f2)--(f1);
        \draw (f3)--(f1);
      \end{tikzpicture}
~
    \tikzsetnextfilename{tikz-hierarchy-recurrent}
\begin{tikzpicture}[sloped, level distance=39pt, font=\small, sibling distance=43pt]
  \tikzstyle{every node}=[]
  \tikzstyle{every path}=[<-]
  \node {\RT}
  child [grow=right]{
    node {\TC}
    edge from parent [->]
    child {
      node {\JAO}
      edge from parent [->]
    }
    child {
      node {\JOA}
      edge from parent [->]
    }
  }
  child [grow=left]{
    node {\TCR}
    child {
      node[right] {\ER}
      child[grow=left] {
        node {\KR}
      }
      child{
        node[right] {\EB}
        child {
          node[right] {\EP}
        }
      }
    }
    child {
      node {\PR}
      child {
        node {\CR}
        child {
          node[right] {\AC}
          child {
            node[right] {\TINT}
          }
        }
      }
    }
  };
\end{tikzpicture}
\caption{\label{fig:hierarchies}Hierarchies from~\cite{CCF09} (left) and~\cite{CFQS12} (right), unified in Figure~\ref{fig:hierarchy-complete}.}
\end{figure}

Due to the reporting nature of this document, the hiearchies from~\cite{CCF09} and~\cite{CFQS12} are reproduced unchanged (up to the names) on Figure~\ref{fig:hierarchies}.
We briefly recall the arguments behind these inclusions and present the new relations. 

\paragraph{Finite properties.} Unless $n=0$, it holds that something true for all nodes is also true for at least one node, therefore $\TC \subseteq \JOA$, $\TC \subseteq \JAO$, and  $\KG \subseteq \EOA$. Because every edge induces a valid (one-hop) journey, it holds that $\EOA \subseteq \JOA$ and $\KG \subseteq \TC$, and (somewhat reversely) $\EOA \subseteq \JAO$. It also holds that $\TCR \subseteq \RT \subseteq \TC$.
These inclusions are actually strict, and so are most of the subsequent ones.

\paragraph{Recurrent properties.} As expected, \TCR is the largest class among those based on recurrent properties. In particular, it contains \ER, which itself contains \EB, which in turn contains \EP. Indeed, periodicity of edges is a special case of bounded recurrence of edges, which is a special case of recurrence of edges. The containment of \ER into \TCR holds because \ER additionally requires that the footprint is connected, thus reappearance of edges transitively create recurrent journeys between {\em all} pairs of nodes. The inclusion is strict, because some graphs in $\TCR$ may contain non-recurrent edges. In fact, $\TCR$ is the set of graphs in which a connected spanning subset of edges is recurrent~\cite{BDKP16} (see also Section~\ref{sec:robustness}). A special case within $\ER$ is when the footprint is complete, yielding class $\KR = \KG \cap \ER$.

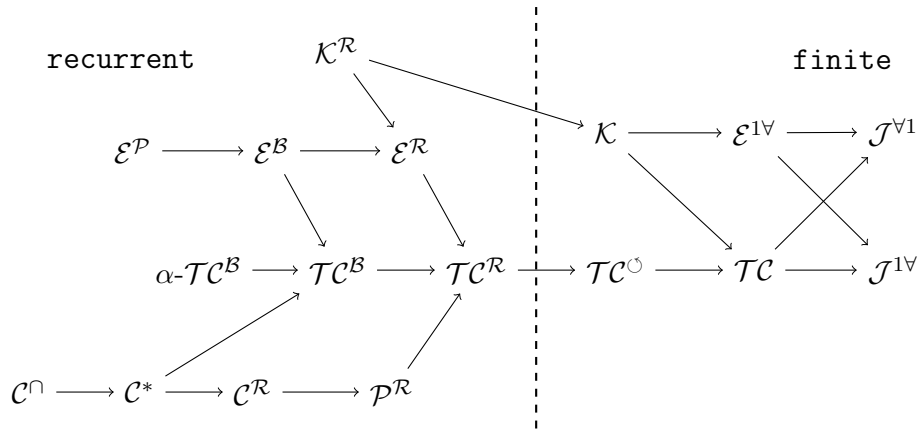
\begin{figure}[h]
  \centering
    \tikzsetnextfilename{tikz-hierarchy-complete}
\begin{tikzpicture}[sloped, level distance=40pt, font=\small, sibling distance=40pt,yscale=1.4,xscale=1.3]
  \tikzstyle{every node}=[]
  \tikzstyle{every path}=[<-]
  \node (RT){\RT}
  child [grow=right]{
    node (TC){\TC}
    edge from parent [->]
    child[grow=right]{
      node (JOA){\JOA}
      edge from parent [->]
    }
    child {
      node[above=14.6pt] (JAO){\JAO}
      edge from parent [->]
    }
  }
  child [grow=left]{
    node (TCR) {\TCR}
    child {
      node[below=11pt,right=16pt] (ER){\ER}
      child {
        node[above=10pt, right=12pt] (KR){\KR}
      }
      child [grow=left] {
        node (EB){\EB}
        child {
          node (EP){\EP}
        }
      }
    }
    child {
      node (TCB){\TCB}
      child {
        node (ATCB){$\alpha$-\TCB}
      }
    }
    child {
      node[above=10pt,right=8pt] (PR){\PR}
      child {
        node (CR){\CR}
        child {
          node[right] (AC){\AC}
          child {
            node[right] (TINT){\TINT}
          }
        }
      }
    }
  };
  \path (TC)+(-1.5,1.3) node (KG){\KG};
  \path (TC)+(0,1.3) node (EOA){\EOA};
  \draw[->] (AC)--(TCB);
  \draw[->] (EB)--(TCB);
  \draw[->] (KG)--(TC);
  \draw[->] (KG)--(EOA);
  \draw[->] (EOA)--(JOA);
  \draw[->] (EOA)--(JAO);
  \draw[bend right=15pt] (KG) -- (KR);
  \tikzstyle{every path}=[]
  \draw[dashed,thick] (-.8,-1.5) -- (-.8,2.5);
  \tikzstyle{every node}=[font=\normalsize]
  \path (-5,2) node {{\tt recurrent}};
  \path (2.3,2) node {{\tt finite}};
\end{tikzpicture}
\caption[Relations of inclusion between classes.]{\label{fig:hierarchy-complete}Updated hierarchy of the main classes of dynamic networks.}
\end{figure}

The repetition of available paths within snapshots eventually creates journeys between nodes, so $\PR \subseteq \TCR$. Since connected snapshots offer paths between all pairs of nodes, it follows that $\CR \subseteq \PR$. Next, what is true in each step must also be true infinitely often, thus $\AC \subseteq \CR$. The case of $\TINT$ is more subtle because the {\em universal} version of this class (see Section~\ref{sec:parametrized}) equals $\AC$, while {\em instantiated} versions of it may be strictly smaller than $\AC$. The relations represented on Figures~\ref{fig:hierarchies} and~\ref{fig:hierarchy-complete} are for the instantiated versions of $\TINT$.

Having a bounded temporal diameter at all times implies that all nodes can recurrently reach each other through journeys, thus $\TCB \subseteq \TCR$. The inclusion is actually strict, as one could design a graph in $\TCR$ whose temporal diameter keeps growing unboundedly with time.

Under mild assumption (including discrete settings and bounded latency continuous settings), it holds that $\AC \subseteq \TCB$. Finally, the existence of $\alpha$-journeys imply bounds on the temporal diameter (by $\alpha(n-1)$ time units, plus latencies), implying that $\alpha$-$\TCB \subseteq \TCB$ (and justifying the $\cal B$ superscript in the name). 
All these relations are depicted in the updated version of the hierarchy on Figure~\ref{fig:hierarchy-complete}.

\force
\unforce

\begin{avenue}
  Investigate the relations between directed analogues of these classes, \ie when directed (non-symmetrical) edges can exist. The impact is significant. For example, with undirected edges, the repetition of journeys from a node $u$ to a node $v$ eventually creates backward journeys from $v$ to $u$, which explains why no recurrent version of \JOA was defined ($\JOA$$^{\cal R}$ would amount to \TCR itself). This type of ``reversibility'' argument does not apply to directed networks.
\end{avenue}

\subsection{Computational Relations}
\label{sec:computational-relations}

Besides set-theoretical relations (\ie inclusions) among classes of graphs, \eg $\EP \subset \EB \subset \ER$, we considered with Flocchini, Mans, and Santoro~\cite{CFMS15} (partially reviewed in Section~\ref{sec:shfafo}), the {\em computational} relations induced by various combinations of these classes with given knowledge, namely the number $n$ of nodes in the network, a bound $\Delta$ on the recurrence time, and the period $p$. In particular, we considered the relations between
$\PP(\ER_n)$, $\PP(\EB_\Delta)$, and $\PP(\EP_p)$, 
where $\PP({\cal C}_k)$ is the set of problems one can solve in class ${\cal C}$ with knowledge $k$. 
We showed that, in the considered model, it holds that
\begin{equation}
\label{eq:computational}
\PP(\ER_n)\subset\PP(\EB_\Delta)\subset\PP(\EP_p)
\end{equation}

In other words, the computational relations between these three contexts form a {\em strict} hierarchy. The fact that $\PP(\ER_n)\subseteq\PP(\EB_\Delta)$ comes from  $\EB \subseteq \ER$ together with the fact that $n$ is learnable in $\EB_\Delta$. In fact, the entire footprint can be learned in $\EB_\Delta$, making it a rather powerful setting. For example, $\EB_\Delta$ can emulate algorithms for synchronous static networks, by confining every round within a $\Delta$ window where all neighbors communicate. 
The strictness of inclusions in Equation~\ref{eq:computational} comes from the existence of at least one problem in each setting which is not solvable in the other, using again the example of foremost, shortest, and fastest broadcast with termination (covered in Section~\ref{sec:shfafo}).

\subsection{Stochastic Comparison}

Some of the classes are incomparable from a set-theoretic (and a fortiori computational) perspective. For example, none of \ER, \TCB, or \CR could be declared more general than the others, and the same holds for \EOA and \TC. While incomparable using set theory, one may look at different ordering relations, in particular stochastic ones.

\begin{avenue}
  Compare the generality of (otherwise incomparable) classes of dynamic networks by measuring how long it takes for a random network (\eg edge-markovian dynamic graph) to satify the corresponding property. For recurrent properties, one may look instead at how frequently the property is satisfied.
\end{avenue}

\subsection{Conclusion} 
\label{sec:comparison}

Several open questions related to the relations between classes have been discussed in this section. In way of conclusion, we will simply mention some of the {\em uses} one could make of such a hierarchy. First, we believe that these relations have the potential to guide an algorithm designer. Clearly, one should seek properties that offer sufficient structure to be exploited, while remaining as general as possible. Second, such hierarchy allows one to compare the requirements of several candidate solutions on a rigorous basis. For example, we reviewed in Section~\ref{sec:conditions} two counting algorithms, one of which requires the graph to be in $\KG$, while the other requires it to be in $\JAO$. Precisely, both algorithms were guaranteed to succeed in $\KG$ and to fail outside of $\JAO$; in between, \ie in $\JAO \setminus \KG$, the first algorithm {\em must} fail, while the second {\em could} succeed, implying that the second is more general.
Third, knowing which class a network belongs to provides immediate information as to what problem can be solved within. Giving again an example based on the model in Section~\ref{sec:conditions}, if a network is (say) in $\TC \cap \JOAST$, then we have immediately that (1) broadcast has some chances of success whatever the emitter (depending on the adversary) and (2) it is guaranteed to succeed for at least one emitter.

This back and forth movement between classes and problems also manifests in more elaborate contexts, as seen above with Equation~\ref{eq:computational}, which together with the gradual feasibility of foremost, shortest, and fastest broadcast in these contexts, implies a hierarchy of difficulty among these problems (in terms of topological requirements). 
We believe hierarchies of this type have the potential to lead more equivalence results and formal comparison between problems, and between algorithms.

\section{Testing properties automatically}
\label{sec:offline}

\articles{SIROCCO'09~\cite{CCF09},
ALGOTEL'14~\cite{BCCJN14-fr,BCCJN14-en},
CIAC'15~\cite{CKNP15}, SIROCCO'17~\cite{CKNP17} (long version in Theory of Computing Systems~\cite{CKNP18}),
SNAMAS'11~\cite{SQFCA11},
PhD Y. Neggaz~\cite{Neggaz16}.
}

This section is concerned with the {\it a posteriori} analysis of connectivity traces using centralized algorithm. While we have not carried ourselves such studies on real data, we have designed a number of algorithms in this perspective.
We first discuss methodological aspects, then describe some of our contributions in this area. In particular, we review various basic reductions of temporal properties in dynamic graphs to standard graph properties introduced in~\cite{CCF09}, then focus on the particular case of computing transitive closures of journeys~\cite{BCCJN14-fr,BCCJN14-en}. Next, we present a general framework for computing parameters of sequence-based dynamic graphs. The framework was introduced in~\cite{CKNP15} and generalized in~\cite{CKNP17} (both being combined in~\cite{CKNP18}). It can be {\em instantiated} with specific operations to address the computation of various parameters without changing the high-level logic. Finally, we review some suggestions made in~\cite{SQFCA11} concerning the analysis of two-fold dynamic phenomena, namely the (long-term) evolution of (short-term) temporal metrics.

\subsection{A methodological framework}

\begin{figure}[h]
  \centering
  \tikzsetnextfilename{tikz-methology}
  \begin{tikzpicture}[xscale=.91, yscale=1.2]
    \tikzstyle{every node}=[font=\scriptsize];
    \path[right] (-1,2) node (algo) {Algorithm};
    \path[right] (-1,1.3) node (mobi) {Mobility Model};
    \path[right] (-1,0.7) node (real) {Real Network};
    \path[right] (7.7,2) node (clas) {Temporal properties};
    \path[right] (7.7,1) node (inst) {Dynamic graphs};
    \path[right] (4.7,2) node (cond) {Conditions};
    \path[right] (4.7,1.3) node (tra1) {Network Traces};
    \path[right] (4.7,0.7) node (tra2) {Network Traces};
    \tikzstyle{every node}=[draw, rectangle, rounded corners=6pt, font=\footnotesize];
    \path (3.05,2) node (anal) {Analysis};
    \path (3.15,1.3) node (gene) {Generation};
    \path (3.15,0.7) node (sens) {Collection};
    \path[right] (11.8,1.5) node (chec) {Property Testing};
    \tikzstyle{every node}=[];
    \path[left] (chec)+(0,-.9) node (yes) {Yes};
    \path[right] (chec)+(0,-.9) node (no) {No};
    \path (chec.west)+(0,.1) coordinate (c1);
    \path (chec.west)+(0,-.1) coordinate (c2);
    \path (inst.west)+(0,.08) coordinate (i1);
    \path (inst.west)+(0,-.08) coordinate (i2);
    \tikzstyle{every path}=[->, draw, thin];
    \draw (algo)--(anal);
    \draw (anal)--(cond);
    \draw (cond)--(clas);
    \draw (mobi)--(gene);
    \draw (gene)--(tra1.west);
    \draw (real)--(sens);
    \draw (sens)--(tra2.west);
    \draw (chec)--(yes);
    \draw (chec)--(no);
    \tikzstyle{every path}=[->, draw, thin, shorten >=1.5pt];
    \draw (clas.east)--(c1);
    \draw (inst.east)--(c2);
    \draw (tra1.east)--(i1);
    \draw (tra2.east)--(i2);
  \end{tikzpicture}
  \caption{\label{fig:checking} Suitability of a distributed algorithm in given mobility contexts.}
\end{figure}
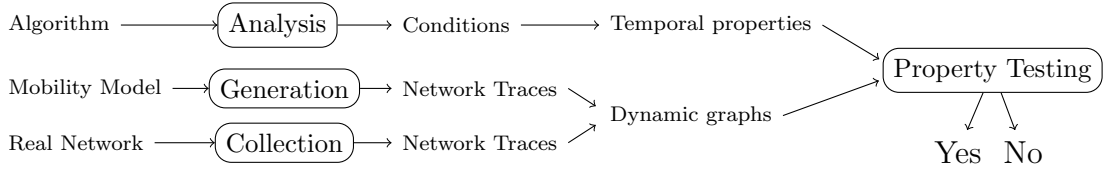

While not a central topic in~\cite{CCF09}, one of the perspectives discussed therein suggest a general approach to connect efforts in the analysis of distributed algorithms with practical considerations (usability of the algorithms). The workflow, depicted in Figure~\ref{fig:checking}, considers temporal properties resulting from the analysis of distributed algorithms, on the one hand, and dynamic graphs resulting from either real-world measures (traces) or artificial generation through mobility models.
The main motivation is to assist the algorithm designer in making realistic assumptions and the algorithm deployer in chosing the appropriate algorithm for the target mobility context.

This architecture has been the guiding principle (motivation) behind most of the work reported in this section.

\subsection{Around transitive closures of journeys (and paths)}

\begin{figure}[h]
\def\wgraph (#1,#2){%
  \tikzstyle{every node}=[draw,fill,circle,inner sep=#1]
  \path (0,0) node (c){};
  \path (c)+(-1,.4) node (a){};
  \path (c)+(1,.4) node (e){};
  \path (c)+(-.6,-.6) node (b){};
  \path (c)+(.6,-.6) node (d){};
  \tikzstyle{every node}=[font=\footnotesize]
  \path (a)+(-#2,0) node (la){$a$};
  \path (b)+(-#2,0) node (lb){$b$};
  \path[above] (c) node (lc){$c$};
  \path (d)+(#2,0) node (ld){$d$};
  \path (e)+(#2,0) node (le){$e$};
}
\centering
  \begin{tikzpicture}[scale=2]
    \wgraph(.8pt,4pt)
    \tikzstyle{every node}=[sloped,below,font=\scriptsize,inner sep=2pt]
    \draw (a)--node{$1$}(b);
    \draw (b)--node{$2$}(d);
    \draw (d)--node{$3,4$}(e);
    \tikzstyle{every node}=[sloped,above,font=\scriptsize,inner sep=1pt]
    \draw (a)--node[inner sep=1.5pt]{$1$}(c);
    \draw (b)--node[inner sep=1.5pt]{$2$}(c);
    \draw (c)--node{$2,3$}(d);
    \draw (c)--node{$3,4$}(e);
    
    \tikzstyle{every node}=[draw, fill, inner sep=.1pt]
    \path (2.6,-.1) coordinate (zz);
    \path (zz)+(18:.6) node (ee) {};
    \path (zz)+(90:.6) node (cc) {};
    \path (zz)+(162:.6) node (aa) {};
    \path (zz)+(234:.6) node (bb) {};
    \path (zz)+(306:.6) node (dd) {};
    \tikzstyle{every node}=[]
    \path (aa)+(-.1,.02) node (laa){$a$};
    \path (bb)+(-.08,-.08) node (laa){$b$};
    \path (cc)+(0,.09) node (laa){$c$};
    \path (dd)+(.08,-.08) node (laa){$d$};
    \path (ee)+(.1,.02) node (laa){$e$};
    
    \tikzstyle{every path}=[>=stealth, shorten >=1.5pt, shorten <=1.5pt]
    \path[<->] (aa) edge [bend right=30] (bb);
    \path[<->] (aa) edge [bend left=30] (cc);
    \path[->] (aa) edge [bend right=10] (dd);
    \path[->] (aa) edge [bend left=10] (ee);
    \path[<->] (bb) edge [bend left=10] (cc);
    \path[<->] (bb) edge [bend right=30] (dd);
    \path[->] (bb) edge [bend right=10] (ee);
    \path[<->] (cc) edge [bend left=10] (dd);
    \path[<->] (cc) edge [bend left=30] (ee);
    \path[<->] (dd) edge [bend right=30] (ee);
    
    \path (zz)+(-1.4,0) coordinate (zzleft);
    \draw[->, semithick, shorten >=50pt] (zzleft)--(zz);
  \end{tikzpicture}
\caption{\label{fig:closure}Transitive closure of journeys (here in discrete time).}
\end{figure}
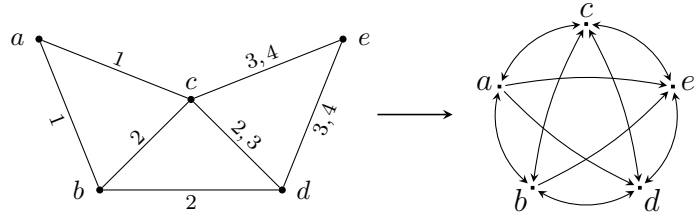

In~\cite{BF03}, Bhadra and Ferreira define the concept of \emph{transitive closure} of journeys as the static directed graph $H = (V,A_{H})$, where $A_{H} = \{(v_i,v_j) : v_i \leadsto v_j)\}$ (see Figure~\ref{fig:closure}). 
We considered in~\cite{CCF09} two versions of this notion, whether the arcs in the closure represent strict or non-strict journeys (see Section~\ref{sec:journeys} for definitions). Testing membership of a given graph to a number of basic classes listed above (based on {\em finite} properties) can be done using static derivated structures like this one.
Precisely, given a graph $\mathcal{G}$, its footprint $G$, its transitive closure $H$, and strict transitive closure $H_{strict}$, membership to the seven classes discussed in Section~\ref{sec:conditions} can be reduced to the following.

\begin{itemize}
\setlength\itemsep{0pt}
\item $\mathcal{G} \in \JOA \Longleftrightarrow$ $H$ contains an out-dominating set of size 1.
\item $\mathcal{G} \in \TC \Longleftrightarrow$ $H$ is a complete graph.
\item $\mathcal{G} \in \JOAST \Longleftrightarrow$ $H_{strict}$ contains an out-dominating set of size 1.
\item $\mathcal{G} \in \TCST \Longleftrightarrow$ $H_{strict}$ is a complete graph.
\item $\mathcal{G} \in \EOA \Longleftrightarrow$ $G$ contains a dominating set of size 1.
\item $\mathcal{G} \in \KG \Longleftrightarrow$ $G$ is a complete graph.
\item $\mathcal{G} \in \JAO \Longleftrightarrow$ $H$ contains an in-dominating set of size 1.
\end{itemize}

\subsubsection{Computing the transitive closure}
\label{sec:compTC}

The problem of computing the transitive closure of a dynamic network varies slightly depending on whether strict or non-strict journeys are considered. We first focus on the case that all journeys must be strict (at most one hop at a time), then we show how to adapt the same solution to non-strict journeys based on a pre-processing trick of the input graph.

Let us first observe that existing algorithms for other problems could be adapted to compute the transitive closure. In particular, one may compute a tree of foremost journeys using the centralized algorithm in Bhadra et al.~\cite{BFJ03}, doing so from {\em every node} and adding arc $(u,v)$ iff $v$ was reached when computing foremost journeys from $u$. Using the data structure from~\cite{BFJ03} (described further down), this amounts to $O(n(m\log k + n\log n))$ time steps, where $k$ is the number of {\em characteristic} dates in the network (times of appearance or disappearance of edges).

In Barjon et al.~\cite{BCCJN14-en} (French version~\cite{BCCJN14-fr}), we proposed an alternative algorithm for computing the transitive closure when the input graph is given as a sequence $\G=\{G_1,G_2,...,G_{k}\}$, with $G_i=(V,E_i)$. The main feature of our algorithm is that it unifies strict and non-strict through a preprocessing step (described later). The algorithm is basic and consists of scanning the graphs in a chronological order, updating for each edge the union of predecessors on both endpoints nodes (to enforce strictness, the union is based on a copy of the predecessors obtained at the end of the previous time). The final set of predecessors is the transitive closure.
Complexity parameters include the number of times $k$, the maximum {\em instant} density $\mu=max(|E_i|)$, and the {\em cumulative} density $m=|\cup E_i|$, yielding an overall cost of $O(k \mu n)$ operations. Depending on the values of the parameters, our algorithm achieves a better or worse complexity than the solution based on foremost trees. In particular, it performs better in sparse scenario where the {\em instant} density and the number of times are low (since we pay a linear dependency on $k$). A comparative table is given in~\cite{BCCJN14-en} for all combination of parameters.

\begin{figure}[h!]
  \def\tgraph{
    \tikzstyle{every node}=[draw, fill, circle, inner sep=.6pt]
    \path (0,0) coordinate (mid);
    \path (mid)+(18:.6) node (e) {};
    \path (mid)+(90:.6) node (c) {};
    \path (mid)+(162:.6) node (a) {};
    \path (mid)+(234:.6) node (b) {};
    \path (mid)+(306:.6) node (d) {};
    \tikzstyle{every node}=[font=\small]
    \path (a) node[left] {$a$};
    \path (b) node[below] {$b$};
    \path (c) node[above] {$c$};
    \path (d) node[below] {$d$};
    \path (e) node[right] {$e$};
    \tikzstyle{every path}=[shorten >=5pt, shorten <=5pt]
  }
  \hspace*{-10pt}
  \begin{minipage}[c]{0.12\linewidth}
  \tikzsetnextfilename{tikz-non-strict-TC}
    \begin{tikzpicture}
      \tikzstyle{every node}=[]
      \draw (0,1) edge[bend right=12pt,->] node[left,text width=1.5cm] {transitive closure of paths} (0,-1);
      \path (0,-2) coordinate (bidon);
    \end{tikzpicture}
  \end{minipage}
  \begin{minipage}[c]{0.8\linewidth}
    \begin{tabular}{@{}c@{\,}|@{\,}c@{\,}|@{\,}c@{\,}|@{\,}c@{\,}||@{\,}c}
      \tikzsetnextfilename{tikz-non-strict-TC-G1}
      \begin{tikzpicture}[scale=1.37]
        \tgraph 
        \draw[->] (a)--(c);
        \draw[<->] (b)--(c);
        \draw[<-] (b)--(d);
        \draw[->] (c)--(d);
        \path (c)+(0,-1.5) node {$G_1$};
      \end{tikzpicture}
      &
      \tikzsetnextfilename{tikz-non-strict-TC-G2}
        \begin{tikzpicture}[scale=1.37]
          \tikzstyle{every node}=[font=\scriptsize]
          \tgraph
          \draw[<-] (a)--(c);
          \draw[<-] (b)--(c);
          \draw[->] (c)--(d);
          \path (c)+(0,-1.5) node {$G_2$};
        \end{tikzpicture}
      &
      \tikzsetnextfilename{tikz-non-strict-TC-G3}
        \begin{tikzpicture}[scale=1.37]
          \tikzstyle{every node}=[font=\scriptsize]
          \tgraph
          \draw[<-] (a)--(b);
          \draw[->] (e)--(d);
          \draw[<-] (c)--(e);

          \path (c)+(0,-1.5) node {$G_3$};
        \end{tikzpicture}
      &
      \tikzsetnextfilename{tikz-non-strict-TC-G4}
        \begin{tikzpicture}[scale=1.37]
          \tikzstyle{every node}=[font=\scriptsize]
          \tgraph
          \draw[->] (c)--(b);
          \draw[<-] (a)--(b);
          \draw[->] (d)--(c);
          \draw[<-] (c)--(e);
          
          \path (c)+(0,-1.5) node {$G_4$};
        \end{tikzpicture}
      &
      \\\hline
      \tikzsetnextfilename{tikz-non-strict-TC-G1b}
      \begin{tikzpicture}[scale=1.37]
        \tikzstyle{every node}=[font=\scriptsize]
        \tgraph
        \draw [->](a) -- (c);
        \draw [->](c) -- (d);
        \draw [->](a) -- (d);
        \draw [->](a) -- (b);
        \draw [<-](b) -- (d);
        \draw [->](c) -- (b);
        \draw [<-](c) -- (b);
        \draw [<-](d) -- (b);
        \path (c)+(0,-1.5) node {$G^*_1$};
      \end{tikzpicture} 
      &
      \tikzsetnextfilename{tikz-non-strict-TC-G2b}
        \begin{tikzpicture}[scale=1.37]
          \tikzstyle{every node}=[font=\scriptsize]
          \tgraph
          \draw [<-](a) -- (c);
          \draw [->](c) -- (d);
          \draw [->](c) -- (b);
          \path (c)+(0,-1.5) node {$G^*_2$};
        \end{tikzpicture}
      &
      \tikzsetnextfilename{tikz-non-strict-TC-G3b}
        \begin{tikzpicture}[scale=1.37]
          \tikzstyle{every node}=[font=\scriptsize]
          \tgraph
          \draw[<-] (a)--(b);
          \draw[<-] (c)--(e);
          \draw[<-] (d)--(e);
          \path (c)+(0,-1.5) node {$G^*_3$};
        \end{tikzpicture}
      &
      \tikzsetnextfilename{tikz-non-strict-TC-G4b}
        \begin{tikzpicture}[scale=1.37]
          \tikzstyle{every node}=[font=\scriptsize]
          \tgraph
          \draw[<-] (a)--(b);
          \draw[<-] (a)--(d);
          \draw[<-] (a)--(c);
          \draw[<-] (a)--(e);
          \draw[<-] (b)--(c);
          \draw[<-] (b)--(e);
          \draw[<-] (b)--(d);
          \draw[<-] (c)--(e);
          \draw[<-] (c)--(d);
          \path (c)+(0,-1.5) node {$G^*_4$};          
        \end{tikzpicture}
      &
      \tikzsetnextfilename{tikz-non-strict-TC-result}
        \begin{tikzpicture}[scale=1.37]
          \tikzstyle{every node}=[font=\scriptsize]
          \tgraph
          \draw[<->] (a)--(b);
          \draw[<->] (a)--(d);
          \draw[<->] (a)--(c);
          \draw[<-] (a)--(e);
          \draw[<->] (b)--(c);
          \draw[<-] (b)--(e);
          \draw[<->] (b)--(d);
          \draw[<-] (c)--(e);
          \draw[<->] (c)--(d);
          \draw[<-] (d)--(e);
          \path (c)+(0,-1.5) node {Result};
        \end{tikzpicture}
      \\
      \multicolumn{4}{c}{
      \tikzsetnextfilename{tikz-non-strict-TC-label}
      \begin{tikzpicture}
        \tikzstyle{every node}=[]
        \path (0,.3) coordinate (bidon);
        \draw (-1.5,0) edge[bend right=5pt,->] node[below=2pt] {transitive closure of journeys} (1.5,0);
      \end{tikzpicture}
      }
    \end{tabular}
  \end{minipage}
  \caption{Computing the transitive closure of both strict and non-strict journeys.}
  \label{fig:non-strict_transitive_closure}
\end{figure}
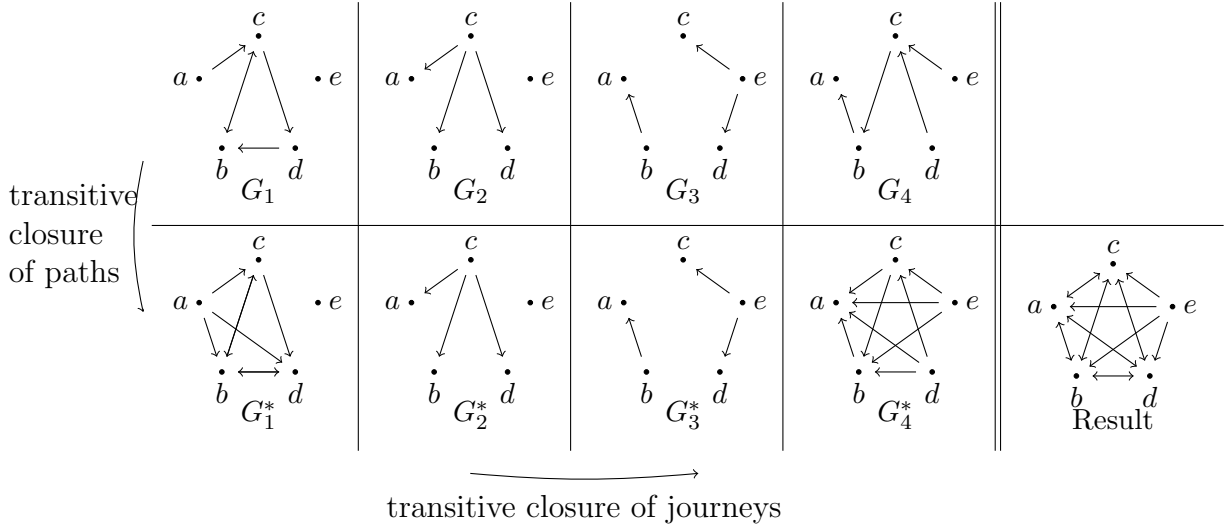

In the case of {\em non-strict} journeys no previous algorithm existed. Our algorithm is based on a double transitive closure: a {\em standard} transitive closure (\ie of paths) applied to each $G_i$, after which the resulting graph can be processed by the algorithm for strict closure (see Figure~\ref{fig:non-strict_transitive_closure}). Interestingly, the pre-processing remains contained within $O(k\mu n)$ operations, resulting in the same running time complexity (up to a constant factor).

\paragraph{More general objects.} A generalization of the concept of transitive closure was proposed in Whitbeck et al.~\cite{WDCG12}, which captures the reachability relative to {\em all} possible initiation times and durations. Once computed, the resulting structure makes it possible to quickly answer queries of the type {\it ``can node $u$ reach node $b$ after time $t$ and within $d$ time units?''}. The complexity of their algorithm (based on an algebra of matrices) is strictly larger than ours~(see~\cite{BCCJN14-en}), but the object is more general.

\boite{Efficiency of data structures}{
  \label{disc:data-structure}
  The algorithms in~\cite{BFJ03},~\cite{WDCG12}, and~\cite{BCCJN14-en} all use different data structures to represent the input network. In~\cite{BFJ03}, the input is a {\em list of temporal adjacencies}, which is similar to an adjacency list augmented with nested lists of presence intervals (one for each edge, see Page 101 of~\cite{Jarry05} for details). In~\cite{WDCG12}, the input is a time-indexed sequence of events of the type $(u,v, \mathit{UP})$, $(v,w,\mathit{DOWN})$. Finally, in~\cite{BCCJN14-en}, the data structure is essentially a sequence of sets of edges $\{E_i\}$. The latter is admittedly a poor choice in terms of performance because it does not exploit stability, the status of an edge being duplicated even when it does not change. Similarly, any change in the network induces a new graph in the sequence. This has tremendous impact on the running time complexity of the above algorithms. Data structures based on sequences of links (link streams) like~\cite{WDCG12} or~\cite{LVM17} often prove more efficient.
}

Despite Discussion~\ref{disc:data-structure}, models based on sequences of graphs may be choices for high-level thinking and algorithmic design, as exemplified by the above pre-processing trick for non-strict journeys that may not be easily expressed in other models. Another illustration is the high-level strategies discussed next, which exploit the specificities of graph sequences.

\begin{open}
  If both strict and non-strict journeys are considered, then some graphs cannot be obtained as transitive closure. For example, the graph $G=(\{u,v,w\},$ $\{(u,v),(v,u),(v,w),(w,v)\})$ (two-hop undirected path) cannot be a transitive closure, because it should include either $(u,w)$ or $(w,u)$ additionally (or both). Characterize the set of graphs which can be obtained as transitive closures.
\end{open}

\subsection{Computing parameters of sequence-based dynamic graphs}
\label{sec:generic}

With Klasing, Neggaz, and Peters, we presented in~\cite{CKNP15} a high-level algorithm for finding the largest $T$ such that a given sequence of graphs $\G$ is $T$-interval connected (see Section~\ref{sec:recurrent-properties} for definitions) and the related decision version for given $T$. The approach is high-level, \ie the graphs in the sequence are considered as atomic  elements that the algorithm manipulates through two abstract operations: {\em intersection} (of two graphs), and {\em connectivity testing} (of one graph). We showed that both the maximization and decision versions of the problem can be solved using only a number of such operations which is linear in the length $\delta$ of the sequence. The technique is based on a walk through a lattice-like meta-graph which we call {\em intersection hierarchy}, the elements of which represent intersections of various sub-sequences of $\G$  (see Figure~\ref{fig:intersection-hierarchy}). Then, the largest T for which the network is T-interval connected corresponds to the row in which the walk terminates ($3$ in both examples).

\begin{figure}[h]
  \centering
  \subfigure[Intersection Hierarchy]{
    \label{fig:intersection-hierarchy}
    \hspace*{-25pt}
    \def\carre (#1,#2,#3){
  \path (#1,#2+1) node[node] (a){};
  \path (#1+1,#2+1) node[node] (b){};
  \path (#1,#2) node[node] (c){};
  \path (#1+1,#2) node[node] (d){};
  \path (#1+.5,#2-.8) node (e){#3};
}
\tikzset{node/.style={draw,circle,fill=none,red,gray,thick,inner sep = 1.6pt}}
\tikzstyle{every node}=[]
    \begin{tikzpicture}[xscale=.35, yscale=.35]
      \node (G) [] 
      at (-2,0.5) {\large{$\G^1$}};
      \node (G) [] 
      at (-0.4,3.4) {\large{$\G^2$}};
      \node (G) [] 
      at (1,6.3) {\large{$\G^3$}};
      \node (G) [] 
      at (2.4,9.2) {\large{$\G^4$}};

      \carre (0,0,$G_{(1,1)}$)
      \draw (a)--(b);
      \draw (a)--(c);
      \draw (c)--(b);
      \draw (c)--(d);
      \draw (b)--(d);
      \carre (3,0,$G_{(2,2)}$)
      \draw (a)--(c);
      \draw (c)--(b);
      \draw (c)--(d);
      \draw (a)--(b);
      \carre (6,0,$G_{(3,3)}$)
      \draw (a)--(c);
      \draw (d)--(b);
      \draw (c)--(d);
      \draw (a)--(b);
      \carre (9,0,$G_{(4,4)}$)
      \draw (a)--(c);
      \draw (b)--(c);
      \draw (d)--(b);
      \draw (c)--(d);
      \draw (a)--(b);
      \carre (12,0,$G_{(5,5)}$)
      \draw (b)--(c);
      \draw (a)--(d);
      \draw (d)--(b);
      \draw (c)--(d);
      \draw (a)--(b);
      \carre (15,0,$G_{(6,6)}$)
      \draw (b)--(c);
      \draw (d)--(b);
      \draw (c)--(d);
      \draw (a)--(b);
      \carre (18,0,$G_{(7,7)}$)
      \draw (b)--(c);
      \draw (c)--(d);
      \draw (a)--(b);
      \carre (21,0,$G_{(8,8)}$)
      \draw (a)--(b);
      \draw (a)--(c);
      \draw (b)--(c);
      \draw (c)--(d);

      \carre (1.5,2.9,$G_{(1,2)}$)
      \draw (a)--(b);
      \draw (a)--(c);
      \draw (c)--(b);
      \draw (c)--(d);
      \carre (4.5,2.9,$G_{(2,3)}$)
      \draw (a)--(c);
      \draw (a)--(b);
      \draw (c)--(d);
      \carre (7.5,2.9,$G_{(3,4)}$)
      \draw (a)--(c);
      \draw (b)--(d);
      \draw (a)--(b);
      \draw (c)--(d);
      \carre (10.5,2.9,$G_{(4,5)}$)
      \draw (b)--(c);
      \draw (b)--(d);
      \draw (a)--(b);
      \draw (c)--(d);
      \carre (13.5,2.9,$G_{(5,6)}$)
      \draw (b)--(c);
      \draw (b)--(d);
      \draw (a)--(b);
      \draw (c)--(d);
      \carre (16.5,2.9,$G_{(6,7)}$)
      \draw (b)--(c);
      \draw (c)--(d);
      \draw (a)--(b);
      \carre (19.5,2.9,$G_{(7,8)}$)
      \draw (a)--(b);
      \draw (b)--(c);
      \draw (c)--(d);

      \carre (3,5.8,$G_{(1,3)}$)
      \draw (a)--(b);
      \draw (a)--(c);
      \draw (c)--(d);
      \carre (6,5.8,$G_{(2,4)}$)
      \draw (a)--(b);
      \draw (a)--(c);
      \draw (c)--(d);
      \carre (9,5.8,$G_{(3,5)}$)
      \draw (a)--(b);
      \draw (b)--(d);
      \draw (c)--(d);
      \carre (12,5.8,$G_{(4,6)}$)
      \draw (b)--(c);
      \draw (b)--(d);
      \draw (a)--(b);
      \draw (c)--(d);
      \carre (15,5.8,$G_{(5,7)}$)
      \draw (b)--(c);
      \draw (a)--(b);
      \draw (c)--(d);
      \carre (18,5.8,$G_{(6,8)}$)
      \draw (b)--(c);
      \draw (a)--(b);
      \draw (c)--(d);

      \carre (4.5,8.7,$G_{(1,4)}$)
      \draw (a)--(b);
      \draw (a)--(c);
      \draw (c)--(d);
      \carre (7.5,8.7,{$G_{(2,5)}$})
      \draw (a)--(b);
      \draw (c)--(d);
      \carre (10.5,8.7,$G_{(3,6)}$)
      \draw (a)--(b);
      \draw (c)--(d);
      \draw (b)--(d);
      \carre (13.5,8.7,$G_{(4,7)}$)
      \draw (a)--(b);
      \draw (c)--(d);
      \draw (b)--(c);
      \carre (16.5,8.7,$G_{(5,8)}$)
      \draw (a)--(b);
      \draw (c)--(d);
      \draw (b)--(c);
    \end{tikzpicture}
  }
  \subfigure[Maximization walk]{
    \label{fig:tmpdiam}
    \hspace*{-20pt}
    \xdefinecolor{bordeaux}{rgb}{0.7,0.10,0.10}
\begin{tikzpicture}[scale=.41]
\tikzstyle{direction}=[->,thick,dashed]
\node (l1) [] at (0,-16.2) {};
\draw[direction] (l1) to[] (3.15,-9.8);

  \tikzstyle{every node}=[draw, circle, inner sep=2.5pt,gray]
  \foreach \j in {9,...,16}{
     \foreach \i in {1,...,\j}{
       \pgfmathsetmacro{\ii}{\i + (16-\j)/2};
       \path (\ii,-\j) node (\i\j){};
     }
  }
  \tikzstyle{every path}=[thin,dashed,gray,dash pattern=on 1pt off 1.5pt]
  \foreach \j in {9,...,15}{
     \foreach \i in {1,...,\j}{
       \pgfmathtruncatemacro{\jj}{\j + 1};
       \pgfmathtruncatemacro{\ii}{\i + 1};
       \draw (\i\jj) -- (\i\j);
       \draw (\ii\jj) -- (\i\j);
     }
  }
  \tikzstyle{every node}=[font=\LARGE, red]
  \path (19) node {$\times$};
  \path (510) node {$\times$};
  \path (711) node {$\times$};
  \path (1012) node {$\times$};
  \path (1113) node {$\times$};
  \tikzstyle{every node}=[draw, circle, inner sep=2.3pt, fill=gray]
  \draw[ultra thick, solid] (816) node {}--(715) node {}--(614) node {}--(513) node {}--(412) node {}--(311) node {}--(210) node {};
  \draw[ultra thick, solid] (916) node {}-- (915) node {}-- (914) node {}-- (913) node {};
  \draw[ultra thick, solid] (1316) node {}--(1215) node {}-- (1114) node {}--(1013) node {}-- (912) node {};
  \draw[ultra thick, solid] (1416) node {}-- (1415) node {};
  \draw[ultra thick, solid] (1414) node {}--(1515) node {}-- (1616) node {};
  \path (1616) node {};

  \tikzstyle{every path}=[]
  \tikzstyle{every node}=[draw, inner sep=5pt, thin, darkgray]
  \path (210) node[circle] {};
  \path (912) node[circle] {};
  \path (1414) node[circle] {};
  \tikzstyle{every node}=[draw, circle, inner sep=2.5pt, fill=bordeaux]
  \path (116) node {};
  \path (115) node {};
  \path (114) node {};
  \path (113) node {};
  \path (112) node {};
  \path (111) node {};
  \path (110) node {};
  \path (210) node {};
  \path (310) node {};
  \path (410) node {};
  \path (611) node {};
  \path (812) node {};
  \path (912) node {};
  \path (1214) node {};
  \path (1314) node {};
  \path (1414) node {};

  \tikzstyle{every path}=[very thick,rounded corners=6pt]
  \draw (116)--(115)--(114)--(113)--(112)--(111)--(110)--(19);
  \draw (19)--(110.north east)--(210)--(310);
  \draw (310)--(410)--(510)--(511.north east)--(611);
  \draw (611)--(711)--(712.north east)--(812);
  \draw (812)--(912)--(1012)--(1013.north east)--(1113);
  \draw (1113)--(1114.north east)--(1214);
  \draw (1214)--(1314)--(1414);

\end{tikzpicture}
    \hspace*{-10pt}
  }
\caption{(a) Example of an intersection hierarchy (the first row is the input sequence itself). (b)~Example of execution of the maximization algorithm. {\it The elements in gray and red are the only ones that need to be effectively computed; the gray ones are referred to as ladders.}}
\end{figure}

Then, with the same co-authors in~\cite{CKNP17}, we generalized this framework for computing other parameters than T-interval connectivity. At a general level, the two operations are referred to as a {\em composition} operation and a {\em test} operation. The framework allows us to compute various parameters of dynamic graphs by simply changing the two operations, while using a high-level strategy for the construction and for the walk that are problem-insensitive (up to having one for maximization, one for minimization). 
The framework is illustrated in~\cite{CKNP17} through three other problems (in addition to $T$-interval connectivity, which are {\em bounded realization of the footprint, temporal diameter,} and {\em round trip temporal diameter}.

\newcommand{\tmpdiam}{{\rmfamily\scshape Temporal-Diameter}\xspace}
\newcommand{\bound}{{\rmfamily\scshape Bounded-Realization-of-the-Footprint}\xspace}
\newcommand{\rtdiam}{{\rmfamily\scshape Round-Trip-Temporal-Diameter}\xspace}
\newcommand{\tmp}{{\rmfamily\scshape Temporal-Connectivity}\xspace}
\newcommand{\stmp}{{\rmfamily\scshape Strict-Temporal-Connectivity}\xspace}

\paragraph{Realization of the Footprint.}
\label{sec:bound}

This property is motivated by Class~\EB. Given a footprint $G$ (whose vertex set is identified with that of $\G$), the problem is to find the smallest duration $b$ such that in any window of length $b$ in $\G$, all edges of $G$ appear at least once. 
Here, the composition operation is the {\em union} of two graphs and the test operation is the {\em equality to $G$}. The row in which the walk terminates is the answer $b$.

\paragraph{Temporal Diameter.}
\label{sec:tmpdiam}

This property is motivated by Classes~\TC and~\TCB.
The problem is to find the smallest duration $d$ such that every pair of node can communicate through a journey in any window of length $d$.
The hierarchy built here is one of {\em transitive closures} of journeys. 
More precisely, the composition operation is the \textit{concatenation} of two transitive closures, and the test operation is {\em equality to the complete graph}. The row in which the walk terminates is the answer $d$. 

\paragraph{Round-trip Temporal Diameter.}
\label{sec:rtdiam}
This property is motivated by Class~\RT. The problem is to find the smallest duration $d$ such that a {\em back-and-forth} journey exists between every two nodes. 
This problem is more complicated than searching two contiguous temporally connected subsequences: we must allow that the arrival of some forth journeys occur after the departure of other back journeys, making the problem more intricate. To solve the problem, we introduce the notion of round-trip transitive closure, in which every present arc $(u,v)$ is additionally labelled with two extra times that indicate the earliest arrival and latest departure of journeys from $u$ to $v$. 
The hierarchy built for this problem is one of {\em round-trip transitive closures}, using a special type of concatenation for composition, and equality to a complete graph for test, with the additional constraint that for the earliest arrival of every edge must be smaller than its latest departure. 

\paragraph{Concluding remarks.}

These algorithms consider strict journeys by default, but they can be adapted to non-strict journeys by using the same pre-processing trick as above (illustrated on Figure~\ref{fig:non-strict_transitive_closure}).
Other results from~\cite{CKNP17} include showing that the algorithms can work online with amortized $\Theta(1)$ composition and test operations for each graph of the sequence
and proving constructively that some of the problems are in {\bf NC} (NC is Nick's class, containing all problems solvable in polylogarithmic time on parallel machines with polynomially many processors).

\subsection{Fine-grain dynamics \vs Coarse-grain dynamics}

Social network analysis is traditionally concerned with measuring parameters related to the network as well as the {\em evolution} of these parameters over time. 
The problem of capturing the evolution of structural parameters (such as paths, connectivity, distances, and clustering coefficients) proceeds usually through aggregation windows. One considers a sequence of graphs $G_1, G_2, ...$ such that each $G_i$ corresponds to the aggregation of interactions (often seen as a union of edges) taking place over some interval $[t_i,t_i']$. In other words, every $G_i$ is the footprint of some $\G_{[t_i,t'_i]}$ and then the main object of study becomes the sequence of these footprints. (Other terminologies in complex systems include {\em aggregated graphs} or {\em induced graphs}.)
Note that choosing the appropriate duration for time windows in this type of investigation is a typically difficult problem (see \eg~\cite{KKB12,CB13,LCF15}), which we have not considered {\it per se}.

The case of temporal parameters like journeys or temporal distances is more complex because two different time scales are involved in their studies: (1) their very definition relies on time and does not survive aggregation, and (2) their average features may be themselves subject to long-term evolution.
For example, questions like {\it how does the temporal distance evolve between nodes over large time scales?} embed two different references to time. Same for questions like {\it how does a network self-organize, optimize, or deteriorate over time in terms of temporal efficiency?}
We introduced in~\cite{SQFCA11} a distinction between {\em fine-grain} and {\em coarse-grain} dynamics to capture this distinction. The coarse-grain evolution of fine-grain dynamics is studied through considering a {\em sequence} of {\em non-aggregated} graphs $\G_1, \G_2, ...$, each one corresponding to $\G_{[a,b]}$ for some interval $[a,b]$ (see Section~\ref{sec:subgraph} for definitions). Then temporal metrics can be mesured on each $\G_i$ and their long term evolution examined classically over the sequence.

\paragraph{Concluding remarks.} This section reviewed some of our contributions around the question of testing properties of dynamic networks, given a trace of the communication links over some period of time. These contributions are of a {\em centralized} nature. However, the recognition of properties from a distributed point of view is a relevant question. In particular, if the nodes of a distributed network are able to learn properties about the underlying dynamic network, then this might help solve further problems. Some of our works reviewed in Section~\ref{sec:shfafo} relate to this topic, for example we explain that the nodes can learn the footprint of the network if the edges reappear within bounded time. From that, they can infer the number of nodes, which helps reduce in turn the message complexity of some problems. We also present in Section~\ref{sec:tclocks} a distributed algorithm for learning temporal distances among the nodes of a periodic network.

\begin{avenue}
  Investigate further distributed problems pertaining to the learning of properties of the dynamic network (self-awareness), with the aim to exploit the corresponding structure and gained knowledge in the solving of other problems.
\end{avenue}

\section{From classes to movements (and back)}
\label{sec:real-world}

This section is more prospective; it discusses several links between real-world mobility and properties of dynamic networks. Fragments of these discussions appeared in a joint report with Flocchini in 2013~\cite{CF13a}; other fragments are recurrent in our own talks; however, most of the ideas were not yet systematically explored (whence the ``prospective'' adjective). In particular, the second part about movement synthesis prefigure a longer-term research program for us, some aspects of which are the object of a starting PhD by Jason Schoeters (since Nov. 2017).

\subsection{Real-world mobility contexts}

Real-world mobility contexts are varied, including for instance sensors, pedestrians, robots, drones, vehicles, or satellites. Each of these networks is of course dynamic, but in its own peculiar way. An interesting question is to understand what properties could be found in each type of network, making it possible to design better distributed algorithms that exploit this structure in more specific ways.

\begin{figure}[h]
  \centering
      \begin{tikzpicture}[sloped, level distance=40pt, darkgray, font=\Large, sibling distance=35pt,xscale=1.4,yscale=1.2]
        \tikzstyle{every path}=[<-]
        \tikzstyle{every node}=[draw, rounded corners=3pt, inner sep=1.5pt]

        \node[grow=left] (f9) {\TCRrep}
          child[grow=left] {
            node[yshift=.8cm] (f10) {\ERrep}
            child [sibling distance=16pt] {
              node[yshift=15pt] (f11) {\EBrep}
              child {
                node (f12) {\EPrep}
              }
            }
            child {
              node[above] (f17) {\KRrep}
            }
          }
          child [grow=left]{
            node[yshift=-.7cm] (f13) {\PRrep}
            child {
              node (f14) {\CRrep}
              child {
                node (f15) {\ACrep}
                child {
                  node (f16) {\TINTrep}
                }
              }
            }
          };

        \tikzstyle{every node}=[inner sep=0pt]
        \tikzstyle{every path}=[brown]
        \path (f10)+(0,1.8) node (smartphone){\includegraphics[width=1.0cm]{./figs-finding/smartphone.png}};
        \draw (smartphone)--(f10);
        \draw (smartphone)edge[dashed](f11);
        \path (f9)+(.2,-1.5) node (voiture){\includegraphics[width=1.0cm]{./figs-finding/voiture.png}};
        \draw (voiture)--(f9);
        \path (f12)+(0,1.5) node[inner sep=-2pt] (satellite){\includegraphics[width=1.0cm]{./figs-finding/satellite.png}};
        \draw (satellite)--(f12);
        \path (f15)+(.95,-1.6) node[left] (robot){\includegraphics[width=1.0cm]{./figs-finding/robot.png}};
        \path (robot)+(-1.15,.2) node[left] (sensor){\includegraphics[width=.6cm]{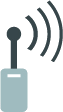}};
        \draw (sensor)--(f15);
        \draw (sensor)edge[dashed](f16);
        \path (robot)+(-.2,.5) node[above]{\large ?};
        \path (f15)+(1.5,-1.4) node[right] (drone){\includegraphics[width=1.0cm]{./figs-finding/drone.png}};
        \path (drone) node[above=10pt]{\large ?};
        \path (drone)+(1,-.1) node[inner sep=-4pt] (soldier){\includegraphics[width=.6cm]{./figs-finding/soldier.png}};
        \path (soldier) node[above=14pt]{\large ?};
        \path (f17)+(-1.2,0) node (canard){\includegraphics[width=1.0cm]{./figs-finding/duck.png}};
        \draw (canard)--(f17);

        \tikzstyle{every path}=[black]
      \end{tikzpicture}
    \caption{\label{fig:mobility-contexts}Properties of real-world mobility contexts.}
\end{figure}

A number of such connections are illustrated in Figure~\ref{fig:mobility-contexts}, using the visual representation of the classes (see Table~\vref{tab:classes}).
For instance, it is generally admitted that satellites have periodic movements (Class $\EP$), while sensor networks remain in general connected at any instant (Class $\AC$). Interaction between smartphones may represent interactions between people in a given context, such as in a small company with bounded recurrence of edges, typically within a week (Class $\EB$), or in a community with unbounded, yet recurrent interactions (Class $\ER$). Vehicular networks exhibit an important range of densities and connectivity patterns, but they still offer recurrent connectivity over time and space (Class $\TCR$). Vehicles also share some traits with pedestrian through having their movements constrained by the environment (roads and pathways).

Through linking mobility contexts to graph properties, we expect to understand better what are the possibilities and limitations of each context, as well as enabling a more systematic transfer of results among the different contexts. It also help understand how networks of different natures could inter-operate.

\begin{avenue}
  Conduct a more systematic study of what property (structure) could be expected in what context, using real mobility traces datasets like the ones from the CRAWDAD initiative~\cite{crawdad}.
\end{avenue}

On a related note, studies measuring specific temporal metrics in such traces have been carried out recently, \eg reachability~\cite{WDCG12} and density of interaction~\cite{gaumont}.

\subsection{Inducing structure through movements}

Some types of mobility contexts like drones or robots have the
convenient feature that the entities {\em control} their movements; that
is, mobility is active rather than passive. Of course, the entities
have a primary mission which determines to a large extent
their patterns of movements. However, it is
possible to {\em influence} their movements in order to force the satisfaction
of certain properties. This general approach is often considered
with respect to classical properties, \eg swarms of mobile entities 
moving while remaining connected all the time. 

Going further, one may typically relax the connectivity constraint,
making it possible for the network to operate in a disconnected mode
with {\em temporal} connectivity constraints. Such properties would
make movements less constrained while preserving the ability to
communicate within the group. Several temporal properties may actually
be considered. Of particular interest are movements preserving a {\em
  bounded temporal diameter} (Class \TCB), because this makes it
possible for the nodes to distinguish between a temporary isolated
node and a crashed node (if the group does not hear from it for some time). 

\begin{avenue}
\label{av:movements}
  Design collective mobility patterns whose resulting graph/network satisfies a number of temporal properties, possibly inspired from the classes reviewed in this chapter.
\end{avenue}

This topic raises in turn a number of fundamental questions about movements, in particular about the way they should be approached in theoretical computer science. A lot of effort has been devoted to movements synthesis in control theory, with approaches ranging from analytical (here, we use ``analytical'' in the mathematical sense of analytic functions) to heuristics based on potential fields. We refer here in particular to movements satisfying {\em kinodynamic} constraints (acceleration, deceleration, and more generally inertia). It seems that little research was conducted in this direction so far from a discrete and combinatorial setting, which is however at the core of our algorithmic culture. (We do not include here a large body of literature about mobile robots in the distributed computing community, where the motivations and constraints are typically different, \eg forming fixed patterns, breaking symmetry, or gathering at a same point.) 

\begin{avenue}
  Explore discrete combinatorial models for movement synthesis which satisfy acceleration constraints and yet are simple enough to serve as a basis for theoretical investigation. Consider trajectory problems typically based on heuristics and simulations (\eg path planning, obstacle avoidance), trying to tackle them with a more algorithmic approach (\eg exact algorithms, reductions, approximations).
\end{avenue}

In a recent work with M. Raffinot and J. Schoeters, we revisited the TSP problem subject to acceleration constraints modelled by a ``Vector Racer''-like model. In such a model, the mobile entities can move by discrete amount (integral vector coordinates), and every next vector cannot diverge by much from the previous one. This approach made it possible for us to design {\em reductions} from and to the TSP with inertia constraints to other versions of TSP. The years to come will likely see the emergence of an algorithmic science of movements based on similar {\it discrete} approaches, motivated by the fact that mobile entities with controlled movements are becoming commonplace. On a more personal note, we find the topic quite exciting and forsee many interesting connections to temporal properties being possibly induced by movements in the resulting network. (Another recent work considering such model for an $s$-$t$ trajectory optimization is~\cite{racetrack}.)

\chapter{Beyond structure}

This chapter reviews several contributions not directly related to the presence of {\em structure} in dynamic networks, which are however fully integrated within the same lines of work. In fact, some sections relate explicitly to the {\em absence} of structure. We first review a series of contributions around the topic of maintaining a distributed spanning forest in a highly-dynamic network when no assumption is made on the underlying dynamics. This topic is a recurrent source of interest to us since the introduction of the abstract scheme in~\cite{Cas06}; it has since then generated many developments with various co-authors (listed below)~\cite{BCCJN18,PCGC10,CCGP13}), covering different aspects of the problem. Next, we present an algorithm called \tclocks, elaborated with Flocchini, Mans, and Santoro through~\cite{CFMS11} and~\cite{CFMS14}, which makes it possible to measure temporal distances in a distributed (and continuous-time) setting. While the algorithm requires no particular structure on the underlying dynamics, it has interesting applications when more structure {\em is} available (\eg if the network is periodic). Then, we present a theoretical investigation about the expressivity of time-varying graphs in terms of automata and languages, carried out with Flocchini, Godard, Santoro, and Yamashita in~\cite{CFGSY12,CFGSY13,CFGSY15}, focusing on the impact of being able to buffer information at intermediate nodes (along journeys). We conclude with a small collection of interesting facts which we have collected over the years, in which the dynamic feature seems to have an impact on the very nature of things.

\section{Spanning forest without assumptions}
\label{sec:forest}

\articles{ICAS'06~\cite{Cas06}, ALGOTEL'10~\cite{PCGC10}, ADHOCNOW'13~\cite{CCGP13}, OPODIS'14~\cite{BCCJN14b} (long version in The Computer Journal~\cite{BCCJN18}), PhD Y. Neggaz~\cite{Neggaz16}, PhD M. Barjon~\cite{barjon}}

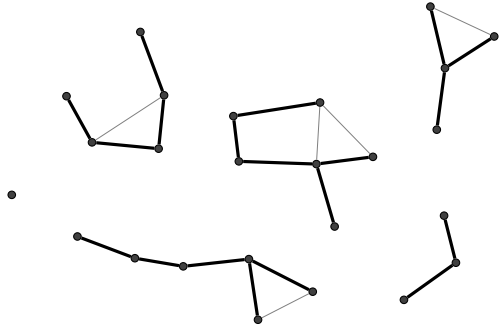
\begin{figure}[h]
  \centering
  \begin{tikzpicture}[scale=.6]
    \tikzstyle{every node}=[draw,fill=darkgray,circle,inner sep=1]
    \path (5.26,7.14) node (v6) {};
    \path (6.78,6.68) node (v9) {};
    \path (6.9,5.68) node (v10) {};
    \path (5.14,5.96) node (v5) {};
    \path (3.68,6.1) node (v3) {};
    \path (3.12,7.12) node (v1) {};
    \path (4.74,8.54) node (v8) {};
    \path (8.68,6.98) node (v15) {};
    \path (8.6,5.62) node (v14) {};
    \path (9.0,4.24) node (v16) {};
    \path (1.92,4.94) node (v0) {};
    \path (11.24,6.38) node (v20) {};
    \path (11.42,7.74) node (v22) {};
    \path (9.84,5.78) node (v17) {};
    \path (12.5,8.44) node (v24) {};
    \path (11.1,9.1) node (v19) {};

    \path (11.4,4.48) node (v21) {};
    \path (10.52,2.62) node (v18) {};
    \path (11.66,3.44) node (v23) {};
    \path (7.12,3.52) node (v11) {};
    \path (3.36,4.02) node (v2) {};
    \path (4.62,3.54) node (v4) {};
    \path (5.68,3.36) node (v7) {};
    \path (8.52,2.8) node (v13) {};
    \path (7.32,2.18) node (v12) {};
    \tikzstyle{every path}=[very thin, gray];
    \draw (v15)--(v14);
    \draw (v13)--(v12);
    \draw (v15)--(v17);
    \draw (v6)--(v3);
    \draw (v24)--(v19);

    \draw (v9)--(v10);
    \draw (v6)--(v5);
    \draw (v3)--(v1);
    \draw (v6)--(v8);
    \draw (v9)--(v15);
    \draw (v10)--(v14);
    \draw (v14)--(v16);
    \draw (v2)--(v4);
    \draw (v11)--(v7);
    \draw (v4)--(v7);
    \draw (v11)--(v13);
    \draw (v11)--(v12);
    \draw (v23)--(v21);
    \draw (v23)--(v18);
    \draw (v3)--(v5);
    \draw (v20)--(v22);
    \draw (v14)--(v17);
    \draw (v22)--(v24);
    \draw (v22)--(v19);
      \tikzstyle{every path}=[very thick];
      \draw (v9)--(v10);
      \draw (v6)--(v5);
      \draw (v3)--(v1);
      \draw (v6)--(v8);
      \draw (v9)--(v15);
      \draw (v10)--(v14);
      \draw (v14)--(v16);
      \draw (v2)--(v4);
      \draw (v11)--(v7);
      \draw (v4)--(v7);
      \draw (v11)--(v13);
      \draw (v11)--(v12);
      \draw (v23)--(v21);
      \draw (v23)--(v18);
      \draw (v3)--(v5);
      \draw (v20)--(v22);
      \draw (v14)--(v17);
      \draw (v22)--(v24);
      \draw (v22)--(v19);
  \end{tikzpicture}
  \caption{\label{fig:spanning-forest}A spanning forest in a disconnected graph.}
\end{figure}

We reviewed in Section~\ref{sec:redefinition} several ways to reformulate the spanning tree problem in highly-dynamic networks. Among the possible ways, one is as the {\em maintenance} of a set of trees, containing only edges that exist at the considered time (see Figure~\ref{fig:spanning-forest} for an illustration). The formulation of spanning trees as a maintenance problem is quite standard in the area of ``dynamic graph algorithms''. In constrast, here, the setting is distributed and the network is typically disconnected, meaning that the best configuration is a single tree per component. Furthermore, the nodes may not be given enough time between the changes to converge to a single tree per component.
In fact, it may even be the case that convergence never occurs. 
This section reviews our attempts at understanding what can still be done in such a context, with a focus on solutions where all decisions are taken {\em immediately}, without consulting the rest of the component.

\subsection{Background}

We introduced in 2006~\cite{Cas06} a high level (abstract) algorithm for this problem, whose initial purpose was to illustrate the automatic generation of code based on relabeling rules (a somewhat unrelated topic). Several versions of this scheme were subsequently studied; Guinand and Pigné realized some experimentations on random and deterministic variants, with or without backtracking, leading to a short joint account of experimental results in~\cite{PCGC10} (in French). 
Then, we established the correctness of the algorithm in the abstract model three years later in~\cite{CCGP13}. More recently, together with Johnen and Neggaz, we showed in~\cite{BCCJN14b} (long version~\cite{BCCJN18}) that this algorithm can be adapted in the synchronous message-passing model, at the cost of substantial sophistications and a new technique for dealing with non-atomicity. The article includes simulation results obtained together with Barjon and Chaumette, showing that the algorithm is relevant in a practical scenario (based on real mobility datasets).

A number of other works in the literature considered spanning trees in dynamic networks, with various limitations on the dynamics~\cite{asynchronous,synchronous,mosbah-tree,Baala03,BBS13,Awerbuch08}, with a special mention to~\cite{Awerbuch08}, in which the restrictions are very mild (and the computed trees are minimum). We refer the reader to~\cite{BCCJN18} for a more detailed description of these works and how they relate to our solution.

\subsection{The algorithm}
\label{sec:principle}

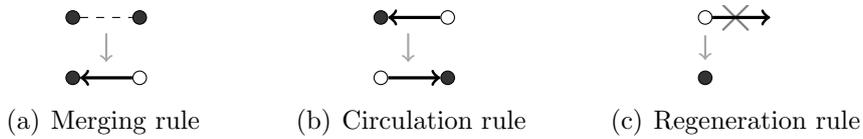
\begin{figure}[h]
  \centering
  \subfigure[Merging rule]{
    \begin{tikzpicture}[scale=.8]
      \clip (-2,-1.2) rectangle (1,.2);
      \tikzstyle{every node}=[draw,circle,fill=black!80,inner sep=1.8pt]
      \path (-1,0) node (v11) {};
      \path (.1,0) node (v12) {};
      \path (-1,-1) node (v21) {};
      \path (.1,-1) node[fill=white] (v22) {};
      \draw[dashed] (v11)-- coordinate[midway](mid1) (v12);
      \draw[very thick,<-] (v21)-- coordinate[midway](mid2)(v22);
      \draw[thick, gray!70, shorten >=6pt, shorten <=6pt] (mid1) edge[->] (mid2);
      \tikzstyle{every node}=[font=\footnotesize,above=2pt]
    \end{tikzpicture}
  }\hspace{1cm}
  \subfigure[Circulation rule]{
    \begin{tikzpicture}[scale=.8]
      \clip (-2.3,-1.2) rectangle (1.3,.2);
      \tikzstyle{every node}=[draw,circle,fill=black!80,inner sep=1.8pt]
      \path (-1,0) node (v11) {};
      \path (.1,0) node[fill=white] (v12) {};
      \path (-1,-1) node[fill=white] (v21) {};
      \path (.1,-1) node (v22) {};
      \draw[very thick,<-] (v11)--(v12);
      \draw[very thick,->] (v21)--(v22);
      \path (-.55,0) coordinate (mid1);
      \path (-.55,-1) coordinate (mid2);
      \draw[thick, gray!70, shorten >=6pt, shorten <=6pt] (mid1) edge[->] (mid2);
      \tikzstyle{every node}=[font=\footnotesize,above=2pt]
    \end{tikzpicture}
  }\hspace{.7cm}
  \subfigure[Regeneration rule]{
    ~
    \begin{tikzpicture}[scale=.8]
      \clip (-3.3,-1.2) rectangle (.3,.2);
      \tikzstyle{every node}=[draw,circle,fill=black!80,inner sep=1.8pt]
      \path (-2,0) node[fill=white] (v11) {};
      \path (-2,-1) node (v21) {};
      \path (-.55,0) coordinate (mid1);
      \path (-.55,-1) coordinate (mid2);
      \tikzstyle{every node}=[]
      \path (-.9,0) node[gray] (v12) {\hspace{-1.1cm} \LARGE $\times$};
      \draw[very thick, ->] (v11)--(v12);
      \draw[thick, gray!70, ->, shorten >=4pt, shorten <=4pt] (v11)--(v21);
    \end{tikzpicture}
    ~
  }
  \caption{\label{fig:scheme} Spanning forest principle (high-level representation). {\it Black nodes are those having a token. Black directed edges denote child-to-parent relationships. Gray vertical arrows represent transitions.}}
\end{figure}

The algorithm in~\cite{Cas06} was specified at a high-level of abstraction, based on a coarse grain interaction model inspired from graph relabeling systems~\cite{LMS99}, where interaction occurs among two neighbors in an atomic way (the model is different from population protocols in that the scheduler does not abstract dynamism, but only communication, see Section~\ref{sec:conditions} for a discussion on these two models). The general principle is as follows. Initially every node has a token, meaning that it is the {\em root} of a tree (initially, its own tree). The first rule (merging rule on Figure~\ref{fig:scheme}) specifies that if two neighbors having a token interact, then their trees are merged and one of the node becomes the parent of the other (which also loses its token). In absence of merging opportunity, the tokens execute a walk (typically random, but not necessarily) {\em within} the tree in the search for other merging opportunities (circulation rule). The circulation of the token flips the direction of relations in the tree, so that the node having the token is always the root of its tree, and the tree is correctly rooted (all relations point towards the root). The fact that the walk uses only edges in the tree is crucial, because it makes it possible to regenerate a token {\em immediately} when an edge of the tree disappears (regeneration rule, on the child side). Indeed, the child side of a disappeared edge knows that its subtree is token free; it can regenerate a token without concertation and in a seemless way for the rest of its subtree. As a result, both mergings and regenerations are purely localized and immediate decisions.

\paragraph{Properties.} We proved in~\cite{CCGP13} that the above scheme guarantees that, at any time, the network remains covered by a set of trees such that 1) no cycle exists, 2) every node belongs to a tree, and 3) exactly one node has a token in every tree and all the edges of the tree are properly directed towards it. These properties hold whatever the chaos of topological changes. One might object here that an algorithm doing basically nothing also satisfies these requirements, which is correct. The above scheme also guarantees that (with slight care given to the implementation), the process eventually converges towards a single tree per connected component if the network stops changing at some point. In some sense, it is a {\em best effort} algorithm. Correctness was much more difficult to establish in the case of the message-passing version; it occupied us for months and was eventually done successfully~\cite{BCCJN18}.

\paragraph{Speed of convergence.} Little is known about the speed of convergence of (any version of) this algorithm. As already discussed, it may so happen that the rate of change is too fast for the algorithm to converge. Furthermore, the network may be partitioned at all time steps. We suggested in~\cite{PCGC10} that an appropriate metric in such cases is the {\em number of trees per connected component}, considered for example in a stationary regime (in case of probabilistic investigation) or over a real-world mobility trace (in case of simulations, which is exactly what we did in~\cite{BCCJN18}).

\begin{open}
Understand the theoretical complexity of this process. Based on discussions with knowledgeable colleagues in related areas (in particular T. Radzik, D. Sohier, and J.-F. Marckert), even a static analogue of this process based only on the merging and circulation operations has not been considered in standard graphs. These operations thus define a new form of coalescing process whose study remains to be done.
\end{open}

We made preliminary observations in this direction in~\cite{CCGP13}, however the problem remains mostly open. Because the algorithmic principle is high-level, an ambiguity arises as to the exact behavior of the process: is it continuous or discrete? is it synchronous or asynchronous? what priority rules apply in case of ties? {\it etc.} These parameters have an impact on the analysis. However, we believe that the choice among them should be precisely driven by simplicity. In the longer term, the objective is to understand the dynamic version (with edge dynamics and regeneration rule) in a stochastic model of dynamic network like edge-markovian dynamic graphs~\cite{CleMMPS08}, characterizing \eg the number of trees per connected components in a stationary regime, as a function of birth and death rates for edges. We refer the reader to~\cite{BZ89,IJ90,AF02,CEOR12,broutin} for references on coalescing processes in general.

\section{Measuring temporal distances distributedly}
\label{sec:temporal-lags}

\articles{IPDPS'11~\cite{CFMS11}, arXiv'12~\cite{CFMS12b}, IEEE Trans. Computers'14~\cite{CFMS14}}

We mentioned in Chapter~\ref{chap:dynamic-networks} a number of temporal concepts, among which that of {\em temporal distance} that accounts for the time it takes for entities of a dynamic network to reach each others.
We present here a joint work with Flocchini, Mans, and Santoro~\cite{CFMS11} in which we design {\em distributed} algorithms for the nodes to learn temporal distances to each other in a unstructured dynamic network. The time setting is {\em continuous}; contacts may have arbitrary durations; and the latency of individual contacts is non-zero, altogether  making the problem more complex. Our solution relies on the definition of an abstraction called \tclocks, which enables to solve other problems in periodically-varying networks.

\subsection{Overview}
The problem of measuring temporal distances in dynamic networks was addressed in a number of works from the field of social network analysis~\cite{Holme05, Kostakos09, KKW08}. 
In particular, Kossinets, Kleinberg, and Watts~\cite{KKW08} flip the point of view, asking how "out-of-date" an entity (node) $v$ is with respect to every other node at a time $t$. They define a concept of {\em view} as the answer table to this question, the indices being the nodes and the values indicating, for every other node $u$, the latest time at which $u$ could send a piece of information arriving at $v$ before $t$ (\ie latest time of influence). The setting is centralized and contacts among entities are punctual, which implies that the views only changes punctually by discrete amounts. Finally, the speed of propagation along a contact (edge latency) is neglecte as well.

In~\cite{CFMS11}, we considered the {\em distributed} version of this problem in a {\em continuous} time setting where contacts can have arbitrary {\em durations}. We also relaxed the assumption that latency is instantaneous, by considering an arbitrary (but fixed) latency. We construct an abstraction called \tclocks, which every node can use to know, at any point in time, how out-of-date it is with respect to every other node. The main difficulty stems from considering continuous time together with non punctual contacts, the combination of which induces {\em continuums} of {\em direct} journeys (direct journeys are journeys realized without pauses at intermediate nodes) which induces continuous evolution of the view (more below). The edge latencies also complicates things by making it possible for an direct journey to be actually slower than an indirect one (indirect journeys are such that a pause is made at least at one intermediate node).

The use of \tclocks is illustrated through solving two problems in periodically varying networks (class $\EP$), namely building reusable {\em foremost broadcast trees} (same article~\cite{CFMS11}), and building {\em fastest broadcast trees} (in~\cite{CFMS12b}), both being combined in a long version~\cite{CFMS14}. The latter reduces to the problems of (i) finding minimum time of {\em temporal eccentricity} (\ie when the duration needed to reach all other nodes is the smallest), then (ii) building a {\em foremost} broadcast tree for this particular time (modulo the period $p$). Next, we briefly discuss some of the key technical aspects.

\subsection{Temporal views}
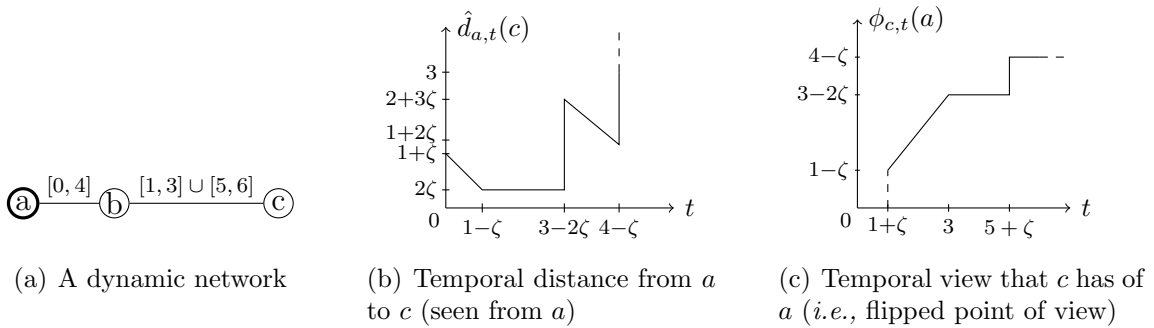
\begin{figure}[h]
  \centering
  \subfigure[A dynamic network]{
    \label{fig:basicgraph}
    \begin{tikzpicture}[scale=1.2]
      \tikzstyle{every node}=[draw,circle, minimum size=11pt, inner sep=0pt]
      \path (0,-.5) coordinate;
      \path (0,0) node[very thick] (a){a};
      \path (a)+(1,0) node (b){b};
      \path (a)+(2.8,0) node (c){c};
      \tikzstyle{every node}=[below,font=\scriptsize,inner sep=1pt]
      \draw (a)--node[above, yshift=1pt]{$[0,4]$}(b);
      \draw (b)--node[above, yshift=1pt]{$[1,3] \cup [5,6]$}(c);
    \end{tikzpicture}
  }
  ~~~~
  \subfigure[Temporal distance from $a$ to $c$ \mbox{(seen from $a$)}]{
    \label{fig:viewplot-d}
    \begin{tikzpicture}[scale=.6]
      \draw[->] (0,0)--(0,4);
      \draw[->] (0,0)--(5,0);
      \draw (0,1.2)--(.8,0.4)--(2.6,0.4)--(2.6,2.4)--(3.8,1.4) -- (3.8,3) edge [dashed](3.8,4);
      \tikzstyle{every node}=[font=\scriptsize, inner sep=3pt]
      \path (0,0) node[below left, inner sep=2pt] {$0$};
      \path (0,0.4) node[left] {$2\zeta$};
      \path (0,1.2) node[left] {$1$$+$$\zeta$};
      \path (0,1.65) node[left] {$1$$+$$2\zeta$};
      \path (0,2.4) node[left] {$2$$+$$3\zeta$};
      \path (0,3) node[left] {$3$};
      \tikzstyle{every node}=[font=\scriptsize, inner sep=4pt]
      \path (.8,0) node[below] {$1$$-$$\zeta$};
      \path (2.6,0) node[below] {$3$$-$$2\zeta$};
      \path (3.8,0) node[below] {$4$$-$$\zeta$};
      \tikzstyle{every node}=[font=\footnotesize, inner sep=4pt]
      \path (0,4) node[right] {$\temp{d}_{a,t}(c)$};
      \path (5,0) node[right] {$t$};
      \draw (.8,-.1)--(.8,.1);
      \draw (2.6,-.1)--(2.6,.1);
      \draw (3.8,-.1)--(3.8,.1);
      \draw (-.08,.4)--(.08,.4);
      \draw (-.08,1.2)--(.08,1.2);
      \draw (-.08,1.5)--(.08,1.5);
      \draw (-.08,2.4)--(.08,2.4);
      \draw (-.08,3)--(.08,3);
    \end{tikzpicture}
  }
  ~~~~
  \subfigure[Temporal view that $c$ has of $a$ \mbox{(\ie flipped point of view)}]{
    \label{fig:viewplot}
    \begin{tikzpicture}[xscale=.4,yscale=.5]
      \draw[->] (0,0)--(0,5);
      \draw[->] (0,0)--(7,0);
      \draw (1,0)edge[dashed](1,1);
      \draw (1,1)--(3,3)--(5,3)--(5,4)--(6,4)
      edge [dashed](7,4);
      \tikzstyle{every node}=[font=\scriptsize, inner sep=3pt]
      \path (0,0) node[below left, inner sep=2pt] {$0$};
      \path (0,1) node[left] {$1$$-$$\zeta$};
      \path (0,3) node[left] {$3$$-$$2\zeta$};
      \path (0,4) node[left] {$4$$-$$\zeta$};
      \path (1,0) node[below] {$1$+$\zeta$};
      \tikzstyle{every node}=[font=\scriptsize, inner sep=4pt]
      \path (3,0) node[below] {$3$};
      \path (5,0) node[below] {$5+\zeta$};
      \tikzstyle{every node}=[font=\footnotesize, inner sep=4pt]
      \path (0,5) node[right] {$\phi_{c,t}(a)$};
      \path (7,0) node[right] {$t$};
      \draw (1,-.1)--(1,.1);
      \draw (3,-.1)--(3,.1);
      \draw (5,-.1)--(5,.1);
      \draw (-.08,1)--(.08,1);
      \draw (-.08,3)--(.08,3);
      \draw (-.08,4)--(.08,4);
    \end{tikzpicture}
    ~~
  }
  \caption[Temporal distance and temporal views.]{\label{fig:overlapping}Temporal distance and temporal views as a function of time {\it (with $\zeta \ll 1$)}.}
\end{figure}

In~\cite{CFMS11,CFMS14} we refer to the concept of view as {\em temporal views} to avoid conflicts with the (unrelated) views of Yamashita and Kameda~\cite{YamK96}. As explained, considering non punctual contacts implies the possible co-existence of indirect journeys, and {\em continuums} of direct journeys on the other hand. While both can be dealt with in the same way in discrete time, their combination in continuous time produces complex patterns of temporal distances among nodes, as illustrated on Figure~\ref{fig:overlapping} (from~\cite{CFMS14}). 

Temporal distance and temporal view actually refer to the same quantity seen from different perspectives: the first is a {\em duration} seen from the emitter, while the second is a time seen from the receiver.

\subsection{The {\sc T-Clocks} abstraction}
\label{sec:tclocks}

\begin{figure}[h]
  \centering
  \tikzsetnextfilename{tikz-tclocks}
\begin{tikzpicture}[yscale=0.99]
  \tikzstyle{every node}=[font=\normalsize]
  \path (0,.5) node (dtn) {Network};
  \path (0,2) node (vc) {\tclocks};
  \path (0,3.8) node[sloped,bend right] (client) {Higher-Level Algorithm};
  \tikzstyle{every node}=[font=\scriptsize, inner sep=1pt]
  \tikzstyle{every path}=[semithick, shorten <=4pt, shorten >=1pt]
  \draw ([xshift=-.25cm]vc.north) edge[->] node[pos=.75,left] {{\tt levelChanged()}} node[pos=.55, right, inner sep=2pt]{}([xshift=-.55cm]client.south);
  \draw ([xshift=.25cm]vc.north) edge[->] node[pos=.75,right] {{\tt dateImproved()}} ([xshift=.55cm]client.south);
  \draw ([xshift=-.2cm]dtn.north) edge[->] node[pos=.75,left] {{\tt onEdgeAppearance()}} node[pos=.6, right]{} ([xshift=-.5cm]vc.south);
  \draw ([xshift=.2cm]dtn.north) edge[->] node[pos=.75,right] {{\tt onEdgeDisappearance()}} ([xshift=.5cm]vc.south);
  \tikzstyle{every path}=[semithick]
  \draw (50:4.8) arc (50:130:4.8);
  \draw (50:2.8) arc (50:130:2.8);
  \draw (50:1.2) arc (50:130:1.2);
\end{tikzpicture}
\caption{\label{fig:relation} \tclocks as an abstraction to track temporal views. Picture from \cite{CFMS14}.}
\end{figure}
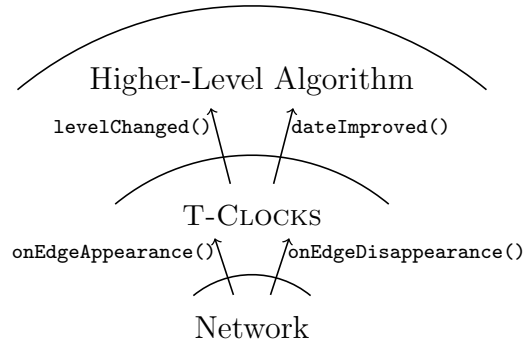

Direct journeys are often faster than indirect ones, but this is not always true. As a result, the temporal view $\phi$ that a node has (of another node) could result from either types of journey. 
The algorithm we introduced tracks the evolution of both types of views separately. While indirect views need only be updated punctually (in a similar way as in~\cite{KKW08}), direct views evolve continuously, and are thus inferred, on demand, from the knowledge of the (currently best) number of hops in contemporaneous journeys from a considered node (say $u$) to the local node $v$. We call this hop distance the {\em level} of $v$ with respect to $u$. The \tclocks algorithm consists of maintaining up-to-date information at each node $v$ relative to its $level$ (for direct views), and latest $date$ of influence (for indirect views) relative to every other node $u$. This work assumed the same computational model as in~\cite{CFMS15} (described in Section~\ref{sec:shfafo}); in particular, it relies on presence oracle to detect the appearance and disappearance of incident edges. The resulting abstraction is illustrated on Figure~\ref{fig:relation}.

\subsection{Using {\sc T-Clocks} in periodically-varying networks}
Using \tclocks, it becomes easier to build foremost or fastest broadcast trees in periodically-varying networks ($\EP$), thereby completing the results presented in Section~\ref{sec:shfafo}. We briefly review how \tclocks were used to build foremost broadcast trees in~\cite{CFMS11} and fastest broadcast trees in~\cite{CFMS12b} (both being combined in~\cite{CFMS14}).

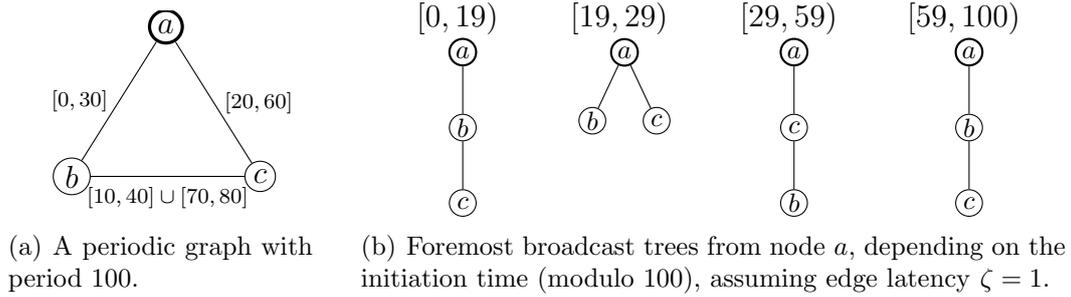
\begin{figure}[h]
  \centering
    \subfigure[A periodic graph with period $100$.]{
    \tikzsetnextfilename{tikz-triangle}
    \label{fig:triangle}
    \begin{tikzpicture}[scale=1.8]
      \clip (-1,-1.4) rectangle (1,.12);
      \tikzstyle{every node}=[draw, circle, inner sep=1.5pt]
      \path (0,0) node[very thick] (a){$a$};
      \path (a)+(-58:1.3) node (c){$c$};
      \path (a)+(-122:1.3) node (b){$b$};
      \tikzstyle{every node}=[font=\scriptsize]
      \draw (a)--node[midway,left]{$[0,30]$}(b);
      \draw (a)--node[midway,right]{$[20,60]$}(c);
      \draw (b)--node[midway,below]{$[10,40]\cup [70,80]$}(c);
    \end{tikzpicture}
  }
  \hspace{10pt}
  \subfigure[Foremost broadcast trees from node $a$, depending on the initiation time (modulo $100$), assuming edge latency $\zeta=1$.]{
    \label{fig:trees}
  \renewcommand{\tabcolsep}{13pt}
    \begin{tabular}[b]{ccccc}
      $[0,19)$&$[19,29)$&$[29,59)$&$[59,100)$\\
    \tikzsetnextfilename{tikz-triangle-tree1}
      \begin{tikzpicture}
        \tikzstyle{every node}=[draw, circle, minimum size=10pt, inner sep=0pt, font=\footnotesize]
        \path (0,0) node[thick] (ta){$a$};
        \path (ta)+(-90:1) node (tb){$b$};
        \path (tb)+(-90:1) node (tc){$c$};
        \tikzstyle{every node}=[font=\scriptsize]
        \draw (ta)--(tb);
        \draw (tb)--(tc);
      \end{tikzpicture}
      &
    \tikzsetnextfilename{tikz-triangle-tree2}
      \begin{tikzpicture}
        \tikzstyle{every node}=[draw, circle, minimum size=10pt, inner sep=0pt, font=\footnotesize]
        \path (0,0) node[thick] (ta){$a$};
        \path (ta)+(-115:1) node (tb){$b$};
        \path (ta)+(-65:1) node (tc){$c$};
        \path (ta)+(-90:2.18) coordinate (bidon);
        \tikzstyle{every node}=[font=\scriptsize]
        \draw (ta)--(tb);
        \draw (ta)--(tc);
      \end{tikzpicture}
      &
    \tikzsetnextfilename{tikz-triangle-tree3}
      \begin{tikzpicture}
        \tikzstyle{every node}=[draw, circle, minimum size=10pt, inner sep=0pt, font=\footnotesize]
        \path (0,0) node[thick] (ta){$a$};
        \path (ta)+(0,-1) node (tc){$c$};
        \path (tc)+(0,-1) node (tb){$b$};
        \tikzstyle{every node}=[font=\scriptsize]
        \draw (ta)--(tc);
        \draw (tc)--(tb);
      \end{tikzpicture}
      &
    \tikzsetnextfilename{tikz-triangle-tree4}
      \begin{tikzpicture}
        \tikzstyle{every node}=[draw, circle, minimum size=10pt, inner sep=0pt, font=\footnotesize]
        \path (0,0) node[thick] (ta){$a$};
        \path (ta)+(-90:1) node (tb){$b$};
        \path (tb)+(-90:1) node (tc){$c$};
        \tikzstyle{every node}=[font=\scriptsize]
        \draw (ta)--(tb);
        \draw (tb)--(tc);
      \end{tikzpicture}
    \end{tabular}
  }
  \caption{\label{fig:time-dependent} Example of foremost broadcast trees from a given node, as a function of the initiation times at the emitter (adapted from~\cite{CFMS14}).}
\end{figure}

{\bf Foremost broadcast trees in \EP.} These trees indicate what structural path (routing choice) a message initiated at a given emitter (here node $a$) should follow to arrive at every other node (here, $b$ and $c$) at the earliest possible time. Examples of such trees are shown on Figure~\ref{fig:time-dependent}, where the intervals correspond to when the message is {\em initiated} at the emitter (rather than when the routing is made). Some edges may not be available directly, in which case the message should be sent upon their next appearance.
Solving this problem with \tclocks boils down to monitoring at each node (here, $b$ and $c$) the evolution of the temporal view with respect to the considered emitter (here, node $a$). Whether due to direct and indirect journeys, \tclocks make it possible to track which local neighbors are responsible for providing the best view of the emitter over a complete period $p$, these neighbors being then selected as foremost parent for the corresponding times (modulo $p$).

{\bf Fastest broadcast trees in $\EP$.}
Fastest journeys minimize the span between first emission and last reception (possibly at the cost of delaying the first emission). Unlike the foremost quality, it is not always possible to find a tree in which {\em every} journey is fastest; this kind of ``optimal substructure'' feature applied to the foremost metric, making the definition of a tree more natural.
We thus define a fastest broadcast tree as one (among the possibly many) which minimizes time between first emission at the emitter and last reception everywhere.

So defined, an interesting observation is that at least one of the {\em foremost} broadcast trees must also be {\em fastest}. In fact, the considered problem reduces to (i) finding the time of minimum temporal eccentricity of the emitter (modulo the period $p$), and (ii) build a foremost broadcast tree for this initiation time. 
Since the construction of foremost broadcast tree is already solved by the algorithm presented above, we focus on the problem of determining the time of minimum temporal eccentricity using \tclocks.
Note that this problem is interesting in its own right, and may be used as a building block for other tasks than broadcasting (\eg electing a leader based on its ability to reach other nodes quickly). 

The algorithm consists of monitoring the evolution of temporal views at each node over a complete period (relative to the given emitter). The resulting information is converted back to temporal distance information, which are then aggreated back to the emitter, which eventually knows its eccentricity over a complete period.
Finally, the emitter choses any time of minimum eccentricity to initiate the construction of a foremost (and thus here fastest) broadcast tree.

\section{The power of waiting}
\label{sec:expressivity}

\articles{PODC'12~\cite{CFGSY12}, FCT'13~\cite{CFGSY13}, TCS'15~\cite{CFGSY15}}

Together with Flocchini, Godard, Santoro, and Yamashita in~\cite{CFGSY15} (brief anouncement~\cite{CFGSY12} and conference version~\cite{CFGSY13}), we explored
 the connections between dynamic networks and computability in terms of formal languages. Focusing on the particular case of time-varying graphs (see Section~\ref{sec:graph-models} for definitions), we showed how the manipulation of a time-varying graph as an {\em automata} makes it possible to study the power of {\em buffering} in dynamic networks (that is, the ability for a node to store and carry information before retransmitting it). 
In summary, we show that the set of languages that can be generated if only direct journeys are allowed (no waiting/buffering) contains all computable languages, whereas the set of languages if waiting is allowed (indirect journeys are possible) it is just the family of {\em regular} languages. In other words, when waiting is allowed, the expressivity of the environment drops drastically from being as powerful as a Turing machine, to becoming that of a Finite-State machine, which gives a (admittedly abstract) idea of the importance of buffering in dynamic networks.

\subsection{Automata based on time-varying graphs}
\label{sec:TVG-automata}

Given a dynamic network modeled as a time-varying graph $\G$, 
a journey  in $\G$ can be viewed as a word in the alphabet of the edge labels.
In this light,
the class of feasible journeys defines the language $L_f(\G)$ expressed by $\G$,
 where $f\in\{wait, nowait\}$ indicates whether or not  indirect journeys are considered
  feasible by the environment. Hence, a TVG $\G$ whose edges are labelled over  $\Sigma$, can be viewed as 
  a TVG-automaton
 $\TVGA({\G})=(\Sigma, S, I, {\cal E}, F)$ where
$\Sigma$ is the input {\em alphabet};
$S=V$ is the set of {\em states};
$I \subseteq S$ is the set of {\em initial states}; $F \subseteq S$ is the set of  {\em accepting states};
and ${\cal E}\subseteq S \times \lifetime \times \Sigma \times S \times \lifetime$ is the set of {\em transitions} such that $(s,t,a,s',t') \in {\cal E}$
 iff $\exists e=(s,s',a)\in E : \rho(e,t)=1,\zeta(e,t)=t'-t$.

Figure~\ref{fig:anbn} shows an example of a deterministic TVG-automaton $\TVGA({\G})$ that recognizes the context-free language $a^n b^n$ for $n \ge 1$ (using only direct journeys). 
The presence and latency of the edges of ${\G}$ are specified in Table \ref{functions}, where $p$ and $q$ are two distinct  prime numbers greater than 1,
 $v_0$ is the initial state, $v_2$ is the accepting state, and reading starts at
  time $t=1$.

\begin{figure}[h]
 \centering
 \begin{tikzpicture}[shorten >=1pt,node distance=2cm,auto, font=\footnotesize]
   \tikzstyle{every node}=[]
   \node[state,initial] (q0) {$v_0$};
   \node[state](q1) [right of= q0] {$v_1$};
   \node[state,accepting](q2) [below right of= q1] {$v_2$};
   
   \path[->] (q0) edge [loop above] node[left,yshift=-6pt] {$e_0$~} node {\tt a} ()
   edge node[below]{$e_1$} node {{\tt b}} (q1)
   (q1) edge [loop above] node[left,yshift=-6pt] {$e_2$~} node {{\tt b}} ()
   edge node[below]{$e_4$} node[inner sep=2pt] {{\tt b}} (q2);
   \path[->] (q0) edge[bend right] node[below]{$e_3$} node[inner sep=2pt] {{\tt b}} (q2);
 \end{tikzpicture}
 \caption[A TVG-automaton.]{\label{fig:anbn}A TVG-automaton $\TVGA({\G})$ such that $L_{nowait}(\G)=\{a^nb^n : n\ge 1\}$.~\cite{CFGSY15}}
\end{figure}

\begin{table}[tbh]
\begin{center}
\begin{tabular}{|c||c|c|c|}
\hline
$e$ & Presence $\rho(e,t)=1$ iff && Latency $\zeta(e,t)\ =$\\
\hline
\hline
$e_0$ &  always true & &  $(p-1)t$ \\
\hline
$e_1$ &       $t>p$ & & $(q-1)t$ \\
\hline
$e_2$&       $t\ne p^iq^{i-1}$,$i>1$ && $(q-1)t$ \\
\hline
$e_3$&       $t=p$ && any \\
\hline
$e_4$&      $t=p^iq^{i-1}$, $i>1$ &&any   \\
\hline
\end{tabular}
\caption{Presence and latency functions for the edges of $\G$.}
\label{functions}
 \end{center}
\end{table}
 
The reader may have noticed the basic principle employed here, which consists of using
latencies as a mean to {\em encode} words into time, and presences as a means to {\em select} them through opening
the appropriate edges at the appropriate time.

\subsection{Summary of the results}

We characterize the two sets of languages  
${\cal L}_{nowait} =\{L_{nowait} (\G) : \G \in {\cal U}\}$
   and     ${\cal L}_{wait} =\{L_{wait} (\G) : \G \in {\cal U}\}$, 
where $\cal U$ is the set of all time-varying graphs; 
that is, the languages expressed when waiting is or is not allowed.
For each of these two sets,
the complexity of  recognizing any language in the set (that is,  the computational power
needed by the accepting automaton)  defines  the level of
difficulty of the environment.
We first study the expressivity of time-varying graphs when waiting is {\em not} 
allowed---that is, the only feasible journeys are direct ones---and prove that 
the set  ${\cal L}_{nowait}$ contains  all computable  languages.
The proof is constructive and shows how any given Turing machine can be simulated by a time-varying graph.

\noindent
{\bf Corollary 3.3 in~\cite{CFGSY15}.}
  For any computable language L, there exists a time-varying graph \G such that $L=L_{nowait}(\G)$.

We next examined the expressivity of time-varying graphs if {\em indirect} journeys are allowed; that is, entities have the choice to wait for future opportunities of interaction rather than seizing only those that are directly available.
In striking contrast with the non-waiting case, the languages ${\cal L}_{wait}$
recognized by TVG-automata are  precisely the set of {\em regular} languages.
In other words, when waiting is no longer forbidden, the power of the accepting automaton  
 (\ie the difficulty of the environment, the power of the adversary), drops drastically from being
at least as powerful as a Turing machine, 
 to becoming that of a finite state machine. 
This gap is a measure of
the computational power of waiting. (This result may appear counterintuitive, but the fact that the power of the environment/adversary drops is in fact a good news for the algorithm, which is a more intuitive way to think of this result.)

\noindent
{\bf Theorem 4.13 in~\cite{CFGSY15}.}
  ${\cal L}_{wait}$ is the set of regular languages.

To better understand the power of waiting, 
we then turn our attention to  {\em bounded waiting}---that is, when 
indirect journeys are considered feasible if the pause 
between consecutive edges in the journeys have a bounded duration $d>0$.
In other words, at each step of the journey, waiting is allowed only for at most $d$ time units. 
We examine  the set  ${\cal L}_{wait[d]}$ of  the languages expressed by TVG in this case and prove the negative result that  the complexity of
the environment is not affected by allowing waiting for a limited amount of time,
that is, for any fixed $d\geq 0$, ${\cal L}_{wait[d]} = {\cal L}_{nowait}$. 
As a result, the power of the adversary is decreased only if it has no control over the
{\em length} of waiting, \ie if the waiting is unpredictable. 

\noindent
{\bf Theorem 5.1 in~\cite{CFGSY15}.}
 For any
fixed $d\geq 0$,
${\cal L}_{wait[d]} = {\cal L}_{nowait}$.

The basic idea of the proof  is to reuse the same technique as for the nowait case, but with a dilatation of time, 
\ie given the bound $d$, the edge schedule is time-expanded by a factor $d$ 
(and thus no new choice of transition is created compared to the no-waiting case).
As a result, the power of the adversary is decreased only if it has no control over the
length of waiting, \ie if the waiting is unpredictable. 

\begin{open}
  Understand the significance of this technical result. Depending on the point of view, say (1) network designer or (2) edge presence adversary, the interpretation of the result is fundamentally different. In (1) it is a good news, since it gives more expressivity to design interactions in the network, but in (2) it implies that a distributed algorithm may be forced to act in a way imposed by the environment. The overall implications are not yet well understood.
\end{open}

\begin{open}
  Do these results have analogues in the area of {\em timed automata}, which are automata where transitions are subject to time constraints. (See \eg~\cite{AD94}).
\end{open}

\section{Collection of curiosities}
\label{sec:curiosities}

We conclude this part of the document with a small collection of observations regarding concepts whose dynamic version (temporal version) seem different {\em in essence} to their static analogues, or whose implications are perhaps not yet well understood. Some of these observations are well-known or direct, others less obvious and some raise interesting questions. 

\subsection{Relation between paths and journeys}
\label{sec:diameter}
\begin{figure}[h]
  \centering
  \tikzsetnextfilename{tikz-footprint-connectivity}
  \begin{tikzpicture}[scale=1.4]
    \tikzstyle{every node}=[draw, circle, minimum size=11pt, inner sep=0pt]
    \path (0,.3) node (a){$a$};
    \path (1, 0) node (b){$b$};
    \path (2.2,0) node (c){$c$};
    \path (3.2,.3) node (d){$d$};
    \tikzstyle{every node}=[font=\scriptsize, inner sep=1pt]
    \draw (a)--node[midway, below, xshift=-2pt, yshift=-2pt]{$0$}(b);
    \draw (b)--node[midway, below]{$1$}(c);
    \draw (c)--node[midway, below, xshift=2pt, yshift=-2pt]{$0$}(d);
  \end{tikzpicture}
  \caption[An example TVG that is not ``connected over time''.]{\label{fig:connectedG} A graph whose footprint is connected, but which is not temporally connected.}
\end{figure}
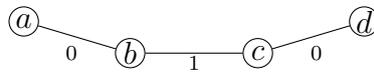

A first and simple fact is that having a connected footprint does not implies that the network is temporally connected, as exemplified by the graph on Figure~\ref{fig:connectedG} (from~\cite{CFQS12}).

\begin{figure}[h]
  \centering
  \begin{tikzpicture}[scale=.8]
    \path (0,0) coordinate (O); 
    \path (4,0) coordinate (O2); 

    \tikzstyle{every node}=[draw]
    \node[cloud, cloud puffs=16, cloud puff arc= 100, 
    minimum width=2.4cm, minimum height=3.0cm, aspect=1] at (O) (cl){};
    \node[cloud, cloud puffs=16, cloud puff arc= 100, 
    minimum width=2.4cm, minimum height=3.0cm, aspect=1] at (O2) (cl2){};
    \tikzstyle{every node}=[]
    \path (cl.south) node[below] {Informed nodes~~~~~};
    \path (cl2.south) node[below] {~~~~~Non-informed nodes};

    \tikzstyle{every node}=[defnode]
    \path (O)+(0:1) node (a){};
    \path (O)+(72:1) node (b){};
    \path (O)+(144:1) node (c){};
    \path (O)+(-144:1) node (d){};
    \path (O)+(-72:1) node (e){};

    \path (O2)+(0:-1) node (a2){};
    \path (O2)+(72:-1) node (b2){};
    \path (O2)+(144:-1) node (c2){};
    \path (O2)+(-144:-1) node (d2){};
    \path (O2)+(-72:-1) node (e2){};

    \draw (a)--(a2);
    \tikzstyle{every path}=[gray, dashed]
    \draw (a)--(b)--(e)--(a)--(c)--(b)--(d)--(c)--(e)--(d)--(a);
    \draw (a2)--(b2)--(e2)--(a2)--(c2)--(b2)--(d2)--(c2)--(e2)--(d2)--(a2);
  \end{tikzpicture}
  \caption{\label{fig:diameter}Small diameter vs. large temporal diameter.}
\end{figure}
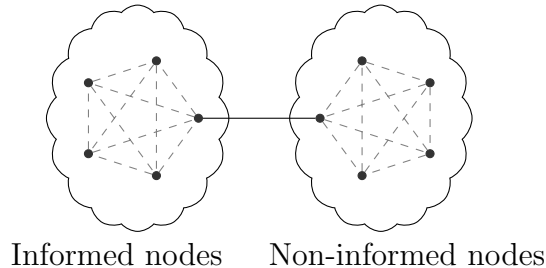

Clementi and Pasquale ask the reader in~\cite{CP11} whether one can define a graph whose temporal diameter is large despite the fact that every snapshot has small diameter.
This is indeed feasible. Using a broadcasting analogy in discrete time, one can design every snapshot as two cliques linked by a single edge, one of which represents informed nodes and the other non-informed nodes. In each step, the construction is updated based on the same rule (see Figure~\ref{fig:diameter}). Then, every snapshot has diameter $3$, while the temporal diameter is $n-1$.
Observe that the temporal diameter cannot exceed $n-1$ in this setting, because at least one new node is informed in each time step. The latter argument applies more generally when all the snapshots are connected. In fact, it is the reason why $\AC \subseteq \TCB$ in Section~\ref{sec:relations}.

\paragraph{Hardness of computing the temporal diameter.}

Godard and Mazauric~\cite{GM14} consider the problem of computing the temporal diameter of dynamic networks in an offline, but uncertain setting. More precisely, the network is described by an {\em unknown} sequence of graphs, each snapshot of which belongs to a {\em known} set of possible graphs. Every next graph in the sequence is independent from the previous (time-homogeneity). In this context, they show that computing the worst-case temporal diameter of the network is hard. In fact, it is inapproximable in general, and becomes approximable (but still NP-Hard) if all the possible graphs contain a common connected spanning subgraph. In contrast, computing the standard diameter of a standard graph has a low-polynomial complexity.

\subsection{Connected components}

There are several ways to consider the temporal analogue of a component in terms of journeys. If the lifetime is {\em infinite}, it seems reasonable to define them as maximal sets of nodes, each of which can reach the others {\em infinitely often} through a journey, either within bounded or unbounded time windows (this is the point of view we adopted in~\cite{GCLL15}, reviewed in Section~\ref{sec:carlos}). If the lifetime is {\em finite}, one may rather define components as maximal sets of nodes, each of which can reach the others {\em at least once}. 

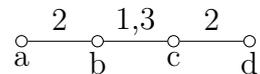
\begin{figure}[h]
  \centering
\begin{tikzpicture}
  \tikzstyle{every node}=[draw,circle,inner sep=1.5pt]
  \path (0,0) node (a) {};
  \path (1,0) node (b) {};
  \path (2,0) node (c) {};
  \path (3,0) node (d) {};
  \tikzstyle{every node}=[]
  \path (a) node[below] {a};
  \path (b) node[below] {b};
  \path (c) node[below] {c};
  \path (d) node[below] {d};
  \draw (a) -- node[above] {2} (b);
  \draw (b) -- node[above=-2pt] {1,3} (c);
  \draw (c) -- node[above] {2} (d);
\end{tikzpicture}
\caption{\label{fig:overlap}Two overlapping maximal components}
\end{figure}

The latter point of view is the one adopted by Bhadra and Ferreira in~\cite{BF03}, which seems to be the first significant study on the subject (2003). Bhadra and Ferreira further distinguish between {\em open} and {\em closed} components, depending on whether the journeys can travel through nodes outside the component. The existence of open components may seem strange, but it follows naturally from the fact that transitivity does not apply to journeys. Bhadra and Ferreira also observe that maximal components may overlap, which we illustrate using a minimal example on Figure~\ref{fig:overlap} where $\{a,b,c\}$ and $\{b,c,d\}$ are two maximal components ($a$ and $d$ cannot reach each other).
Again, this contrasts with static graphs, in which maximal components {\em partition} the network (\ie they define equivalence classes among the nodes). 

\subsubsection{Exponential number?}

In a static network (directed or undirected alike), the number of component cannot exceed $O(n)$ because components partition the nodes. The overlapping nature of components in dynamic graphs (see \eg Figure~\ref{fig:overlap}) raises the question of their number. In particular, can there be exponentially many maximal components?

\begin{figure}[h]
  \centering
  \subfigure[Moon and Moser graph]{
    \label{fig:moon-moser}
    ~~~
    \includegraphics[width=3.5cm]{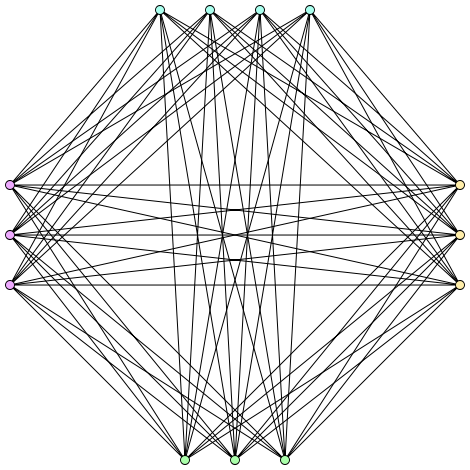}
    ~~~
  }
  \subfigure[Semaphore gadget to replace every edge]{
    ~~~
    \begin{tikzpicture}
      \tikzstyle{every node}=[defnode]
      \path (0,0) node (vv){};
      \path (0,2) node (uu){};

      \path (2,0) node (v){};
      \path (2,2) node (u){};
      \path (1.75,1) node (v'){};
      \path (2.25,1) node (u'){};

      \tikzstyle{every node}=[inner sep=4pt]

      \path (.3,1) node[right] {$\longrightarrow$};

      \path (u) node[above] {$u$};
      \path (v) node[below] {$v$};
      \path (uu) node[above] {$u$};
      \path (vv) node[below] {$v$};
      \path (u') node[right] {$u'$};
      \path (v') node[left] {$v'$};
      \draw (uu)--(vv);
      \draw (u)-- node[right]{1}(u');
      \draw (u')-- node[right]{2}(v);
      \draw (v)-- node[left]{1}(v');
      \draw (v')-- node[left]{2}(u);
    \end{tikzpicture}
    ~~~
  }
  \caption{\label{fig:exponential}Dynamic networks can have super-polynomially many maximal components. The argument holds for strict and non-strict journeys alike.}
\end{figure}
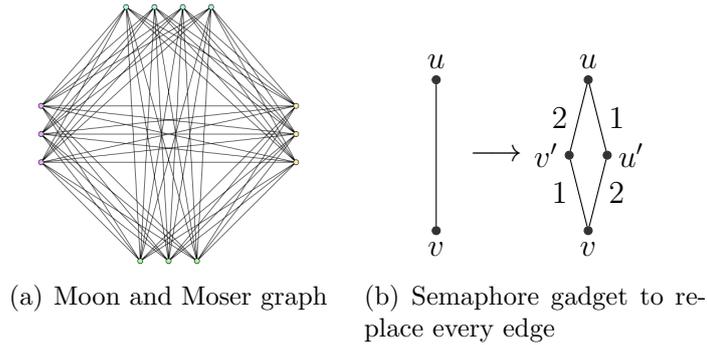

Let us first examine the case that only strict journeys are allowed (\ie, latency is positive). Take a footprint that contains exponentially many maximal cliques, such as a Moon and Moser graph (Figure~\ref{fig:moon-moser}), and make all of its edges appear at the same time for a single time unit of time (or $\zeta$ time). Then, only one-hop journeys may exist, implying that every maximal clique in the footprint induces a maximal component in the dynamic network. (This idea is inspired from a similar construct in~\cite{KKK00} used to show that {\em every} edge of the footprint might be necessary to achieve temporal connectivity.)

This argument for exponentially many connected components does not hold if non-strict journeys are possible (latency is neglected). With Klasing, Neggaz, Peters, and Renault~\cite{CKNPD16}, we designed a slightly more complex gadget that generalizes this argument to all types of journeys. Take again a Moon and Moser graph and replace every edge $(u,v)$ with four edges $(u,u')$, $(u',v)$, $(v,v')$, $(v',u)$, where $u'$ and $v'$ are new nodes (see Figure~\ref{fig:exponential}). Give these edges presence times $1,2,1,2$, respectively. (We call this a {\em semaphore} gadget.)
Then, the arrival at $u$ from $v$ (resp. at $v$ from $u$) is now too late for crossing another edge, turning original cliques into maximum components again. Moon and Moser graphs can have up to $3^{n/3}=2^{\Theta(n)}$ maximal cliques. Our solution induces a quadratic blow up in the number of vertices (two extra vertices per edge), leading to $2^{\Theta{(\sqrt{n})}}$ maximal components. 

\subsubsection{Computational complexity}

Bhadra and Ferreira prove in~\cite{BF03} that finding a maximum size component in a dynamic network is NP-hard, based on a reduction from the clique problem. The reduction gadgets are different from our constructs above and no counting argument is invocated. Note that, while finding the largest component is hard, testing if a given set of nodes are temporally connected is clearly not and can be inferred directly from the transitive closure of journeys (itself efficiently computable, see Section~\ref{sec:compTC}). 

On a related note, Viard et al.~\cite{VML17} look at the enumeration of temporal cliques, which in that case refer to cliques in the footprint of all $\G_{[t,t+\Delta]}$ for some $\Delta$ (sliding window), which are also exponentially many. This is however less surprisingly, since cliques can already be in exponential number in static graphs (unlike components).

\subsubsection{How about recurrent components?}

Interestingly, if we consider the above {\em recurrent} version of connected components, which requires that journeys exist infinitely often among the nodes, then transitivity comes back in the picture and every two overlapping components collapse into one.
Connected components in this case amount to (standard) components in the {\em eventual footprint}. Recurrent components are in this sense much closer to their classical analogue than non-recurrent ones.

\begin{avenue}
  Make a more systematic study of these questions.
\end{avenue}

\subsection{Menger's theorem}
\begin{figure}[h]
  \centering
  \subfigure[A static networks]{
    ~~~~
    \label{fig:menger-static}
    \begin{tikzpicture}[xscale=1.3]
      \tikzstyle{every node}=[defnode]
      \path (0,0) node (s){};
      \path (1,1) node (v1){};
      \path (.6,-1) node (v2){};
      \path (1.4,-1) node (v3){};
      \path (2,0) node (t){};
      
      \tikzstyle{every node}=[inner sep=2pt]
      \path (s) node[left=3pt] {$s$};
      \path (t) node[right=3pt] {$t$};
      \path (v1) node[above=2pt] {$v_1$};
      \path (v2) node[below left=0pt] {$v_2$};
      \path (v3) node[below right=0pt] {$v_3$};
      \draw (s) -- (v1);
      \draw (s) -- (v2);
      \draw (v1) -- (t);
      \draw (v1) -- (v2);
      \draw (v1) -- (v3);
      \draw (v2) -- (v3);
      \draw (v3) -- (t);
    \end{tikzpicture}
    ~~~~
  }
  \subfigure[A dynamic networks]{
    ~~~~
    \label{fig:menger-dynamic}
    \begin{tikzpicture}[xscale=1.3]
      \tikzstyle{every node}=[defnode]
      \path (0,0) node (s){};
      \path (1,1) node (v1){};
      \path (.6,-1) node (v2){};
      \path (1.4,-1) node (v3){};
      \path (2,0) node (t){};
      
      \tikzstyle{every node}=[inner sep=2pt]
      \path (s) node[left=3pt] {$s$};
      \path (t) node[right=3pt] {$t$};
      \path (v1) node[above=2pt] {$v_1$};
      \path (v2) node[below left=0pt] {$v_2$};
      \path (v3) node[below right=0pt] {$v_3$};
      \draw (s) -- node[above left] {$5$}(v1);
      \draw (s) -- node[below left] {$1$}(v2);
      \draw (v1) -- node[above right] {$3$}(t);
      \draw (v1) -- node[left] {$2$}(v2);
      \draw (v1) -- node[right] {$6$}(v3);
      \draw (v2) -- node[below] {$4$}(v3);
      \draw (v3) -- node[below right] {$7$}(t);
    \end{tikzpicture}
    ~~~~
  }
  \caption{\label{fig:menger} Illustration of Menger's theorem in both a static and a dynamic network (adapted from~\cite{KKK00})}
\end{figure}
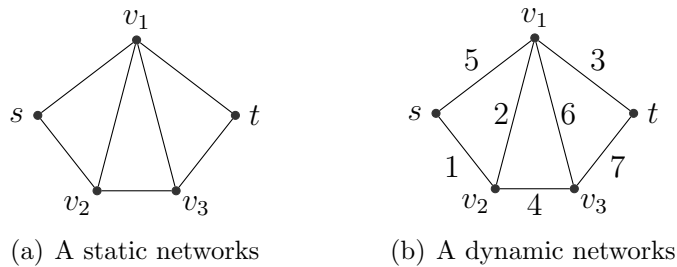

In the context of static networks, Menger's theorem~\cite{Menger27} states that the maximum number of node-disjoint paths between any two nodes $s$ and $t$ is equal to the minimum number of nodes needed to separate them. For example, in the {\em static} network depicted on Figure~\ref{fig:menger-static}, two disjoint paths exist between $s$ and $t$, namely $(s,v_1,t)$ and $(s,v_2,v_3,t)$, and indeed at least two nodes must be removed to disconnect $s$ and $t$ (\eg $v_1$ and $v_2$). In~\cite{Berman96}, Berman observe that the natural analogue of Menger's theorem does not hold in dynamic networks. We reproduce here an example from~\cite{KKK00} (Figure~\ref{fig:menger-dynamic}), showing a dynamic network in which the theorem fails: any two journeys must share a node, and yet two nodes are needed to separate $s$ and~$t$.
 Intuitively, the times are used to {\em force} a journey to use certain edges, which prevents journeys whose underlying paths are the same as above. Note that some {\em footprints} (characterized in~\cite{KKK00}) have the nice property that the resulting network remains Mengerian whatever the presence schedule of the edges.

\subsection{Degrees and density}
Bramas and Tixeuil~\cite{BT16} study the problem of data aggregation in a dynamic wireless sensor network where collisions may occur. The goal is to find a collision-free schedule that aggregates data from all the nodes to a single designated node in minimum time. This problem is known to be NP-complete in static networks, and so, even when the degree is restricted to 4~\cite{Chen05} (improved to 3 in~\cite{BT16}). Interestingly, Bramas and Tixeuil prove in the same paper that although the problem is trivial in static networks with degree at most 2, it is NP-hard in dynamic networks whose degree is never more than 2 at a single time. This result again tells us something about essential differences between static and dynamic contexts. In this case, however, it must be pointed out that the {\em footprint} of the graph used in their reduction does have degree more than 2. 

\begin{open}
\label{open:degree}
See if a different reduction exists that limits the degree to $2$ even in the footprint, making an even more essential separation between static and dynamic networks.
\end{open}

\paragraph{Cumulated \vs instant degrees.}
On a related note, with Barjon {\it et al.}, we characterized in~\cite{BCCJN14-en} the complexity of an algorithm for dynamic networks based on two types of densities: the maximum {\em instant} density (largest amount of edges existing at a same time) and the {\em cumulative} density (number of edges in the footprint), both being possibly much different.
The same distinction applies between both types of {\em degrees}, and in some sense, the above question (Open question~\ref{open:degree}) embodies this distinction.

\subsection{Optimality of journeys prefixes}

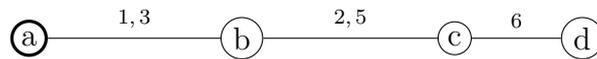
\begin{figure}[h]
  \centering
  \tikzsetnextfilename{tikz-fastest-prefix}
  \begin{tikzpicture}[xscale=2.8]
    \tikzstyle{every node}=[circle, draw, inner sep=1.8pt]
    \path (0,0) node[very thick] (a) {a};
    \path (1,0) node (b) {b};
    \path (2,0) node (c) {c};
    \path (2.6,0) node (d) {d};
    
    \tikzstyle{every node}=[font=\scriptsize, above]
    \draw (a) -- node {$1,3$} (b);
    \draw (b) -- node {$2,5$} (c);
    \draw (c) -- node {$6$} (d);
  \end{tikzpicture}
  \caption{\label{fig:line}A graph in which prefixes of fastest journeys are not themselves fastest.}
\end{figure}

Bhadra et al. observed in~\cite{BFJ03} that {\em foremost} journeys have the convenient feature that all prefixes of a foremost journey are themselves foremost journeys.
This feature allows us to define a distributed version of the problems in~\cite{BFJ03} in terms of foremost broadcast {\em trees}, all parts of which are themselves foremost (reviewed in Sections~\ref{sec:shfafo} and~\ref{sec:temporal-lags}). Unfortunately (and interestingly), this feature does not apply to the {\em fastest} metric, pushing us to define fastest broadcast trees in a different way (minimizing the overall span between first emission and last reception). We give  in Figure~\ref{fig:line} a concrete example where the fastest quality is not prefix-stable, which adapts a continuous-time example we gave in~\cite{CFMS14} into the discrete time setting (for simplicity).
To see this, consider the fastest journey from node $a$ to node $d$, namely $\{(ab,3),(bc,5),(cd,6)\}$. One of the prefixes of this journey, namely $\{(ab,3),(bc,5)\}$ is not itself a fastest journey, since $\{(ab,1),(bc,2)\}$ is faster. (Note that using the latter as an alternative prefix towards $d$ would lead to a non-fastest journey.)


\begin{thebibliography}{100}

\bibitem{AKM14}
Eric Aaron, Danny Krizanc, and Elliot Meyerson.
\newblock Dmvp: foremost waypoint coverage of time-varying graphs.
\newblock In {\em International Workshop on Graph-Theoretic Concepts in
  Computer Science}, pages 29--41. Springer, 2014.

\bibitem{mosbah-tree}
Sheila Abbas, Mohamed Mosbah, and Akka Zemmari.
\newblock Distributed computation of a spanning tree in a dynamic graph by
  mobile agents.
\newblock In {\em Proc. of IEEE Int. Conference on Engineering of Intelligent
  Systems (ICEIS)}, pages 1--6, 2006.

\bibitem{flows}
Eleni~C Akrida, Jurek Czyzowicz, Leszek Gasieniec, Lukasz Kuszner, and Paul~G
  Spirakis.
\newblock Temporal flows in temporal networks.
\newblock In {\em International Conference on Algorithms and Complexity}, pages
  43--54. Springer, 2017.

\bibitem{AGMS15}
Eleni~C Akrida, Leszek Gasieniec, George~B Mertzios, and Paul~G Spirakis.
\newblock On temporally connected graphs of small cost.
\newblock In {\em International Workshop on Approximation and Online Algorithms
  (WAOA)}, pages 84--96. Springer, 2015.

\bibitem{AMSZ18}
Eleni~C. Akrida, George~B. Mertzios, Paul~G. Spirakis, and Viktor Zamaraev.
\newblock Temporal vertex cover with a sliding time window.
\newblock {\em CoRR}, abs/1802.07103, 2018.

\bibitem{AF02}
David Aldous and Jim Fill.
\newblock Reversible markov chains and random walks on graphs, 2002.

\bibitem{padec}
Karine Altisen, Pierre Corbineau, and St{\'{e}}phane Devismes.
\newblock A framework for certified self-stabilization.
\newblock {\em Logical Methods in Computer Science}, 13(4), 2017.

\bibitem{AD94}
Rajeev Alur and David~L Dill.
\newblock A theory of timed automata.
\newblock {\em Theoretical computer science}, 126(2):183--235, 1994.

\bibitem{ACFQS11}
Fr\'ed\'eric Amblard, Arnaud Casteigts, Paola Flocchini, Walter Quattrociocchi,
  and Nicolas Santoro.
\newblock {On the Temporal Analysis of Scientific Network Evolution}.
\newblock In {\em {International Conference on Computational Aspects of Social
  Networks (CASoN)}}, pages 169--174, 2011.

\bibitem{AAD+06}
D.~Angluin, J.~Aspnes, Z.~Diamadi, M.~Fischer, and R.~Peralta.
\newblock Computation in networks of passively mobile finite-state sensors.
\newblock {\em Distributed Computing}, 18(4):235--253, 2006.

\bibitem{AE84}
B.~Awerbuch and S.~Even.
\newblock Efficient and reliable broadcast is achievable in an eventually
  connected network.
\newblock In {\em Proceedings of the 3rd ACM symposium on Principles of
  distributed computing (PODC)}, pages 278--281, Vancouver, Canada, 1984. ACM.

\bibitem{Awerbuch08}
Baruch Awerbuch, Israel Cidon, and Shay Kutten.
\newblock Optimal maintenance of a spanning tree.
\newblock {\em J. ACM}, 55(4):18:1--18:45, September 2008.

\bibitem{Baala03}
Hichem Baala, Olivier Flauzac, Jaafar Gaber, Marc Bui, and Tarek El-Ghazawi.
\newblock A self-stabilizing distributed algorithm for spanning tree
  construction in wireless ad hoc networks.
\newblock {\em Journal of Parallel and Distributed Computing}, 63:97--104,
  2003.

\bibitem{robots-coq}
Thibaut Balabonski, Pierre Courtieu, Lionel Rieg, S{\'{e}}bastien Tixeuil, and
  Xavier Urbain.
\newblock Certified gathering of oblivious mobile robots: Survey of recent
  results and open problems.
\newblock In {\em joint 22nd Int. Workshop on Formal Methods for Industrial
  Critical Systems - and - 17th Int. Workshop on Automated Verification of
  Critical Systems (FMICS-AVoCS)}, pages 165--181, 2017.

\bibitem{kuhn}
Philipp Bamberger, Fabian Kuhn, and Yannic Maus.
\newblock Local distributed algorithms in highly dynamic networks.
\newblock {\em arXiv preprint arXiv:1802.10199}, 2018.

\bibitem{BZ89}
Judit Bar-Ilan and Dror Zernik.
\newblock Random leaders and random spanning trees.
\newblock In {\em Workshop on Distributed Algorithms (WDAG)}, volume 392 of
  {\em Lecture Notes in Computer Science}, pages 1--12. Springer Berlin
  Heidelberg, 1989.

\bibitem{barjon}
Matthieu Barjon.
\newblock {\em Autour des groupes tol{\'{e}}rants aux d{\'{e}}lais dans les
  flottes mobiles communicantes. (On Delay-Tolerant Groups in Communicating
  Mobile Fleets)}.
\newblock PhD thesis, University of Bordeaux, France, 2016.

\bibitem{BCCJN14b}
Matthieu Barjon, Arnaud Casteigts, Serge Chaumette, Colette Johnen, and
  Y.~Neggaz.
\newblock Maintaining a spanning forest in highly dynamic networks: The
  synchronous case.
\newblock In {\em 18th Int. Conference on Principles of Distributed Systems
  (OPODIS)}, volume 8878 of {\em LNCS}, pages 277--292, 2014.

\bibitem{BCCJN18}
Matthieu Barjon, Arnaud Casteigts, Serge Chaumette, Colette Johnen, and
  Y.~Neggaz.
\newblock Maintaining a distributed spanning forest in highly dynamic networks.
\newblock {\em The Computer Journal}, 2018.

\bibitem{BCCJN14-en}
Matthieu Barjon, Arnaud Casteigts, Serge Chaumette, Colette Johnen, and
  Yessin~M. Neggaz.
\newblock Testing temporal connectivity in sparse dynamic graphs.
\newblock {\em CoRR}, abs/1404.7634, 2014.

\bibitem{BCCJN14-fr}
Matthieu Barjon, Arnaud Casteigts, Serge Chaumette, Colette Johnen, and
  Yessin~M. Neggaz.
\newblock Un algorithme de test pour la connexité temporelle dans les graphes
  dynamiques de faible densité.
\newblock In {\em 16e Rencontres Francophones sur les Aspects Algorithmiques de
  T\'el\'ecommunications (ALGOTEL)}, 2014.

\bibitem{BCF09}
H.~Baumann, P.~Crescenzi, and P.~Fraigniaud.
\newblock {Parsimonious flooding in dynamic graphs}.
\newblock In {\em Proceedings of the 28th ACM Symposium on Principles of
  Distributed Computing (PODC)}, pages 260--269, Calgary, Canada, 2009. ACM.

\bibitem{racetrack}
Michael~A Bekos, Till Bruckdorfer, Henry F{\"o}rster, Michael Kaufmann, Simon
  Poschenrieder, and Thomas St{\"u}ber.
\newblock Algorithms and insights for racetrack.
\newblock In {\em Proc. of 8th Int. Conf. on {FUN with algorithms},
  LIPIcs-Leibniz}, volume~49. Schloss Dagstuhl-Leibniz-Zentrum fuer Informatik,
  2016.

\bibitem{Berman96}
K.A. Berman.
\newblock {Vulnerability of scheduled networks and a generalization of Menger's
  Theorem}.
\newblock {\em Networks}, 28(3):125--134, 1996.

\bibitem{BBS13}
Thibault Bernard, Alain Bui, and Devan Sohier.
\newblock Universal adaptive self-stabilizing traversal scheme: Random walk and
  reloading wave.
\newblock {\em J. Parallel Distrib. Comput.}, 73(2):137--149, 2013.

\bibitem{BF03}
S.~Bhadra and A.~Ferreira.
\newblock Complexity of connected components in evolving graphs and the
  computation of multicast trees in dynamic networks.
\newblock In {\em Proceedings of the 2{nd} International Conference on Ad Hoc,
  Mobile and Wireless Networks (AdHoc-Now)}, pages 259--270, Montreal, Canada,
  2003. Springer.

\bibitem{BDP17}
Marjorie Bournat, Swan Dubois, and Franck Petit.
\newblock Computability of perpetual exploration in highly dynamic rings.
\newblock In {\em Distributed Computing Systems (ICDCS), 2017 IEEE 37th
  International Conference on}, pages 794--804. IEEE, 2017.

\bibitem{BMT16}
Quentin Bramas, Toshimitsu Masuzawa, and S{\'e}bastien Tixeuil.
\newblock {Distributed Online Data Aggregation in Dynamic Graphs}.
\newblock Research report, {Sorbonne Universit{\'e}s, UPMC Univ Paris 06, CNRS,
  LIP6 UMR 7606, 4 place Jussieu 75005 Paris. ; Osaka University, Japan},
  January 2016.

\bibitem{BT16}
Quentin Bramas and S{\'e}bastien Tixeuil.
\newblock The complexity of data aggregation in static and dynamic wireless
  sensor networks.
\newblock {\em Information and Computation}, 2016.

\bibitem{BDKP16}
Nicolas Braud-Santoni, Swan Dubois, Mohamed-Hamza Kaaouachi, and Franck Petit.
\newblock The next 700 impossibility results in time-varying graphs.
\newblock {\em International Journal of Networking and Computing}, 6(1):27--41,
  2016.

\bibitem{broutin}
Nicolas Broutin and Jean-Fran{\c{c}}ois Marckert.
\newblock A new encoding of coalescent processes: applications to the additive
  and multiplicative cases.
\newblock {\em Probability Theory and Related Fields}, 166(1-2):515--552, 2016.

\bibitem{BFJ03}
B.~{Bui-Xuan}, A.~Ferreira, and A.~Jarry.
\newblock Computing shortest, fastest, and foremost journeys in dynamic
  networks.
\newblock {\em International Journal of Foundations of Computer Science},
  14(2):267--285, April 2003.

\bibitem{asynchronous}
Janna Burman and Shay Kutten.
\newblock Time optimal asynchronous self-stabilizing spanning tree.
\newblock In Andrzej Pelc, editor, {\em Distributed Computing}, volume 4731 of
  {\em Lecture Notes in Computer Science}, pages 92--107. Springer Berlin
  Heidelberg, 2007.

\bibitem{CB13}
Rajmonda~Sulo Caceres and Tanya Berger-Wolf.
\newblock Temporal scale of dynamic networks.
\newblock In {\em Temporal Networks}, pages 65--94. Springer, 2013.

\bibitem{Cas06}
Arnaud Casteigts.
\newblock Model driven capabilities of the {DA-GRS} model.
\newblock In {\em Proc. of 1st Intl. Conference on Autonomic and Autonomous
  Systems (ICAS)}, pages 24--32. IEEE, 2006.

\bibitem{Cas07}
Arnaud Casteigts.
\newblock {\em Contribution \`a l'Algorithmique Distribu\'ee dans les R\'eseaux
  Mobiles Ad Hoc - Calculs Locaux et R\'e\'etiquetages de Graphes Dynamiques}.
\newblock PhD thesis, University of Bordeaux, 2007.

\bibitem{Cas18-hdr}
Arnaud Casteigts.
\newblock {\em A Journey Through Dynamic Networks (with Excursions)}.
\newblock Habilitation \`a diriger des recherches, University of Bordeaux, June
  2018.

\bibitem{CB10}
Arnaud Casteigts and Louise Bouchard.
\newblock Intégration de la dimension temporelle dans l'analyse des réseaux
  de santé en français.
\newblock Technical report, Institut de Santé des Populations d'Ottawa, 2010.

\bibitem{CCF09}
Arnaud Casteigts, Serge Chaumette, and Afonso Ferreira.
\newblock Characterizing topological assumptions of distributed algorithms in
  dynamic networks.
\newblock In {\em Proc. of 16th Intl. Conference on Structural Information and
  Communication Complexity (SIROCCO)}, volume 5869 of {\em LNCS}, pages
  126--140. Springer, 2009.

\bibitem{CCGP13}
Arnaud Casteigts, Serge Chaumette, Fr{\'e}d{\'e}ric Guinand, and Yoann
  Pign{\'e}.
\newblock Distributed maintenance of anytime available spanning trees in
  dynamic networks.
\newblock In {\em Proc. of 12th conf. on Adhoc, Mobile, and Wireless Networks
  (ADHOC-NOW)}, LNCS, pages 99--110, 2013.

\bibitem{CDPR17}
Arnaud Casteigts, Swan Dubois, Franck Petit, and John~Michael Robson.
\newblock Robustness in highly dynamic networks.
\newblock {\em CoRR}, abs/1703.03190, 2017.

\bibitem{CF13a}
Arnaud Casteigts and Paola Flocchini.
\newblock Deterministic algorithms in dynamic networks: Formal models and
  metrics.
\newblock Technical report, Commissioned by Defense Research and Development
  Canada (DRDC), 2013.

\bibitem{CF13b}
Arnaud Casteigts and Paola Flocchini.
\newblock Deterministic algorithms in dynamic networks: Problems, analysis, and
  algorithmic tools.
\newblock Technical report, Commissioned by Defense Research and Development
  Canada (DRDC), 2013.

\bibitem{CFGSY12}
Arnaud Casteigts, Paola Flocchini, E.~Godard, Nicolas Santoro, and Masafumi
  Yamashita.
\newblock Brief announcement: Waiting in dynamic networks.
\newblock In {\em Proceedings of the 31th ACM Symposium on Principles of
  Distributed Computing (PODC)}, 2012.

\bibitem{CFGSY15}
Arnaud Casteigts, Paola Flocchini, E.~Godard, Nicolas Santoro, and Masafumi
  Yamashita.
\newblock On the expressivity of time-varying graphs.
\newblock {\em Theoretical Computer Science}, 590:27--37, 2015.

\bibitem{CFGSY13}
Arnaud Casteigts, Paola Flocchini, Emmanuel Godard, Nicolas Santoro, and
  Masafumi Yamashita.
\newblock Expressivity of time-varying graphs.
\newblock In {\em 19th International Symposium on Fundamentals of Computation
  Theory (FCT)}, 2013.

\bibitem{CFMS10}
Arnaud Casteigts, Paola Flocchini, Bernard Mans, and Nicolas Santoro.
\newblock Deterministic computations in time-varying graphs: Broadcasting under
  unstructured mobility.
\newblock In {\em 6th IFIP International Conference on Theoretical Computer
  Science (TCS)}, pages 111--124, 2010.

\bibitem{CFMS11}
Arnaud Casteigts, Paola Flocchini, Bernard Mans, and Nicolas Santoro.
\newblock Measuring temporal lags in delay-tolerant networks.
\newblock In {\em 25th IEEE Intl. Parallel \& Distributed Processing Symposium
  (IPDPS)}, pages 209--218, 2011.

\bibitem{CFMS12b}
Arnaud Casteigts, Paola Flocchini, Bernard Mans, and Nicolas Santoro.
\newblock Building fastest broadcast trees in periodically-varying graphs.
\newblock {\em CoRR}, abs/1204.3058, 2012.

\bibitem{CFMS14}
Arnaud Casteigts, Paola Flocchini, Bernard Mans, and Nicolas Santoro.
\newblock Measuring temporal lags in delay-tolerant networks.
\newblock {\em IEEE Transactions on Computers}, 63(2), 2014.

\bibitem{CFMS15}
Arnaud Casteigts, Paola Flocchini, Bernard Mans, and Nicolas Santoro.
\newblock Shortest, fastest, and foremost broadcast in dynamic networks.
\newblock {\em Int. Journal of Foundations of Computer Science},
  26(4):499--522, 2015.

\bibitem{CFQS11}
Arnaud Casteigts, Paola Flocchini, Walter Quattrociocchi, and Nicolas Santoro.
\newblock Time-varying graphs and dynamic networks.
\newblock In {\em 10th International Conference on Ad Hoc Networks and Wireless
  (ADHOC-NOW)}, pages 346--359, 2011.

\bibitem{CFQS12}
Arnaud Casteigts, Paola Flocchini, Walter Quattrociocchi, and Nicolas Santoro.
\newblock Time-varying graphs and dynamic networks.
\newblock {\em International Journal of Parallel, Emergent and Distributed
  Systems}, 27(5):387--408, 2012.

\bibitem{CKNP15}
Arnaud Casteigts, R.~Klasing, Yessin Neggaz, and J.~Peters.
\newblock Efficiently testing {T-Interval} connectivity in dynamic graphs.
\newblock In {\em 9th Int. Conference on Algorithms and Complexity (CIAC)},
  volume 9079 of {\em LNCS}, pages 89--100, 2015.

\bibitem{CKNP17}
Arnaud Casteigts, Ralf Klasing, Yessin Neggaz, and Joseph Peters.
\newblock A generic framework for computing parameters of sequence-based
  dynamic graphs.
\newblock In {\em Proc. of 24th Intl. Conference on Structural Information and
  Communication Complexity (SIROCCO)}, 2017.

\bibitem{CKNP18}
Arnaud Casteigts, Ralf Klasing, Yessin Neggaz, and Joseph Peters.
\newblock Computing parameters of sequence-based dynamic graphs.
\newblock {\em Theory of Computing Systems}, 2018.

\bibitem{CKNPD16}
Arnaud Casteigts, Ralf Klasing, Yessin Neggaz, Joseph Peters, and David
  Renault.
\newblock Private communication, 2016.

\bibitem{CMM11}
Arnaud Casteigts, Bernard Mans, and Luke Mathieson.
\newblock On the feasibility of maintenance algorithms in dynamic graphs.
\newblock {\em CoRR}, abs/1107.2722, 2011.

\bibitem{coq09}
P.~Cast{\'e}ran, V.~Filou, and M.~Mosbah.
\newblock Certifying distributed algorithms by embedding local computation
  systems in the coq proof assistant.
\newblock In {\em Proc. of Symbolic Computation in Software Science (SCSS'09)},
  2009.

\bibitem{coq11}
Pierre Cast{\'{e}}ran and Vincent Filou.
\newblock Tasks, types and tactics for local computation systems.
\newblock {\em Stud. Inform. Univ.}, 9(1):39--86, 2011.

\bibitem{ChMMD08}
A.~Chaintreau, A.~Mtibaa, L.~Massoulie, and C.~Diot.
\newblock The diameter of opportunistic mobile networks.
\newblock {\em Communications Surveys \& Tutorials}, 10(3):74--88, 2008.

\bibitem{chalopin}
J{\'e}r{\'e}mie Chalopin.
\newblock {\em Algorithmique distribu{\'e}e, calculs locaux et homomorphismes
  de graphes}.
\newblock PhD thesis, Bordeaux 1, 2006.

\bibitem{paulusma}
J{\'{e}}r{\'{e}}mie Chalopin and Dani{\"{e}}l Paulusma.
\newblock Graph labelings derived from models in distributed computing: {A}
  complete complexity classification.
\newblock {\em Networks}, 58(3):207--231, 2011.

\bibitem{matthias}
Bernadette Charron-Bost, Matthias F{\"u}gger, and Thomas Nowak.
\newblock Approximate consensus in highly dynamic networks: The role of
  averaging algorithms.
\newblock In {\em International Colloquium on Automata, Languages, and
  Programming}, pages 528--539. Springer, 2015.

\bibitem{squad}
Bernadette Charron-Bost and Shlomo Moran.
\newblock The firing squad problem revisited.
\newblock In {\em LIPIcs-Leibniz International Proceedings in Informatics},
  volume~96. Schloss Dagstuhl-Leibniz-Zentrum fuer Informatik, 2018.

\bibitem{heard-of}
Bernadette Charron-Bost and Andr{\'e} Schiper.
\newblock The heard-of model: computing in distributed systems with benign
  faults.
\newblock {\em Distributed Computing}, 22(1):49--71, 2009.

\bibitem{Chen05}
Xujin Chen, Xiaodong Hu, and Jianming Zhu.
\newblock Minimum data aggregation time problem in wireless sensor networks.
\newblock In {\em International conference on mobile ad-hoc and sensor
  networks}, pages 133--142. Springer, 2005.

\bibitem{CleMMPS08}
A.~Clementi, C.~Macci, A.~Monti, F.~Pasquale, and R.~Silvestri.
\newblock {Flooding time in edge-markovian dynamic graphs}.
\newblock In {\em Proceedings of the 27th ACM Symposium on Principles of
  distributed computing (PODC)}, pages 213--222, Toronto, Canada, 2008. ACM.

\bibitem{CP11}
A.~Clementi and F.~Pasquale.
\newblock Information spreading in dynamic networks: An analytical approach.
\newblock In S.~Nikoletseas and J.D.P. Rolim, editors, {\em Theoretical Aspects
  of Distributed Computing in Sensor Networks}, chapter~19, pages 591--619.
  Springer, 2011.

\bibitem{CEOR12}
Colin Cooper, Robert Els{\"a}sser, Hirotaka Ono, and Tomasz Radzik.
\newblock Coalescing random walks and voting on graphs.
\newblock In {\em Proceedings of the 31st ACM symposium on Principles of
  distributed computing (PODC)}, pages 47--56. ACM, 2012.

\bibitem{coulouma}
{\'{E}}tienne Coulouma, Emmanuel Godard, and Joseph~G. Peters.
\newblock A characterization of oblivious message adversaries for which
  consensus is solvable.
\newblock {\em Theor. Comput. Sci.}, 584:80--90, 2015.

\bibitem{DKP15}
Swan Dubois, Mohamed{-}Hamza Kaaouachi, and Franck Petit.
\newblock Enabling minimal dominating set in highly dynamic distributed
  systems.
\newblock In {\em 17th International Symposium on Stabilization, Safety, and
  Security of Distributed Systems ({SSS})}, pages 51--66, 2015.

\bibitem{Graphstream}
A.~Dutot, F.~Guinand, D.~Olivier, and Y.~Pign{\'e}.
\newblock {Graphstream: A tool for bridging the gap between complex systems and
  dynamic graphs}.
\newblock {\em EPNACS: Emergent Properties in Natural and Artificial Complex
  Systems}, 2007.

\bibitem{eppstein90}
David Eppstein, Giuseppe~F. Italiano, Roberto Tamassia, Robert~E. Tarjan,
  Jeffery~R. Westbrook, and Moti Yung.
\newblock {Maintenance of a minimum spanning forest in a dynamic planar graph}.
\newblock In {\em Proc. 1st Symp. Discrete Algorithms}, pages 1--11. ACM and
  SIAM, January 1990.

\bibitem{FatenGeneral}
Faten Fakhfakh, Mohamed Tounsi, Ahmed~Hadj Kacem, and Mohamed Mosbah.
\newblock Towards a formal model for dynamic networks through refinement and
  evolving graphs.
\newblock In {\em Software Engineering, Artificial Intelligence, Networking and
  Parallel/Distributed Computing 2015}, pages 227--243. Springer, 2016.

\bibitem{FatenForest}
Faten Fakhfakh, Mohamed Tounsi, Mohamed Mosbah, Ahmed~Hadj Kacem, and Dominique
  Mery.
\newblock A formal approach for maintaining forest topologies in dynamic
  networks.
\newblock In {\em 16th IEEE/ACIS International Conference on Computer and
  Information Science (ICIS 2017)}, 2017.

\bibitem{AMMZ12}
Antonio Fern{\'a}ndez~Anta, Alessia Milani, Miguel~A. Mosteiro, and Shmuel
  Zaks.
\newblock Opportunistic information dissemination in mobile ad-hoc networks:
  the profit of global synchrony.
\newblock {\em Distributed Computing}, 25(4):279--296, 2012.

\bibitem{Fer04}
A.~Ferreira.
\newblock Building a reference combinatorial model for {MANETs}.
\newblock {\em IEEE Network}, 18(5):24--29, 2004.

\bibitem{Fer02}
Afonso Ferreira.
\newblock On models and algorithms for dynamic communication networks: The case
  for evolving graphs.
\newblock In {\em In Proc. ALGOTEL}, 2002.

\bibitem{FV02}
Afonso Ferreira and Laurent Viennot.
\newblock A note on models, algorithms, and data structures for dynamic
  communication networks.
\newblock Technical Report 4403, INRIA, 2002.

\bibitem{FKMS12a}
P.~Flocchini, M.~Kellett, P.~Mason, and N.~Santoro.
\newblock Searching for black holes in subways.
\newblock {\em Theory of Computing Systems}, 50(1):158--184, 2012.

\bibitem{FKMS12d}
P.~Flocchini, M.~Kellett, P.C. Mason, and N.~Santoro.
\newblock Searching for black holes in subways.
\newblock {\em Theory of Computing Systems}, 50(1):158--184, 2012.

\bibitem{FMS09}
P.~Flocchini, B.~Mans, and N.~Santoro.
\newblock Exploration of periodically varying graphs.
\newblock In {\em Proceedings of 20th International Symposium on Algorithms and
  Computation (ISAAC)}, pages 534--543, 2009.

\bibitem{FMS13}
Paola Flocchini, Bernard Mans, and Nicola Santoro.
\newblock On the exploration of time-varying networks.
\newblock {\em Theoretical Computer Science}, 469:53--68, 2013.

\bibitem{nieder2}
Till Fluschnik, Hendrik Molter, Rolf Niedermeier, and Philipp Zschoche.
\newblock Temporal graph classes: A view through temporal separators.
\newblock {\em arXiv preprint arXiv:1803.00882}, 2018.

\bibitem{frederickson85}
Greg~N Frederickson.
\newblock Data structures for on-line updating of minimum spanning trees, with
  applications.
\newblock {\em SIAM Journal on Computing}, 14(4):781--798, 1985.

\bibitem{gaumont}
No{\'e} Gaumont, Cl{\'e}mence Magnien, and Matthieu Latapy.
\newblock Finding remarkably dense sequences of contacts in link streams.
\newblock {\em Social Network Analysis and Mining}, 6(1):87, 2016.

\bibitem{GVFWL16}
No{\'e} Gaumont, Tiphaine Viard, Rapha{\"e}l Fournier-S’Niehotta, Qinna Wang,
  and Matthieu Latapy.
\newblock Analysis of the temporal and structural features of threads in a
  mailing-list.
\newblock In {\em Complex Networks VII}, pages 107--118. Springer, 2016.

\bibitem{godard}
Emmanuel Godard.
\newblock {\em Distributed Computability in Communication Networks}.
\newblock Habilitation à diriger des recherches en mathématiques et en
  informatique, Université Aix-Marseille, 2015.

\bibitem{GM14}
Emmanuel Godard and Dorian Mazauric.
\newblock Computing the dynamic diameter of non-deterministic dynamic networks
  is hard.
\newblock In {\em Algorithms for Sensor Systems - 10th International Symposium
  on Algorithms and Experiments for Sensor Systems, Wireless Networks and
  Distributed Robotics, {ALGOSENSORS} 2014, Wroclaw, Poland, September 12,
  2014, Revised Selected Papers}, pages 88--102, 2014.

\bibitem{GCLL15}
Carlos G\'omez-Calzado, Arnaud Casteigts, Mikel Larrea, and Alberto Lafuente.
\newblock A connectivity model for agreement in dynamic systems.
\newblock In {\em 21st Int. Conference on Parallel Processing (EUROPAR)},
  volume 9233 of {\em LNCS}, pages 333--345, 2015.

\bibitem{GAS11}
F.~Greve, L.~Arantes, and P.~Sens.
\newblock What model and what conditions to implement unreliable failure
  detectors in dynamic networks?
\newblock In {\em Proceedings of the 3rd International Workshop on Theoretical
  Aspects of Dynamic Distributed Systems}, pages 13--17, Rome, Italy, 2011.
  Springer.

\bibitem{HG97}
F.~Harary and G.~Gupta.
\newblock {Dynamic graph models}.
\newblock {\em Mathematical and Computer Modelling}, 25(7):79--88, 1997.

\bibitem{Holme05}
P.~Holme.
\newblock {Network reachability of real-world contact sequences}.
\newblock {\em Physical Review E}, 71(4):46119, 2005.

\bibitem{H12}
Petter Holme and Jari Saram{\"a}ki.
\newblock Temporal networks.
\newblock {\em Physics reports}, 519(3):97--125, 2012.

\bibitem{IW11}
D.~Ilcinkas and A.~Wade.
\newblock On the power of waiting when exploring public transportation systems.
\newblock {\em Proceedings of the 15th International Conference on Principles
  of Distributed Systems (OPODIS)}, pages 451--464, 2011.

\bibitem{IJ90}
Amos Israeli and Marc Jalfon.
\newblock Token management schemes and random walks yield self-stabilizing
  mutual exclusion.
\newblock In {\em Proceedings of the ninth annual ACM symposium on Principles
  of distributed computing}, pages 119--131. ACM, 1990.

\bibitem{JMR10}
P.~Jacquet, B.~Mans, and G.~Rodolakis.
\newblock {Information propagation speed in mobile and delay tolerant
  networks}.
\newblock {\em IEEE Transactions on Information Theory}, 56(1):5001--5015,
  2010.

\bibitem{Jarry05}
Aubin Jarry.
\newblock {\em Connexit{\'{e}} dans les r{\'{e}}seaux de
  t{\'{e}}l{\'{e}}communications. (Connectivity in telecommunication
  networks)}.
\newblock PhD thesis, University of Nice Sophia Antipolis, France, 2005.

\bibitem{Kel12}
M.~Kellett.
\newblock {\em Black hole search in the network and subway models}.
\newblock PhD thesis, University of Ottawa, 2012.

\bibitem{KKK00}
D.~Kempe, J.~Kleinberg, and A.~Kumar.
\newblock {Connectivity and inference problems for temporal networks}.
\newblock In {\em Proceedings of 32nd ACM Symposium on Theory of Computing},
  pages 504--513, Portland, USA, 2000. ACM.

\bibitem{KerO09}
A.~Ker{\"a}nen and J.~Ott.
\newblock {DTN over aerial carriers}.
\newblock In {\em Proceedings of 4th ACM Workshop on Challenged Networks},
  pages 67--76, Beijing, China, 2009. ACM.

\bibitem{KKW08}
G.~Kossinets, J.~Kleinberg, and D.~Watts.
\newblock {The structure of information pathways in a social communication
  network}.
\newblock In {\em Proceedings of 14th International Conference on Knowledge
  Discovery and Data Mining (KDD)}, pages 435--443, Las Vegas, USA, 2008. ACM.

\bibitem{Kostakos09}
V.~Kostakos.
\newblock {Temporal graphs}.
\newblock {\em Physica A}, 388(6):1007--1023, 2009.

\bibitem{synchronous}
Alex Kravchik and Shay Kutten.
\newblock Time optimal synchronous self stabilizing spanning tree.
\newblock In Yehuda Afek, editor, {\em Distributed Computing}, volume 8205 of
  {\em Lecture Notes in Computer Science}, pages 91--105. Springer Berlin
  Heidelberg, 2013.

\bibitem{KKB12}
Gautier Krings, M{\'a}rton Karsai, Sebastian Bernhardsson, Vincent~D Blondel,
  and Jari Saram{\"a}ki.
\newblock Effects of time window size and placement on the structure of an
  aggregated communication network.
\newblock {\em EPJ Data Science}, 1(1):4, 2012.

\bibitem{KLO10}
Fabian Kuhn, Nancy Lynch, and Rotem Oshman.
\newblock Distributed computation in dynamic networks.
\newblock In {\em Proceedings of the 42nd ACM symposium on Theory of computing
  (STOC)}, pages 513--522. ACM, 2010.

\bibitem{KO11}
Fabian Kuhn and Rotem Oshman.
\newblock Dynamic networks: models and algorithms.
\newblock {\em ACM SIGACT News}, 42(1):82--96, 2011.

\bibitem{Laplace12}
R{\'e}mi Laplace.
\newblock {\em Applications et services DTN pour flotte collaborative de
  drones}.
\newblock PhD thesis, Universit{\'e} de Bordeaux, 2012.

\bibitem{LVM17}
Matthieu Latapy, Tiphaine Viard, and Cl{\'{e}}mence Magnien.
\newblock Stream graphs and link streams for the modeling of interactions over
  time.
\newblock {\em CoRR}, abs/1710.04073, 2017.

\bibitem{LCF15}
Yannick L{\'e}o, Christophe Crespelle, and Eric Fleury.
\newblock {Non-Altering Time Scales for Aggregation of Dynamic Networks into
  Series of Graphs}.
\newblock In {\em {11th International Conference on emerging Networking
  EXperiments and Technologies -- CoNEXT 2015}}, 11th International Conference
  on emerging Networking EXperiments and Technologies -- CoNEXT 2015,
  Heidelberg, Germany, 2015.

\bibitem{LMS99}
I.~Litovsky, Y.~M\'etivier, and E.~Sopena.
\newblock {\em Graph Relabelling Systems and Distributed Algorithms}.
\newblock {\em H. Ehrig, H.J. Kreowski, U. Montanari and G. Rozenberg (Eds.),
  {\em Handbook of Graph Grammars and Computing by Graph Transformation}},
  pages 1--53, 1999.

\bibitem{LW09b}
C.~Liu and J.~Wu.
\newblock Scalable routing in cyclic mobile networks.
\newblock {\em IEEE Transactions on Parallel and Distributed Systems},
  20(9):1325--1338, 2009.

\bibitem{LDFS16}
Giuseppe Antonio~Di Luna, Stefan Dobrev, Paola Flocchini, and Nicola Santoro.
\newblock Live exploration of dynamic rings.
\newblock In {\em 36th {IEEE} International Conference on Distributed Computing
  Systems, {ICDCS} 2016, Nara, Japan, June 27-30, 2016}, pages 570--579, 2016.

\bibitem{permanent}
Subhrangsu Mandal and Arobinda Gupta.
\newblock Approximation algorithms for permanent dominating set problem on
  dynamic networks.
\newblock In {\em International Conference on Distributed Computing and
  Internet Technology}, pages 265--279. Springer, 2018.

\bibitem{treewidth}
Bernard Mans and Luke Mathieson.
\newblock On the treewidth of dynamic graphs.
\newblock {\em Theoretical Computer Science}, 554:217--228, 2014.

\bibitem{MG12}
F.~Marchand~de Kerchove and F.~Guinand.
\newblock Strengthening topological conditions for relabeling algorithms in
  evolving graphs.
\newblock Technical report, Universit{\'e} du Havre, 2012.

\bibitem{Menger27}
Karl Menger.
\newblock Zur allgemeinen kurventheorie.
\newblock {\em Fundamenta Mathematicae}, 10(1):96--115, 1927.

\bibitem{MS18}
Othon Michail and Paul~G Spirakis.
\newblock Elements of the theory of dynamic networks.
\newblock {\em Communications of the ACM}, 61(2):72--72, 2018.

\bibitem{NS93}
M.~Naor and L.~Stockmeyer.
\newblock What can be computed locally?
\newblock In {\em Proceedings of the 25th annual ACM symposium on Theory of
  computing}, pages 184--193. ACM, 1993.

\bibitem{Neggaz16}
Yessin~M. Neggaz.
\newblock {\em Automatic Classification of Dynamic Graphs}.
\newblock PhD thesis, University of Bordeaux, 2016.

\bibitem{OW05}
R.~O'Dell and R.~Wattenhofer.
\newblock Information dissemination in highly dynamic graphs.
\newblock In {\em Proceedings of the Joint Workshop on Foundations of Mobile
  Computing (DIALM-POMC)}, pages 104--110, Cologne, Germany, 2005. ACM.

\bibitem{PS96}
Alessandro Panconesi and Aravind Srinivasan.
\newblock On the complexity of distributed network decomposition.
\newblock {\em Journal of Algorithms}, 20(2):356--374, 1996.

\bibitem{PCGC10}
Yoann Pign\'e, Arnaud Casteigts, Fr\'ed\'eric Guinand, and Serge Chaumette.
\newblock Construction et maintien d'une for\^et couvrante dans un r\'eseau
  dynamique.
\newblock In {\em 12e Rencontres Francophones sur les Aspects Algorithmiques de
  T\'el\'ecommunications (ALGOTEL)}, 2010.

\bibitem{tredan}
Yvonne~Anne Pignolet, Matthieu Roy, Stefan Schmid, and Gilles Tredan.
\newblock The many faces of graph dynamics.
\newblock {\em Journal of Statistical Mechanics: Theory and Experiment},
  2017(6):063401, 2017.

\bibitem{RBK07}
R.~Ramanathan, P.~Basu, and R.~Krishnan.
\newblock Towards a formalism for routing in challenged networks.
\newblock In {\em Proceedings of 2nd ACM Workshop on Challenged Networks
  (CHANTS)}, pages 3--10, 2007.

\bibitem{message}
Michel Raynal.
\newblock Message adversaries.
\newblock {\em Encyclopedia of Algorithms}, pages 1--6, 2014.

\bibitem{RSCW14}
Michel Raynal, Julien Stainer, Jiannong Cao, and Weigang Wu.
\newblock A simple broadcast algorithm for recurrent dynamic systems.
\newblock In {\em 28th {IEEE} International Conference on Advanced Information
  Networking and Applications, {AINA} 2014, Victoria, BC, Canada, May 13-16,
  2014}, pages 933--939, 2014.

\bibitem{reif87}
John~H Reif.
\newblock A topological approach to dynamic graph connectivity.
\newblock {\em Information Processing Letters}, 25(1):65--70, 1987.

\bibitem{RC17}
G.~{Rossetti} and R.~{Cazabet}.
\newblock {Community Discovery in Dynamic Networks: a Survey}.
\newblock {\em ArXiv e-prints}, July 2017.

\bibitem{Santoro16}
Nicola Santoro.
\newblock Time to change: On distributed computing in dynamic networks
  (keynote).
\newblock In {\em LIPIcs-Leibniz International Proceedings in Informatics},
  volume~46. Schloss Dagstuhl-Leibniz-Zentrum fuer Informatik, 2016.

\bibitem{widmayer}
Nicola Santoro and Peter Widmayer.
\newblock Time is not a healer.
\newblock In {\em Annual Symposium on Theoretical Aspects of Computer Science},
  pages 304--313. Springer, 1989.

\bibitem{SQFCA11}
Nicolas Santoro, Walter Quattrociocchi, Paola Flocchini, Arnaud Casteigts, and
  Fr\'ed\'eric Amblard.
\newblock Time-varying graphs and social network analysis: Temporal indicators
  and metrics.
\newblock In {\em 3rd AISB Social Networks and Multiagent Systems Symposium
  (SNAMAS)}, pages 32--38, 2011.

\bibitem{crawdad}
James Scott, Richard Gass, Jon Crowcroft, Pan Hui, Christophe Diot, and
  Augustin Chaintreau.
\newblock Crawdad dataset cambridge/haggle (v. 2009-05-29).
\newblock {\em CRAWDAD wireless network data archive}, 2009.

\bibitem{SE81}
Yossi Shiloach and Shimon Even.
\newblock An on-line edge-deletion problem.
\newblock {\em Journal of the ACM (JACM)}, 28(1):1--4, 1981.

\bibitem{TSM+09}
J.~Tang, S.~Scellato, M.~Musolesi, C.~Mascolo, and V.~Latora.
\newblock {Small-world behavior in time-varying graphs}.
\newblock {\em Phys. Rev. E}, 81(5), 2010.

\bibitem{Trudeau13}
Richard~J Trudeau.
\newblock {\em Introduction to graph theory}.
\newblock Courier Corporation, 2013.

\bibitem{VML17}
Tiphaine Viard, Cl{\'e}mence Magnien, and Matthieu Latapy.
\newblock Enumerating maximal cliques in link streams with durations.
\newblock {\em arXiv preprint arXiv:1712.06970}, 2017.

\bibitem{WD98}
Duncan~J Watts and Steven~H Strogatz.
\newblock Collective dynamics of small-world networks.
\newblock {\em nature}, 393(6684):440--442, 1998.

\bibitem{WDCG12}
John Whitbeck, Marcelo Dias~de Amorim, Vania Conan, and Jean-Loup Guillaume.
\newblock Temporal reachability graphs.
\newblock In {\em Proc. of MOBICOM}, pages 377--388. ACM, 2012.

\bibitem{YamK96}
Masafumi Yamashita and Tiko Kameda.
\newblock Computing on anonymous networks: Part {I} and {II}.
\newblock {\em IEEE Trans. on Par. and Distributed Systems}, 7(1):69 -- 96,
  1996.

\bibitem{nieder}
Philipp Zschoche, Till Fluschnik, Hendrik Molter, and Rolf Niedermeier.
\newblock The computational complexity of finding separators in temporal
  graphs.
\newblock {\em arXiv preprint arXiv:1711.00963}, 2017.

\end{thebibliography}
\end{document}